\documentclass[twocolumn,secnumarabic,amssymb, nobibnotes, aps, prd]{revtex4-2}
\usepackage{amsmath}
\usepackage{amsfonts}
\usepackage{amssymb}
\usepackage{graphics}
\usepackage{graphicx}
\providecommand{\U}[1]{\protect\rule{.1in}{.1in}}

\begin{document}
	
\title{Teleportation and Entanglement Swapping of Continuous Quantum Variables of Microwave Radiation}
\author{Baleegh Abdo}
\author{William Shanks}
\author{Oblesh Jinka}
\author{J. R. Rozen}
\author{Jason Orcutt}
\affiliation{IBM Quantum, IBM Research Center, Yorktown Heights, New York 10598, USA.}
\date{\today}

\begin{abstract}
	Quantum communication is needed to build powerful quantum computers and establish reliable quantum networks. At its basis lies the ability to generate and distribute entanglement to separate quantum systems, which can be used to run remote quantum operations on them or teleport quantum states from one system to another with the help of classical channels. To this end, it is useful to harness the resource of continuous-variable (CV) entanglement since it can be efficiently and unconditionally produced by squeezing light in a nonlinear medium and can be easily manipulated, distributed, and measured using standard components. While various aspects of CV-based quantum communication have been successfully demonstrated in the optical domain, some key capabilities, such as entanglement swapping, have been lacking in the microwave domain. Here, we demonstrate three key elements of CV-based microwave quantum communication, (1) a Josephson mixer operating as nondegenerate two-mode entangler with maximum measured logarithmic negativity $E_N=1.5$, (2) a quantum teleportation apparatus, capable of  teleporting vacuum and coherent states with a maximum fidelity of $73\%$, which exceeds the $50\%$ classical limit and is mainly limited by intermediate losses in the setup, and (3) an entanglement swapping system which generates entanglement between two remote noninteracting modes via entanglement swapping operations applied to input vacuum and coherent states with maximum measured logarithmic negativity $E_N=0.53$. Such hardware-efficient CV entanglement building blocks that are based on nondegenerate Josephson mixers could enable wide-ranging applications in modular quantum computation, quantum cryptography, and quantum communication.
\end{abstract}

\maketitle
\newpage
\section{Introduction}
Building superconducting quantum computers capable of running error correction codes \cite{LDPCIBM} and solving applied computational problems that are intractable using classical computation \cite{AppSurvey,FactoringResources} would require the formation of a hierarchy of quantum networks, which integrate thousands of quantum modules spread over multiple distances \cite{FutureQC}. In particular those that range between several millimeters to tens of meters. However, for such a large distributed architecture to work, it must have the capability of forming fast, reliable, high-fidelity remote entanglement between various modules of the system, where ideally the performance of the entangling operations is unaffected by the relative locations of the interacting modules. 

Recently, there has been tremendous progress in demonstrating remote entanglement in the short range, i.e., millimeters to meters. Most of the high-fidelity methods deployed in this range use direct gates \cite{ModularRouter,RemoteCR,RemoteCZ}, parametric devices \cite{SNAILswitch,SNAILrouter,OndemandQSTJJ}, or exchange discrete quantum variables (DV), such as single photons and qubit excitation, between quantum nodes connected via quantum links \cite{
	DetMultiQubitEnt,RemoteGmonPRL,DetBiComm,PlugPlay}. However, it remains an outstanding challenge to extend this level of success to the medium range \cite{DetQuantTelDistSCqubits,LoopholefreeBell}, i.e., tens of meters, where, for example, the processors are in separate fridges interconnected by superconducting lines passing through cryogenic links \cite{CryogenicMwLink,IntracityQuantComm}, and the long range, i.e., tens of kilometers, where, for example, the modules are in distant fridges interconnected by optical fibers.

While the long-range connectivity challenge continues to be hindered by the lack of fast, high-efficiency, low-noise optical-to-microwave quantum converters to date \cite{MwOptReview}, both medium and long ranges are expected to benefit from recent theoretical proposals that show potential upsides for using entanglement-based schemes especially those that rely on continuous quantum variables (CV) \cite{MwOptEntQuantTrans,DetMwOptTransQuantTel,LongDistNDampl} or a hybrid of CV and DV \cite{HybEntDist}, where CV refer to the continuous  quadrature amplitudes of the quantized electromagnetic field.    

Among the useful advantages of CV-based entanglement schemes \cite{CVreview,GaussQInforeview} are, (1) the existence of nonlinear devices, i.e., parametric amplifiers, which can reliably and efficiently produce CV entanglement by squeezing light \cite{MwOptEnt,ObsTwoModeSq,SqfluxdrivenJparam,MultiEntJPA,PathEntCVQuanMw}, (2) the entanglement generated by these sources is unconditional and gets produced every inverse bandwidth time \cite{CVreview}, and (3) the entanglement distribution does not strongly depend on the length of the quantum link due to its reliance on propagating coherent states rather than on stationary or resonant modes \cite{
	DetMultiQubitEnt,RemoteGmonPRL,DetBiComm,PlugPlay}. On the other hand, two of the main vulnerabilities of such schemes are, sensitivity to intermediate losses and imperfect entanglement due to the finite squeezing that can be achieved in practice \cite{CVreview}.

Since the discovery of quantum teleportation by Bennett \textit{et al.} \cite{TeleportationBennet}, which remarkably enables the disembodiment of an unknown quantum state at Alice and its full reconstruction at Bob, using shared nonclassical EPR correlations and transmission of purely classical information, teleportation has been successfully demonstrated in various DV and CV settings, including single-photon polarization states \cite{ExpRealTel,ExpRealTel2}, coherent \cite{UncondQuantTel} and squeezed \cite{ExpDemQuantTelSqState} states in the optical domain, and coherent states \cite{ExpQuantTelMw,QuantTelPropMw} in the microwave domain. Moreover, a key requirement of quantum repeaters, i.e., entanglement swapping, in which half of an entangled state is teleported, has been experimentally shown using single-photon polarization states \cite{ExpEntSwapPhotonPolarization} and developed for optical coherent states \cite{UncondTelCVEnt}.  

\begin{figure*}
	[tb]
	\begin{center}
		\includegraphics[
		width=2\columnwidth 
		]%
		{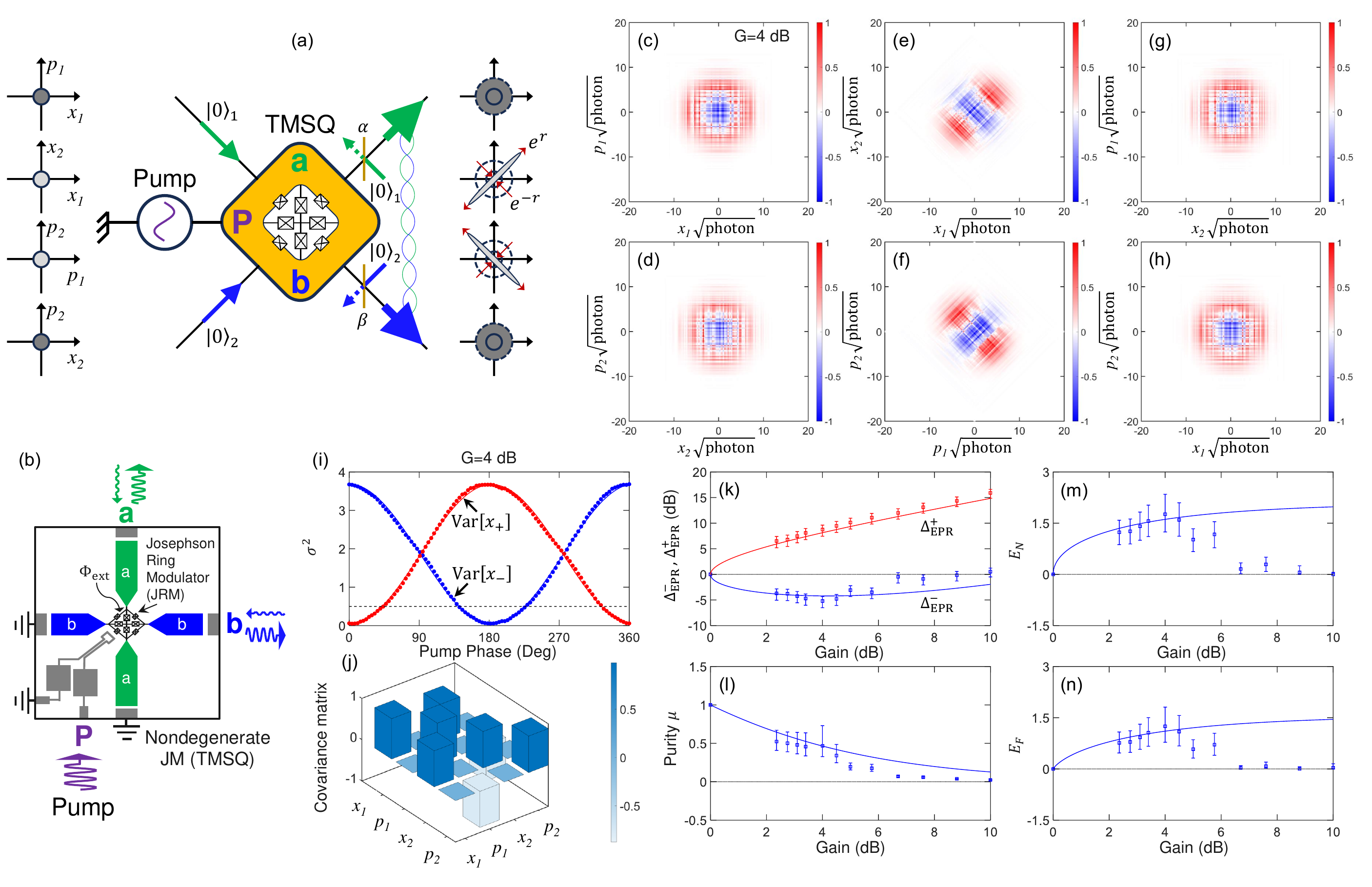}
		\caption{Demonstration of two-mode squeezing and entanglement generated by a nondegenerate three-wave Josephson mixer. (a) Illustration of two-mode squeezing at the output of a three-wave Josephson mixer. When driving the Josephson mixer with a monochromatic pump tone at the frequency sum of modes a and b, input vacuum noise characterized by the orthogonal quadrature pairs $x_{1}=\left( a+a^{\dagger}\right)/2$, $p_{1}=\left( a-a^{\dagger}\right)/2i$ and $x_{2}=\left( b+b^{\dagger}\right)/2$, $p_{2}=\left( b-b^{\dagger}\right)/2i$ get reflected and amplified at the output, mimicking a thermal state as shown in the top and bottom cases on the right side. Two-mode squeezing and entanglement between the modes emerge, however, upon examining cross correlations between the generalized position and momentum of the two modes as shown in the two middle cases on the right. Potential losses within or following the TMSQ device are represented by beam-splitter interactions with the thermal baths at the output of modes a and b, having power reflectivity (i.e., loss) of $\alpha$ and $\beta$, respectively. (b) Schematic representation of the nondegenerate Josephson mixer which serves as a two-mode squeezer. (c)-(h) Histogram difference between pump on and off plotted for the six pairs of dimensionless position and momentum coordinates of modes a and b measured at the output of the TMSQ device at $G=4$ dB. Squeezing below the vacuum level is observed for the EPR-like coordinates of relative position $x_{-}=x_2-x_1$ and total momentum $p_{+}=p_2+p_1$ as seen in plots (e) and (f). (i) The variance of EPR-like coordinates $x_{-}$ (blue filled circles) and $x_{+}$ (red filled circles) as a function of the pump phase. The pump phase axis is shifted to yield a minimum of $\rm{Var}[x_{-}]$ at $180$ degrees. Solid blue and red lines represent sinusoidal fits. Dashed black line represents the variance of vacuum noise. (j) Bar graph representing the covariance matrix elements of the TMSQ device for the minimum $\rm{Var}[x_{-}]$ working point in (i). (k)-(n) exhibit various measured figures of merit for the two mode entanglement as a function of the device gain. (k) shows Duan's criterion for entanglement. (l)-(n) display the state purity $\mu$, logarithmic negativity $E_N$, and entanglement of formation $E_F$, respectively. The solid curves in (k)-(n) are calculated fits based on the TMSQ model with asymmetric power transmission $\bar{\alpha}=0.62$, $\bar{\beta}=1$.         
		}
		\label{TwoModeSq}
	\end{center}
\end{figure*}

In this work, we demonstrate three key capabilities of CV-based quantum communication in the microwave domain, (1) a fast, low-loss, two-mode squeezer based on nondegenerate Josephson mixers (JMs), which are typically used as quantum-limited amplifiers for high-fidelity qubit readout \cite{hybridLessJPC,microstripJPC,JPCreview,Roch,JPCnature}. In particular, we show that such devices can serve as two-mode entangling sources, which are analogous to EPR-particle sources \cite{GenEntMwRadoverTL,QuantumNode,EPRwithNDparam,ExpCharacGaussCh}, (2) a high-fidelity quantum teleportation apparatus for coherent states that employs two JMs, one serving as an EPR source while the other as a which-path-information eraser, and (3) a fast and unconditional entanglement swapping scheme that utilizes three JMs, where the third serves as a second EPR source. In both latter setups, we demonstrate teleportation and entanglement swapping on vacuum and coherent states.  
   
\section{Two-mode squeezing}
To produce two-mode squeezed states, we pump a JM, i.e., nondegenerate parametric amplifier with $\chi^{\left( 2\right) }$ nonlinearity. The input and output bosonic field operators of the amplifier are given by the scattering relations \cite{GenEntMwRadoverTL}
  
\begin{align}
  	\begin{array}
  		[c]{cc}%
  		a_{\rm{out}}=S^{\dagger}a_{\rm{in}}S=\cosh\left(r \right) a_{\rm{in}}+e^{i\varphi_p}\sinh\left(r \right)b^{\dagger}_{\rm{in}}  ,\\
  		b^{\dagger}_{\rm{out}}=S^{\dagger}b^{\dagger}_{\rm{in}}S=\cosh\left( r\right) b^{\dagger}_{\rm{in}}+e^{-i\varphi_p}\sinh\left( r\right) a_{\rm{in}},
  	\end{array}
  	\label{a_b_out_squeezer_main}
\end{align}
  
\noindent where $S=\exp\left(re^{i\varphi_p}a^{\dagger}b^{\dagger}-re^{-i\varphi_p}ab \right) $ is the unitary two-mode squeeze operator, $a$ and $b$ are the annihilation operators associated with modes a and b of the amplifier and $re^{i\varphi_p}$ is the complex squeezing parameter, where $\varphi_p$ is the phase of the pump and $r\geq0$ depends on the pump amplitude and sets the amplifier gain $G=\cosh^{2}\left(r\right)$. In particular, when acting on input vacuum states, the amplifier produces two-mode squeezed vacuum state $\left|  \rm{TMS}\right\rangle=S\left| 0\right\rangle_{a}\left| 0\right\rangle_{b}=\cosh^{-1}\left( r \right)\sum\limits_{n}\left(\tanh \left( r\right)  \right)^n\left| n\right\rangle_{a}\left| n\right\rangle_{b}$, which represents a superposition of twin Fock states with  distinct frequencies traveling on separate transmission lines.  

The operation of the two-mode squeezing (TMSQ) device is illustrated in Fig.\,\ref{TwoModeSq}(a). Input vacuum states $\left| 0 \right\rangle_j$ that lie within the bandwidths of mode a ($j=1$) and b ($j=2$), get amplified and reflected towards the output. In the dimensionless quadrature representation of the fields expressed as conjugate position and momentum variables satisfying the commutation relations $[x_j,p_j]=i/2$, where $x_1=\rm{Re}\left( \textit{a}\right)$, $p_1=\rm{Im}\left( \textit{a}\right)$, and $x_2=\rm{Re}\left( \textit{b}\right)$,  $p_2=\rm{Im}\left( \textit{b}\right)$, the output fields for each mode, when viewed separately, mimic amplified thermal states, however nonlocal two-mode squeezing emerges when inspecting combinations of output field quadratures, i.e.,   $x_{\rm{out,1}}-x_{\rm{out,2}}=e^{-r}\left( x_{\rm{in,1}}-x_{\rm{in,2}}\right)$, and  $p_{\rm{out,1}}+p_{\rm{out,2}}=e^{-r}\left( p_{\rm{in,1}}+p_{\rm{in,2}}\right)$, which are analogous to the relative position and total momentum of a pair of EPR particles. Thus, while the noise variance of each output mode $\left\langle  x^{2}_{\rm{j}}  \right\rangle+\left\langle   p^{2}_{\rm{j}} \right\rangle=\cosh\left( 2r\right)/2 $ increases with $r$, the output noise variance for certain orthogonal linear quadrature combinations undergoes either squeezing below the vacuum noise level, e.g., $\rm{Var}\left[ x_{-}\right]= \left\langle \left(x_1-x_2\right) ^2\right\rangle=e^{-2r}/2$ or anti-squeezing, e.g., $\rm{Var}\left[ x_{+}\right]= \left\langle \left(x_1+x_2\right) ^2\right\rangle=e^{2r}/2$ \cite{CVreview}.   

As illustrated in Fig.\,\ref{TwoModeSq}(a), losses in the TMSQ device or at its output are modeled as beamsplitter interactions with uncorrelated cold thermal baths. Thus, the output fields can be written as 

\begin{align}
	\begin{array}
		[c]{cc}%
		a_{\rm{out'}}=\sqrt{\bar{\alpha}}a_{\rm{out}}+\sqrt{\alpha}a_{\rm{th}}  , \\
		b_{\rm{out'}}=\sqrt{\bar{\beta}}b_{\rm{out}}+\sqrt{\beta}b_{\rm{th}},
	\end{array}
	\label{a_b_out_BS_main}
\end{align} 

\noindent where $a_{\rm{th}}$, $b_{\rm{th}}$ represent bosonic modes of the thermal baths at frequencies $f_1$ and $f_2$, $\alpha$, $\beta$ are the effective power losses, while  $\bar{\alpha}=1-\alpha$, $\bar{\beta}=1-\beta$. 

Moreover, since the two-mode squeezed state belongs to the class of Gaussian states, its Wigner functions is a bipartite normalized Gaussian distribution given by \cite{CVreview}

\begin{equation}
	W\left( \xi\right)=\dfrac{1}{4{\pi}^2\sqrt{\det{V}}}\exp{\left\lbrace-\dfrac{1}{2}\xi V^{-1} {\xi}^{\rm{{T}}} \right\rbrace }, \label{Wigner1_main}
\end{equation}
 
\noindent which is fully characterized by its covariance matrix $V$ whose elements are given by $V_{ij}=\left\langle \xi_i \xi_j+ \xi_j \xi_i\right\rangle/2-\left\langle \xi_i\right\rangle \left\langle \xi_j\right\rangle$, where $\xi_{i}$, $\xi_{j}$ are components of the vector $\xi=\left(x_1,p_1,x_2,p_2 \right)$.  

The circuit diagram of the JM, which implements the TMSQ, is shown in Fig.\,\ref{TwoModeSq}(b). It consists of two orthogonal half-wavelength microstrip resonators denoted a and b with different resonance modes, i.e., $f_b>f_a$. The two resonators are joined at the center by a Josephson ring modulator (JRM) that is flux tunable and serves as a dispersive three-wave mixing element \cite{
	hybridLessJPC,GenEntMwRadoverTL}. Resonators a and b are capacitively coupled to external feedlines that carry the incoming and outgoing fields. A strong coherent microwave tone, i.e., the pump, is applied to the JM through a third filtered on-chip line that is capacitively coupled to the JRM nodes. When applying a pump at the frequency sum $f_p=f_a+f_b$, the JM acts as a nondegenerate quantum-limited amplifier for quantum signals \cite{microstripJPC,JPCnature} and a two-mode squeezer \cite{GenEntMwRadoverTL}.

\begin{figure}
	[tb]
	\begin{center}
		\includegraphics[
		width=\columnwidth 
		]%
		{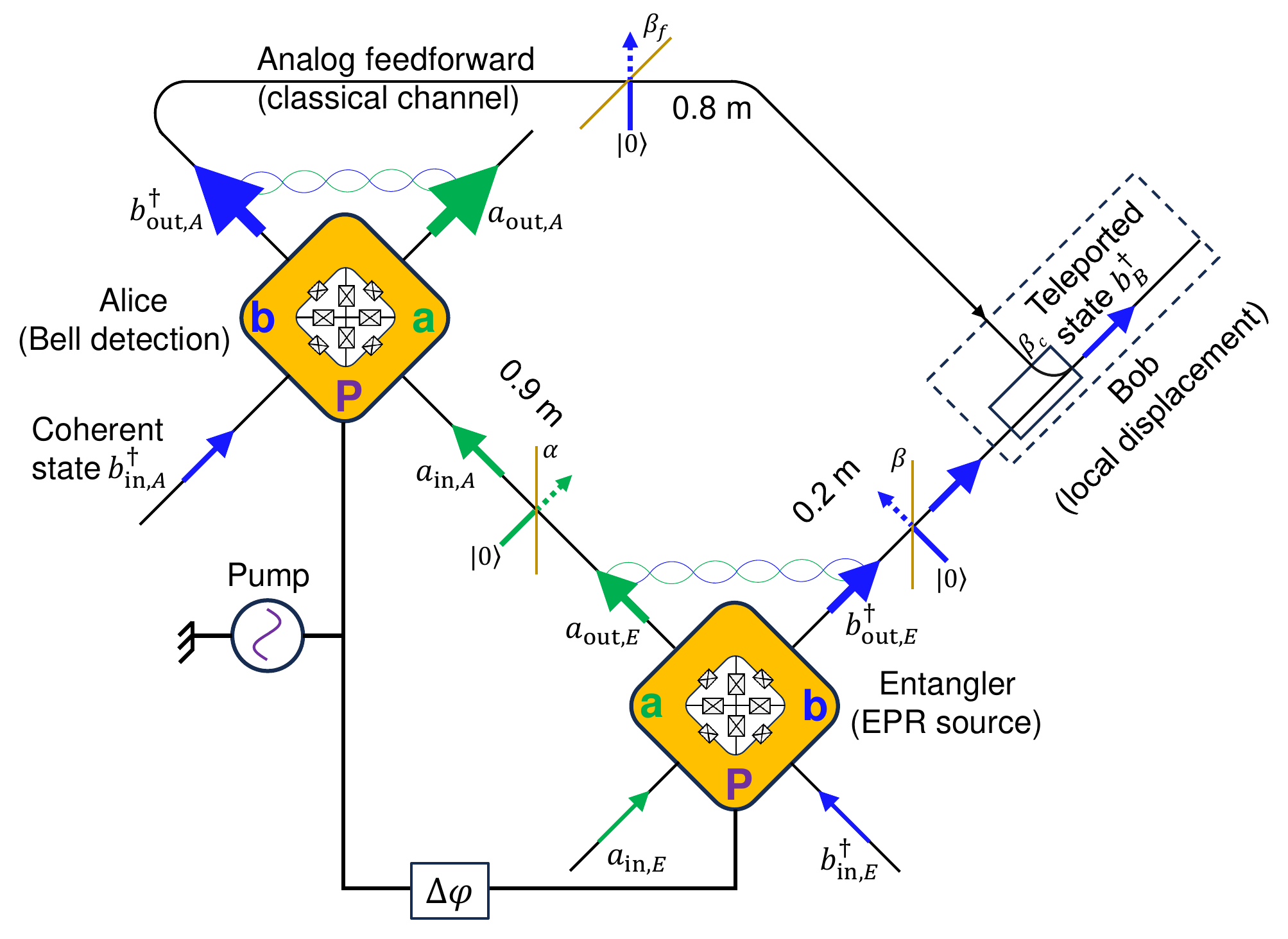}
		\caption{Quantum teleportation scheme for microwave coherent states using two nondegenerate Josephson mixers. A nondegenerate JM, serving as an EPR source, sends entangled continuous-variable beams to Alice and Bob. Alice employs a second nondegenerate JM, acting as a which-path information eraser, to perform a Bell-state like measurement on the incoming coherent input state to be teleported to Bob, i.e., $b^{\dagger}_{\rm{in},A}$, and on half of the entangled state received from the EPR source, i.e., $a_{\rm{in},A}$. Following the `measurement', Alice sends an analog feedforward signal to Bob containing the measurement result via a classical channel. Bob completes the teleportation process by locally displacing his half of the entangled state received from the EPR source using the feedforward signal sent by Alice. Approximate cable lengths between the parties are listed next to the connecting paths. Losses in the setup are modeled using beamsplitter interactions with thermal (i.e., vacuum noise) baths, and given by $\bar{\alpha}=0.62$, $\bar{\beta}=0.93$, $\bar{\beta_f}=0.4$, and $\beta_c=0.1$, where $\beta_c$ is the power coupling factor of the directional coupler. The gain on Alice is set to about $14$ dB to match the total attenuation in the feedforward channel and directional coupler.  
		}
		\label{TeleportationScheme}
	\end{center}
\end{figure}

In Fig.\,\ref{TwoModeSq}(c)-(h) we exhibit the histogram difference between pump on ($G=4$ dB) and off ($G=0$ dB) measured for the six unique pairs of the position and momentum quadratures of the two modes at the output of the TMSQ. In this experiment, we flux tune the JM to have $f_a=7.231$ GHz, $f_b=9.695$ GHz, and apply a pump tone at $f_p=16.926$ GHz. As expected, when inspecting mixed quadrature pairs of position and momentum of the same mode (Fig.\,\ref{TwoModeSq}(c),(d)) or belonging to different modes (Fig.\,\ref{TwoModeSq}(g),(h)), we observe a uniform amplified vacuum noise in the former case and uncorrelated amplified noise in the latter. Whereas, when considering cross correlations between position (Fig.\,\ref{TwoModeSq}(e)) and momentum (Fig.\,\ref{TwoModeSq}(f)) quadratures of the two modes, we observe squeezing of the noise variance below the vacuum level for EPR-like coordinates $x_{-}$ and $p_{+}$, and anti-squeezing for the orthogonal joint coordinates $x_{+}$ and $p_{-}$. In panel (i), we show the periodic dependence of the variance of the EPR-like coordinates $x_{-}$ and $x_{+}$ on the pump phase. As seen in the plot, two-mode squeezing occurs when $\rm{Var}\left[ x_{-}\right]$ (or $\rm{Var}\left[ x_{+}\right]$) goes below the vacuum noise variance of $1/2$, indicated by the dashed black line. Maximum squeezing of $9.2$ dB is attained at the minima of $\rm{Var}\left[ x_{-}\right]$. Furthermore, using the quadrature histograms data measured for $G=4$ dB, we reconstruct the covariance matrix $V$ of the TMSQ device as outlined in Appendix A and display its elements in Fig.\,\ref{TwoModeSq}(j) using a bar graph format. 

In the remaining panels (Fig.\,\ref{TwoModeSq}(k)-(n)), we plot four key entanglement measures, derived from the reconstructed covariance matrix, which characterize the two entangled modes of the TMSQ device as a function of its gain. Namely, we plot the EPR measures $\Delta^{\mp}_{\rm{EPR}}=2(V_{11}+V_{33}\mp2V_{13})$ (Eq.\,(\ref{EPRmin}) and (\ref{EPRmax})), the state purity $\mu=1/(16\sqrt{\det(V)})$ (Eq.\,\ref{purity}) \cite{SympInvMeasGaussStates,LossAsymTWPA}, the logarithmic negativity $E_{N}=\max\left[  0,-\log_{2}\left( 4\nu_{-}\right) \right]$ \cite{GenEntMwRadoverTL,LossAsymTWPA,ObsTwoModSqTWPA,CompMeasEnt}, and the entanglement of formation (EOF) $E_{F}=\max\left[  0, h \left( 4\nu_{-} \right) \right]$ \cite{GenEntMwRadoverTL,EF,EFArbtwomodeGauss} in Fig.\,\ref{TwoModeSq}(k), (l), (m), (n), respectively, where $\nu_{-}$ is the smallest symplectic eigenvalue of the partially transposed covariance matrix \cite{QuantumNode}, given by Eq.\,(\ref{numinus}), and the function $h(x)$ is given by Eq.\,(\ref{hx}). In addition, we depict in Fig.\,\ref{TwoModeSq}(k)-(n) using solid curves the calculated entanglement measures of the TMSQ whose theoretical model incorporates the effect of the asymmetric losses experienced by modes a and b.

\begin{figure*}
	[tb]
	\begin{center}
		\includegraphics[
		width=2\columnwidth 
		]%
		{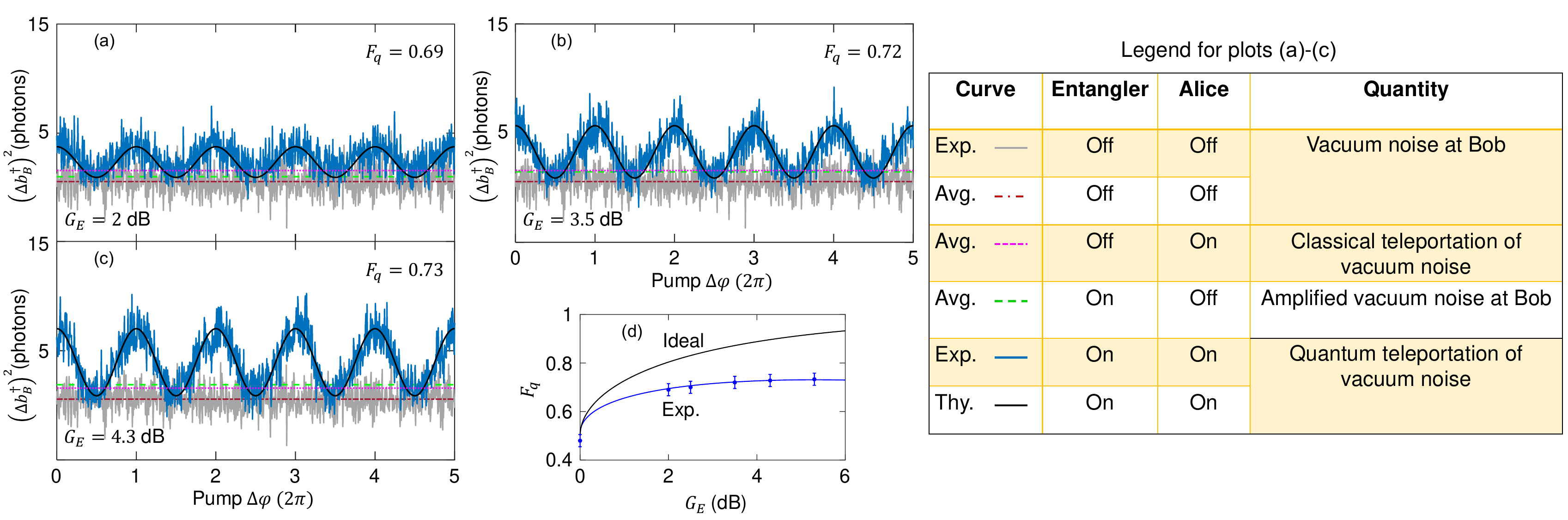}
		\caption{Quantum and classical teleportation of vacuum noise. (a)-(c) exhibit quantum teleportation of vacuum noise at Bob (solid dark blue) measured as a function the pump phase difference between the pumps at the Entangler and Alice $\Delta \varphi$, plotted for ascending power gain on the Entangler $G_E$. The solid black curves are theoretical fits to the data. The gray curves correspond to the measured vacuum noise at Bob without teleportation whose average is plotted as a dashed-dotted dark red line. The dashed green line depicts the amplified vacuum noise at Bob received from the Entangler. Whereas, the dashed magenta line shows the average level for the classically teleported vacuum noise (when the Entangler is off). (d) depicts the measured quantum teleportation fidelity for vacuum noise as a function of the Entangler gain. The solid blue curve represents a theoretical model calculation based on the estimated losses in the setup. In comparison, the solid black curve depicts the calculated quantum teleportation fidelity for a lossless setup.    
		}
		\label{TelQuantVac}
	\end{center}
\end{figure*}

In the case of $\Delta^{\mp}_{\rm{EPR}}$, which also characterize the amount of two-mode squeezing and anrisqueezing, the violation of the inequality $\Delta^{-}_{\rm{EPR}}\geq1$ for a bipartite Gaussian system implies the inseparability of the two subsystems, and hence their entanglement. This inequality can be derived from the general criterion for entanglement proposed by Duan \textit{et al.} \cite{DuanSeparability,CVreview}. In the case of $E_{N}$, its positivity, i.e., $E_{N}>0$, serves a dual purpose. It constitutes a necessary and sufficient condition for entanglement and gives an upper bound on the number of equivalent entangled bits (ebits) that can be distilled from the entangled state. Similarly, $E_{F}$ gives the minimum number of ebits required to create the quantum state as an ensemble of pure states \cite{EF}. 

In particular, in our case (as seen in panels (k)-(n)), we find that the JM exhibits two-mode entanglement up to about $G=8$ dB, with optimal entanglement values observed around $4$ dB, i.e., $\Delta^{-}_{\rm{EPR}}=-5$ dB, $E_{N}=1.5$, and 
$E_{F}=1.25$, followed by a rapid decline above $5$ dB. 
While the calculated entanglement measures, e.g. $E_N$ and $E_F$, yield a good agreement with the measured data in the low gain range, they notably do not capture the decline observed above $5$ dB. Although the origin of such decline in our system remains an open question, it underscores the increased fragility of entangled states as they grow larger in size (i.e., become more ''nonclassical") in the presence of dissipation \cite{EntFragile}. It also seems to follow the decline of two-mode squeezing, e.g., $\Delta^{-}_{\rm{EPR}}$, which is known to be sensitive to asymmetric losses \cite{LossAsymTWPA}. 

Another useful measure for characterizing the generated entanglement is the entanglement bit rate \cite{GenEntMwRadoverTL}, given by $R=B \times E_F$, where $B$ is the dynamical bandwidth of the device. At $4$ dB, we obtain an estimated maximum rate of $70$ Mebit/sec, where $E_F$ is maximal and $B=56$ MHz. 

\section{Teleportation of coherent states}

Equipped with this valuable quantum resource of two-mode entanglement which is produced by nondegenerate JMs, allows us to readily demonstrate two important quantum communication capabilities in the microwave domain, i.e., quantum teleportation and entanglement swapping. 

However, prior to reviewing our teleportation apparatus scheme shown in Fig.\,\ref{TeleportationScheme}, it is instructive to highlight the four key elements that are typically used for the teleportation of coherent states in the optical domain \cite{TelCVProposal,BroadTel}, namely (1) an EPR source which generates two-mode entanglement (for example by interfering two orthogonal single-mode squeezed light on a 50:50 beamsplitter), half of which is sent to Alice and the other to Bob, (2) a 50:50 beamsplitter combined with photon detectors at Alice, which allow her to perform a joint Bell-state measurement on her entangled mode and the input state she wishes to teleport, (3) a classical communication channel between Alice and Bob through which Alice communicates the measurement results to Bob, and (4) a local displacement capability at Bob which he applies to his entangled mode based on the results he receives from Alice, which in turn reconstructs the original state destroyed by Alice's measurement.

\begin{figure*}
	[tb]
	\begin{center}
		\includegraphics[
		width=2\columnwidth 
		]%
		{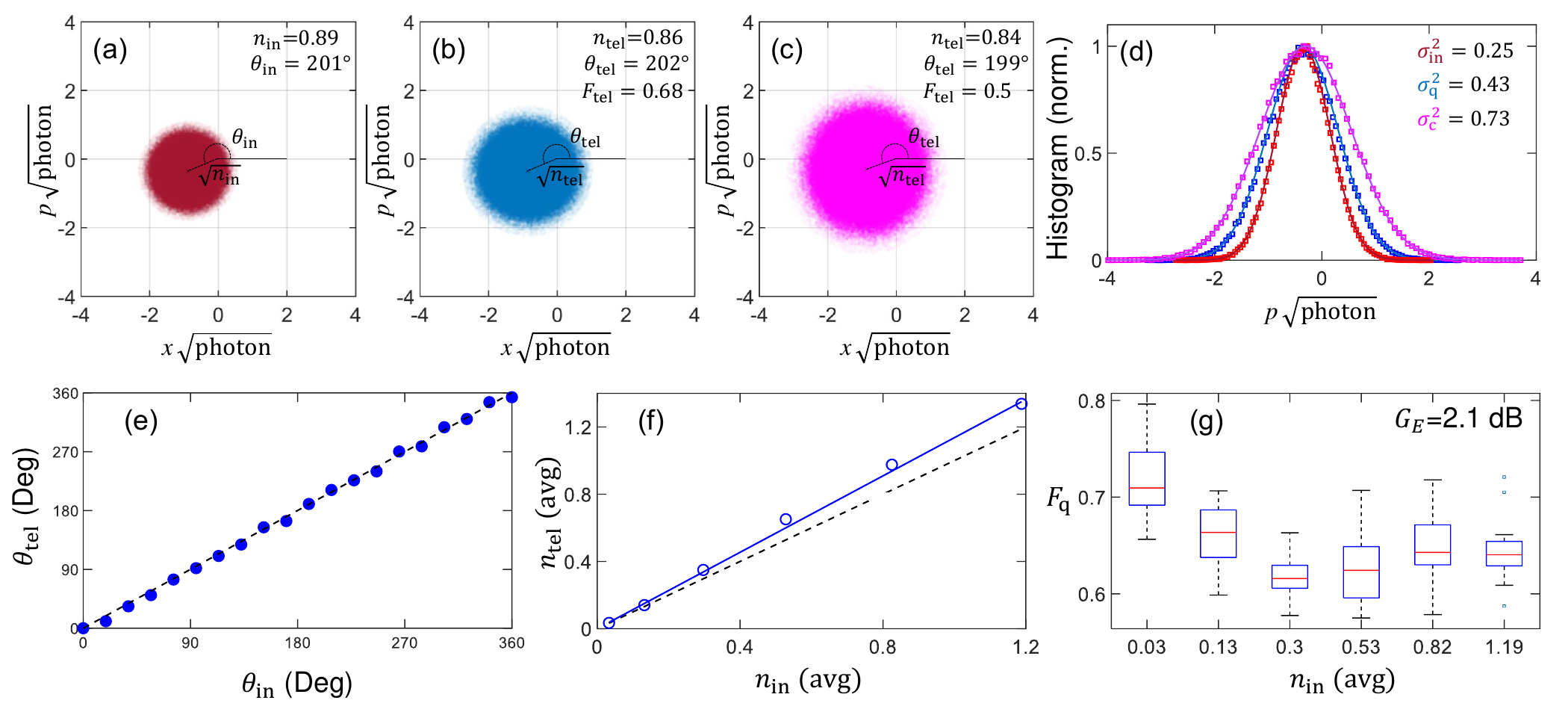}
		\caption{Quantum and classical teleportation of coherent microwave states. (a) \textit{I-Q} (\textit{x-p}) quadrature representation of an applied coherent state at Alice's input port b (to be teleported to Bob) whose center is specified by the polar coordinates ($\sqrt{n_{\rm{in}}}$, $\theta_{\rm{in}}$). (b) and (c) depict the coherent state measured at Bob's output upon applying a quantum and classical teleportation, respectively. (d) Normalized histograms of the measured input coherent state (red), quantum mechanically-teleported (blue) and classically-teleported (magenta) coherent states along the $p$ quadrature. The solid curves in the plot correspond to Gaussian fits. (e) The measured phase of the quantum-mechanically teleported coherent state $\theta_{\rm{tel}}$ plotted versus $\theta_{\rm{in}}$ (filled blue circles). Dashed black line corresponds to the ideal relation $y=\theta_{\rm{in}}$. (f) The average photon number of the  quantum-mechanically teleported coherent state $n_{\rm{tel}}$ versus $n_{\rm{in}}$ (open blue circles). The solid blue line represents a fit to the data, while the dashed black line is a guide to the eye for the ideal relation $y=n_{\rm{in}}$. (g) Distribution of quantum teleportation fidelities measured over 20 input coherent state phases $\theta_{\rm{in}}$ spanning one period taken for different average input photon number $n_{\rm{in}}$. The inner red lines mark the mean fidelities, while the box edges represent the 25th and 75th percentiles and the whiskers indicate the lowest and highest data points. The quantum teleportation in this experiment is performed using an Entangler gain of $2.1$ dB.      
		}
		\label{TelQuantCoh}
	\end{center}
\end{figure*}

Analogously, the scheme of Fig.\,\ref{TeleportationScheme} includes microwave devices that fulfill similar roles. The JM, denoted as the Entangler, plays the role of the EPR source which produces two-mode entangled vacuum states at mode a and b and sends them to Alice and Bob, respectively. The second JM, marked as Alice, effectively plays the role of the Bell-state detector, though with three key differences compared to the beamsplitter and photon detectors used in the optical domain. First, it amplifies the combined signals. But since the amplification is phase preserving, the JM acts in the high-gain limit as a which-path information eraser \cite{WhichPathEraser} for the input state and the entangled mode. Second, the amplified output signal at port b of Alice forms an analog feedforward signal that can be directly sent to Bob without any further processing. However, since the feedforward signal is amplified and the teleportation process requires a unity-gain classical signal, Bob uses a directional coupler to simultaneously attenuate the feedforward signal and displace his entangled mode with it. Third, since the scheme in Fig.\,\ref{TeleportationScheme} relies on wave-interference, successful teleportation only occurs for relative pump phases between the Entangler and Alice, denoted $\Delta\varphi$, that minimize the noise at Bob's output (see Appendix B and C for further details). Note that Fedorov \textit{et al.} \cite{ExpQuantTelMw} have successfully teleported coherent states using a similar microwave scheme based on single-mode squeezers, i.e., Josephson parametric amplifiers. However, the scheme of Fig.\,\ref{TeleportationScheme} which instead employs nondegenerate JMs has the advantage of cutting the cryogenic hardware overhead of the teleportation setup (both parametric and passive devices) by more than half, which in turn leads to additional benefits such as, higher stability, simpler tuneup process, and fewer matching requirements between the various components.    

Using the teleportation scheme of Fig.\,\ref{TeleportationScheme}, we first demonstrate the quantum teleportation of vacuum states which represent Gaussian states with zero mean. In this experiment and what follows, we independently flux tune the different JMs to have the same resonance frequencies $f_a=7.23$ GHz, $f_a=9.707$ GHz and applied pump tone $f_p=16.937$ GHz. In Fig.\,\ref{TelQuantVac}(a)-(c), we plot for different Entangler gains ($G_E$), the normalized noise power (solid dark blue), which we measure on resonance at Bob's output as a function $\Delta \varphi$. The solid black curves represent theory fits based on Eq.\,(\ref{bOutatBobNoisePow2}). As seen in Fig.\,\ref{TelQuantVac}(a)-(c), the measured noise is periodic in $\Delta \varphi$ and its oscillation amplitude grows with $G_E$. Quantum teleportation occurs in these measurements at the minima points which asymptotically approach the average of the measured vacuum noise level at Bob without pumps (i.e., $0.51$) represented by the dashed-dotted dark red line. For reference, we show in gray the measured noise at Bob without pumps. We also display, for comparison, the average noise level (i.e., $1.56$), obtained via classical teleportation (dashed magenta line) when no entanglement is shared between Alice and Bob, i.e., $G_E=0$ dB, and the average amplified noise level at Bob when only the Entangler is on (dashed green line), which increases with $G_E$ as expected. At the bottom and top corners of Fig.\,\ref{TelQuantVac}(a)-(c), we list the $G_E$ employed in each measurement and the resultant quantum teleportation fidelity $F_q$, respectively, where the teleportation fidelity quantifies the amount of overlap between the input and teleported states (Eq.\,(\ref{F_Uhlmann})). In Fig.\,\ref{TelQuantVac}(d), we plot the quantum teleportation fidelity (filled blue circles) measured for input vacuum noise versus $G_E$. The blue curve represents a theoretical model calculation that accounts for the losses in the setup. The black curve, serving as a reference, represents the ideal $F_q$ achievable in a lossless setup (Eq.\,(\ref{Quant_F_reduced})). It is worth noting that the maximum measured teleportation fidelity $F_q=0.73$ exceeds the classical limit $F_c=1/2$.      
      
In Fig.\,\ref{TelQuantCoh}, we further use the scheme of Fig.\,\ref{TeleportationScheme} to perform quantum and classical teleportation of general coherent states (displaced vacuum). In this experiment, we apply a $1$ $\mu$s pulse of a coherent tone on resonance to port b of Alice and measure the teleported result at Bob. The reconstructed input coherent signal is depicted in Fig.\,\ref{TelQuantCoh}(a) using the standard $x$-$p$ ($I$-$Q$) quadrature representation, where the distance between the origin and the 2D Gaussian histogram center corresponds to $\sqrt{n_{\rm{in}}}$, where $n_{\rm{in}}$ is the average photon number of the coherent state and the angle with the $x$ axis $\theta_{\rm{in}}$ represents the coherent state phase. Using the same representation, we plot in Fig.\,\ref{TelQuantCoh}(b) and Fig.\,\ref{TelQuantCoh}(c), the 2D Gaussian histograms of the measured output field at Bob following quantum ($G_E=2.1$ dB) and classical ($G_E=0$ dB) teleportation which yield teleportation fidelities $F_{\rm{tel}}$ of $0.68$ and $0.5$, respectively, where only the former surpasses, beyond the margin of error $\pm0.01$, the $50\%$ barrier which bounds the fidelity of classical teleportation of coherent states \cite{QuantvsClassTelCV}. We also list in the inset the measured average photon number $n_{\rm{tel}}$ and phase $\theta_{\rm{tel}}$ of the teleported signal. In Fig.\,\ref{TelQuantCoh}(d), we plot histogram cut along the $p$ quadrature of the input state (red), the quantum-mechanically teleported state (blue) and the classically teleported state (magenta), which clearly show a noise reduction in the quantum teleportation case versus the classical one. The solid curves in Fig.\,\ref{TelQuantCoh}(d) are Gaussian fits whose variances are listed in the inset. In Fig.\,\ref{TelQuantCoh}(e) we keep $n_{\rm{in}}$ constant (as in Fig.\,\ref{TelQuantCoh}(a)) and measure $\theta_{\rm{tel}}$ of the quantum-mechanically teleported coherent signal as we vary $\theta_{\rm{in}}$ over one period. As indicated by the unity-slope dashed line, $\theta_{\rm{tel}}$ follows $\theta_{\rm{in}}$ quite well. Similarly, in Fig.\,\ref{TelQuantCoh}(f) we compare $n_{\rm{tel}}$ versus $n_{\rm{in}}$ for a fixed $\theta_{\rm{in}}$. The solid blue line represents a fit to the data, while the dashed black line is a guide to the eye with a unity slope. Based on the fit, we obtain a deviation of up to $13.6\%$ between $n_{\rm{tel}}$ and $n_{\rm{in}}$. In Fig.\,\ref{TelQuantCoh}(g), we plot for different $n_{\rm{in}}$ the distribution of quantum teleportation fidelities measured for $20$ input phases $\theta_{\rm{in}}$ spanning one period. As expected, the measurement shows that $F_q$ is largely independent of $\theta_{\rm{in}}$, and with the exception of $n_{\rm{in}}=0.03$, $F_q$ mostly lies in the range $0.6$-$0.7$ for the various $n_{\rm{in}}$. Importantly, such ability to teleport random coherent states with high enough fidelities showcases the viability of utilizing JM-based teleporters in quantum cryptography applications that employ CVs for quantum key distribution  \cite{ExpQuantTelMw,CVquantCrypto}.

\begin{figure*}
	[tb]
	\begin{center}
		\includegraphics[
		width=1.85\columnwidth 
		]%
		{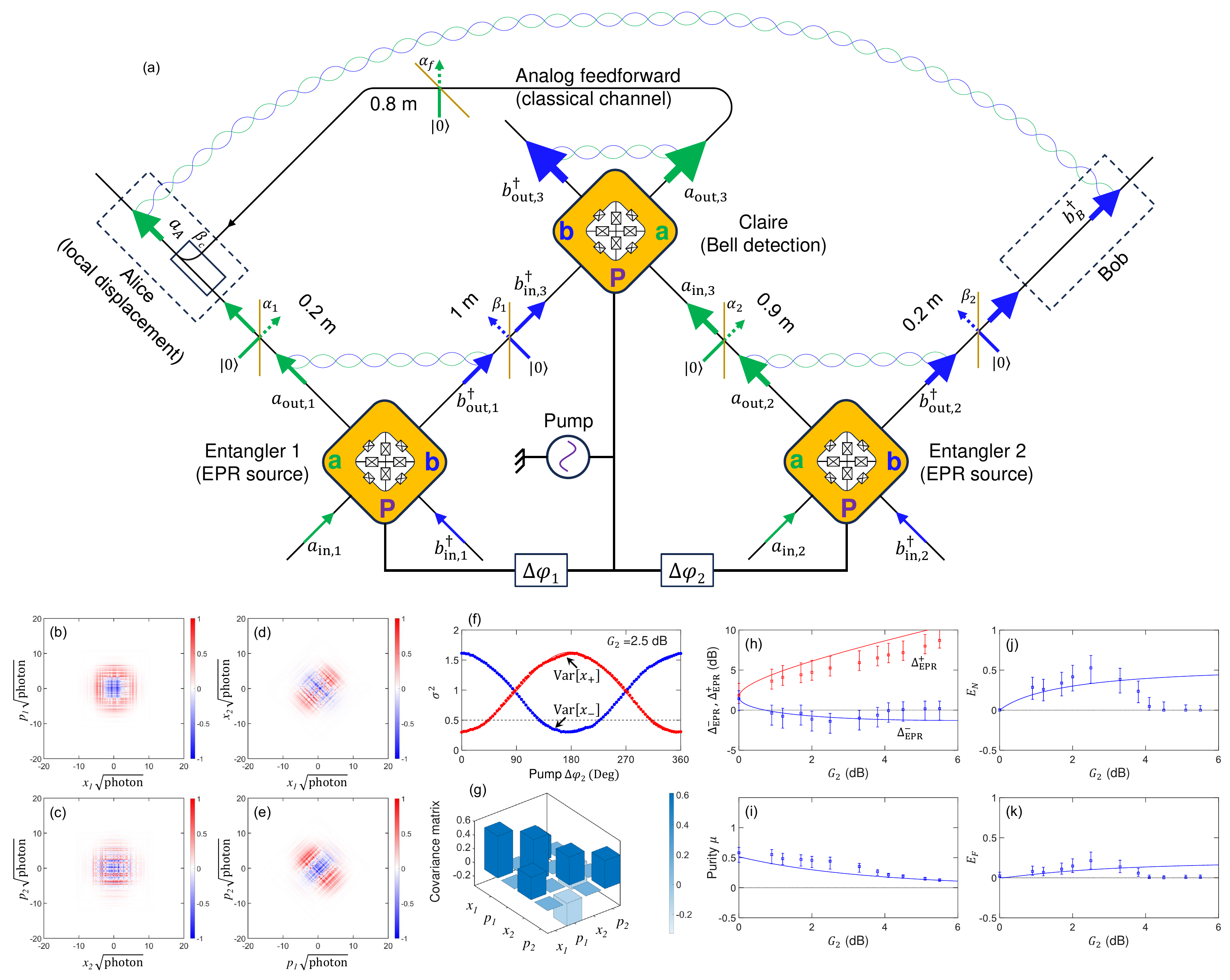}
		\caption{Entanglement swapping of continuous quantum variables with nondegenerate Josephson mixers. (a) Entanglement swapping scheme between two remote nodes, i.e., Alice and Bob, mediated by a `Bell-state' detector, i.e., Claire. Two EPR sources, Entangler 1 and 2, generate two-mode CV entanglement. One-half of the entangled states generated by the Entanglers are sent to Alice and Bob, while the other halves are sent to Claire, who erases their which-path information and forwards the mixing result to Alice through a classical channel. Alice then completes the entanglement swapping process by locally displacing the one-half of the entangled state in her possession using the feedforward signal. Approximate cable lengths between the parties are listed next to the connecting paths. In (b)-(k) Entangler 1 gain is fixed at $G_1=1.4$ dB. (b)-(e) Quadrature histogram difference between pump turned on ($G_2=2.5$ dB) and off plotted for the four main pairs of dimensionless position and momentum quadratures belonging to mode a (i.e., $x_1$, $p_1$) and b (i.e., $x_2$, $p_2$) measured at Alice and Bob. Squeezing below the vacuum level is observed for the EPR-like coordinates of relative position $x_{-}$ and total momentum $p_{+}$ as seen in plots (d) and (e). (f) The variance of $x_{-}$ (blue filled circles) and $x_{+}$ (red filled circles) versus the relative pump phase of Entangler 2. The relative phase between Claire and Entangler 1 ($\Delta\varphi_1$) is set to maximize the teleportation fidelity of Claire's input at port a. The relative pump phase axis is shifted to yield a minimum of $\rm{Var}[x_{-}]$ at $180$ degrees. The solid blue and red lines represent sinusoidal fits. The dashed black line represents the variance of vacuum noise. (g) Bar graph representing the covariance matrix elements of the modes a and b measured at Alice and Bob for the minimum $\rm{Var}[x_{-}]$ working point in (f). (h)-(k) exhibit various measured figures of merit of the remote entanglement between Alice and Bob versus $G_2$. (h) Duan's criterion for entanglement extended to the entanglement swapping case. (i)-(k) display  $\mu$, $E_N$, and $E_F$, characterizing the two-mode entanglement between Alice and Bob generated via entanglement swapping. The solid curves in (h)-(k) are calculated fits based on an entanglement swapping scheme model, which incorporates intermediate losses characterized by $\bar{\alpha}_1=0.9$, $\bar{\alpha}_2=0.72$, $\bar{\beta}_1=0.62$, $\bar{\beta}_2=0.97$, $\bar{\alpha}_f=0.85$, and $\beta_c=0.1$. In this setup, Claire's gain is set to about $10.7$ dB.
		}
		\label{EntSwapQuanVac}
	\end{center}
\end{figure*}

\section{Entanglement swapping}

Here, we show that our quantum teleportation scheme is capable of not only teleporting coherent states, but also genuine quantum states such as the entanglement source itself, i.e., one-half of an entangled state, also known as entanglement swapping, which is a prerequisite for realizing quantum repeaters that are vital in long-range communications \cite{QuantRepeaterReview}. Using the extended teleportation setup shown in Fig.\,\ref{EntSwapQuanVac}(a) we preform entanglement swapping operation which results in the entanglement of two remote coherent fields at Alice and Bob that have never directly interacted. The setup consists of three nondegenerate Josephson mixers, two of which function as two-mode entanglement sources denoted as Entangler 1 and 2, while the third, denoted as Claire, is located between Alice and Bob and mimics the role of a Bell-state detector \cite{UncondTelCVEnt,ConfEntCVQuantTel}. In this scheme, Entangler 1 (2) sends one-half of its entangled state, i.e., mode a (mode b), to Alice (Bob), while sending the other half, i.e., mode b (mode a), to Claire. By operating her nondegenerate Josephson mixer in the high-gain limit, Claire erases the which-path information of her inputs and forwards the amplified output mixed signal to Alice through a classical channel. Alice then completes the teleportation of one-half of the entangled state generated by Entangler 2 (mode a) by attenuating the classical signal she receives from Claire down to unity transmission level and displacing with it her own half of the entangled state (received from Entangler 1). Thus, although modes a and b received by Alice and Bob from Entangler 1 and 2 have not directly interacted, they become entangled following the entangling operation performed by Claire on the other entangled modes sent by the Entanglers. Similar to the teleportation setup of Fig.\,\ref{TeleportationScheme}, we match in this experiment the resonance frequencies of the three mixers (using magnetic flux), phase lock the pumps feeding the Josephson mixers, and fix the relative pump phase $\Delta \varphi_1$ such that the noise at Alice's output is minimized when both Entangler 1 and Claire's JM are on and Entangler 2 is off. 

The results exhibited in Fig.\,\ref{EntSwapQuanVac}(b)-(k) are measured for vacuum states input on Entangler 1 and 2. They reveal two-mode vacuum squeezing and entanglement generated at Alice and Bob due to the entanglement swapping operation. The gain on Entangler 1 is fixed at $G_1=1.4$ dB, while the gain on Entangler 2 is set to $G_2=2.5$ dB in Fig.\,\ref{EntSwapQuanVac}(b)-(g) and varied in Fig.\,\ref{EntSwapQuanVac}(h)-(k). In Fig.\,\ref{EntSwapQuanVac}(b)-(e) we display the quadrature histogram difference with the pump applied to Entangler 2 turned on and off, plotted for the four main pair combinations of $x_1$, $x_2$, $p_1$, $p_2$, where $x_1$ ($x_2$) and $p_1$ ($p_2$) are the position and momentum quadratures measured at Alice (Bob). As seen in Fig.\,\ref{EntSwapQuanVac}(b),(c), when inspecting the quadrature histograms of modes a and b separately, we observe amplified thermal states. However, when inspecting histograms of cross position and momentum quadratures as shown in Fig.\,\ref{EntSwapQuanVac}(d),(e), strong correlations and squeezing below the vacuum noise level are revealed along EPR-like coordinates, i.e., $x_{-}$ and $p_{+}$. In Fig.\,\ref{EntSwapQuanVac}(f), we plot the variance of EPR-like coordinates $x_{-}$ (blue filled circles) and $x_{+}$ (red filled circles) versus the pump phase difference between Entangler 2 and Claire $\Delta \varphi_2$, which shows TMSQ around $180$ degrees as $\rm{Var}[x_{-}]$ drops below the variance of vacuum noise indicated by the dashed horizontal line. Similar to the single JM case, maximum squeezing of $
2.2$ dB is obtained at the minima of $\rm{Var}\left[ x_{-}\right]=0.3$. In Fig.\,\ref{EntSwapQuanVac}(g), we plot a bar graph representation of the covariance matrix elements of modes a (Alice) and b (Bob) taken at the minima working point of Fig.\,\ref{EntSwapQuanVac}(f). Lastly, in Fig.\,\ref{EntSwapQuanVac}(h)-(k), we depict four key entanglement measures for Alice's and Bob's fields as a function of $G_2$, i.e., $\Delta^{\pm}_{\rm{EPR}}$, $\mu$, $E_N$, and $E_F$. The solid curves represent theoretical calculations of the corresponding measures based on the covariance matrix of the entanglement swapping scheme, which incorporates the effect of losses in the system (see Appendix D).

\begin{figure*}
	[tb]
	\begin{center}
		\includegraphics[
		width=1.5\columnwidth 
		]%
		{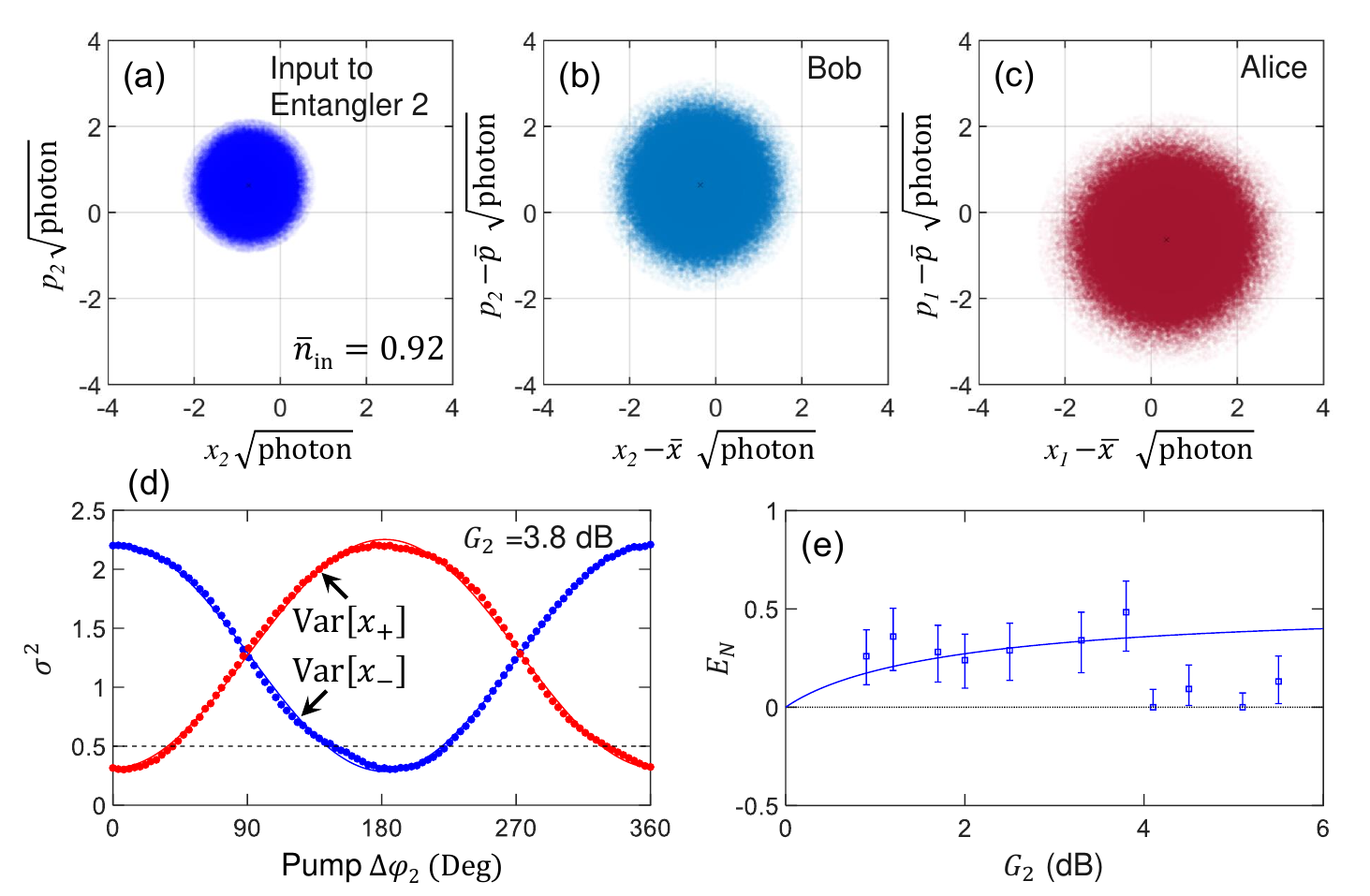}
		\caption{Generating remote entangled coherent states at Alice and Bob using the entanglement swapping setup. (a) \textit{I-Q} (\textit{x-p}) quadrature representation of a coherent state applied to the input b of Entangler 2. (b) and (c) The resulting entangled coherent states measured at Bob (mode b) and Alice (mode a), obtained for $1.2$ dB and $3.8$ dB of gain on Entangler 1 and 2, respectively.  The measured coherent states at Bob and Alice are plotted relative to their collective averages of the position $\bar{x}=\left( \bar{x}_1+\bar{x}_2\right) /2$ and momentum $\bar{p}=\left( \bar{p}_1+\bar{p}_2\right) /2$ quadratures. (d) The variance of EPR-like coordinates $x_{-}$ (blue filled circles) and $x_{+}$ (red filled circles) plotted as a function of the relative pump phase of Entangler 2. The relative phase between Claire and Entangler 1 ($\Delta\varphi_1$) is set to maximize the teleportation fidelity of Claire's input at port a. The relative pump phase axis is shifted to yield a minimum of $\rm{Var}[x_{-}]$ at $180$ degrees. Solid blue and red lines represent sinusoidal fits. Dashed black line represents the variance of vacuum noise. (e) The logarithmic negativity of the coherent states at Alice and Bob measured while fixing the gain on Entangler 1 at 1.2 dB and varying the gain on Entangler 2.      
		}
		\label{EntSwapQuantCoh}
	\end{center}
\end{figure*}

By comparing the entanglement measures obtained using the entanglement swapping system shown in Fig.\,\ref{EntSwapQuanVac}(h)-(k), to those measured using the single-stage nondegenerate Josephson mixer shown in Fig.\,\ref{TwoModeSq}(k)-(n), we find that: (1) like the single-stage JM, the entanglement swapping system exhibits two-mode entanglement when the Entanglers are turned on, i.e., $G_{1,2}>0$, as evidenced by $E_N$, $E_F$, and Simon's criterion for entanglement shown in Fig.\,\ref{SimonCr}. This means that in principle the entanglement swapping system can be deployed to perform coherent state teleportation using the scheme of Fig.\,\ref{TeleportationScheme} with the single-stage Entangler replaced by the entanglement swapping system. (2) The entanglement swapping system exhibits an unavoidable  degradation in the quality of the entanglement as indicated by the lower values achieved for $E_N$, $E_F$, and $\mu$ compared to the single-stage JM. (3) Similar to the JM case, the measured two-mode entanglement for the entanglement swapping case deteriorates and vanishes above certain Entangler gains, e.g., $G_2>4$ dB, which yields a deviation from the theoretical model prediction beyond the margin of error. (4) Unlike the case of JM, in the entanglement swapping system, due to the finite gain on Entangler 1, i.e., $G_1=1.4$ dB, $\Delta^{-}_{\rm{EPR}}$ and $\Delta^{+}_{\rm{EPR}}$ lie above $0$ dB for $G_2=0$ dB as seen in Fig.\,\ref{EntSwapQuanVac}(h), and in turn $\Delta^{-}_{\rm{EPR}}$ drops below $0$ dB above a certain threshold $G_2=0.7$ dB (see Eq.\,(\ref{EPRmEntSwap})), and reaches a minimum of $-1.4$ dB at $G_2=2.5$ dB. 

Moreover, similar to the single JM case, we evaluate the effective maximum entanglement bit rate for the entanglement swapping system. To get a conservative estimate, we consider the smallest dynamical bandwidth of the three JMs (belonging to Claire) of about $B=32$ MHz at the maximal $E_F=0.21$ working point, which together yield $R=6.7$ Mebits/sec.

As yet another demonstration of the entanglement capability of the entanglement swapping system, we set $G_1=1.2$ dB, $G_2=3.8$ dB and apply a coherent state with $\bar{n}_{\rm{in}}=0.92$ to port b of Entangler 2, plotted in Fig.\,\ref{EntSwapQuantCoh}(a) in the $x-p$ quadrature plane, and measure the resultant amplified and entangled output fields at Bob and Alice shown in Fig.\,\ref{EntSwapQuantCoh}(b) and \ref{EntSwapQuantCoh}(c), respectively. Note that the field measured at Bob is largely a phase-preserving amplified version of the input signal reflecting off port b of Entangler 2, while the one measured at Alice represents a teleported state of the entangled amplified signal at the output of port a of Entangler 2. Also, similar to the vacuum-state input case (i.e., Fig.\,\ref{EntSwapQuanVac}(f) and \ref{EntSwapQuanVac}(j)), we show that TMSQ (Fig.\,\ref{EntSwapQuantCoh}(d)) and two-mode entanglement (Fig.\,\ref{EntSwapQuantCoh}(e)) can also be attained for applied coherent states in the entanglement swapping setup.

\section{Discussion}
Following the above demonstrations, we highlight three key aspects that can be improved in order to enhance the overall performance of our CV entanglement setups. First, is reducing or eliminating the intermediate losses between the JMs and between the JMs and the directional couplers. Improving this aspect would be impactful since our theoretical modeling and fits to the data show that the various measured entanglement metrics and teleportation fidelities are primarily limited by such losses. One concrete example which shows the detrimental effect of loss on CV entanglement can be seen in Fig.\,\ref{TelQuantVac}(d). While the calculated and measured teleportation fidelities are limited to about $0.73$ due to intermediate losses in the system, the predicted fidelity in the lossless case exceeds $0.95$ for $G_E=6$ dB. However, eliminating these losses can be challenging. This is because considerable portions of the intermediate losses in the teleportation and entanglement swapping setups originate from commercial ferrite-based circulators which are needed to separate incoming and outgoing signals propagating on the same transmission lines and provide enough isolation between the different JMs (see Appendix E). Thus, eliminating these losses would require the development and deployment of lossless on-chip circulators \cite{CircImplement,CircOctaveDesign} and nondegenerate through Josephson amplifiers that are capable of generating TMSQ in transmission rather than reflection \cite{NDTA}. A similar functionality to the latter device could possibly be achieved using traveling-wave parametric amplifiers \cite{ObsTwoModSqTWPA,BroadbandCVKerrJMetamaterial,DualPumpTWPA} and lossless diplexers designed to filter out the pump tones while allowing entangled modes at different frequencies, generated by the amplifiers, to travel on separate transmission lines. Second, is enhancing the saturation power of JMs, especially those that fulfill the role of which-path information erasers such as, Alice and Claire in the teleportation and entanglement swapping setups, respectively. Having JMs in these roles with high saturation powers enable them to process larger coherent states and tolerate higher power gains on the Entanglers, which ideally correspond to larger squeezing and entanglement. While the present saturation power of JMs ranges between $-140$ dBm and $-130$ dBm at $20$ dB (which amounts to a few photons), has not considerably limited the teleportation fidelity and entanglement swapping metrics achieved in this work, it limited the gains on the erasers to less than $15$ dB, which in turn limited the maximum allowable attenuation of the directional couplers to $10$ dB. It also limited the maximum input power of coherent states that can be teleported or subjected to entanglement-swapping operations. One promising venue for enhancing the saturation power of JMs is to optimize certain inductance ratios that characterize their JRMs and resonators \cite{OptJPC}. Third, is minimizing the arrival time difference of the various analogue signals in the setup by optimizing the distances or cable lengths between the different parties, namely Alice, Bob, Claire, and the Entanglers. In our experiments, these distances were predominantly set by the relative locations of the various components inside the fridge and result in a few nanosecond difference in the propagation time along the different paths. Another likely contributor to arrival time differences is the signal processing time at the which-path information erasers, i.e., Alice and Claire in the teleportation and entanglement swapping setups, respectively, which is inversely proportional to their bandwidths and therefore falls in the range of a few tens of nanonsecond. To minimize the effect of such timing differences in our experiment, we measure the output signals in steady state averaged over $1$ to $2$ microseconds.

Furthermore, it is worth noting that using CV-based entanglement methods which employ broadband two-mode squeezers such as TWPAs or large-bandwidth JMs realized via impedance matching techniques \cite{JPAimpedanceEng,StrongEnvCoupling,SynParamCouplNet}, could offer a key scalability advantage over DV-based methods, in particular if they enable frequency-multiplexed entanglement operations between arrays of qubits, which result in fewer links between modules than DV-based methods which typically require one quantum link for each pair of qubits. Such reduction in the number of links could help to alleviate possible connectivity bottlenecks in large systems.
 
Finally, it is important to emphasize that to generate high-fidelity remote qubit entanglement using CV entanglers similar to the ones demonstrated in this work requires the realization of efficient, high success rate CV to DV conversion and distillation schemes. To this end, several promising theoretical protocols have been proposed recently \cite{HybEntDist,LongDistNDampl,JPCRemoteEnt}. Further innovations will likely be necessary though to realize high-performance CV to DV distillation processes that enable large distributed quantum computers as well as experimental realizations of such schemes to uncover the practical limitations on the rate and fidelity of the distilled DV pairs. 

\section{Summary}
In this work, we show that JMs can generate fast, high-quality CV entanglement between two-different microwave modes traveling on separate transmission lines. At the optimal working point, we measure for the entangled modes logarithmic negativity of $1.5$ and generation rate of $70$ Mebits/sec. By taking advantage of this built-in nondegenerate property, we utilize JMs in their roles as two-mode entanglers and which-path information erasers to form hardware-efficient apparatus capable of performing quantum teleportation and entanglement swapping of input vacuum and coherent states. With the teleportation scheme, we achieve average quantum teleportation fidelities in the range $0.62-0.73$ for coherent states having up to $1.2$ input photons and a maximum fidelity of $0.73$ for vacuum states, which in both cases surpass the classical teleportation limit of $1/2$ and are primarily limited by intermediate losses in the setup. Likewise, we demonstrate with the entanglement swapping apparatus, entanglement generation between two remote modes that lack any prior interaction with one another. At the optimal working point, we measure for the entangled modes produced by the entanglement swapping setup logarithmic negativity of $0.53$ and effective generation rate of $6.7$ Mebits/sec.

In addition to potentially enabling quantum communication in the microwave domain and medium-range remote entanglement between quantum modules, the CV entangling building blocks presented here could serve as a verification and optimization testbed for a variety of theoretical protocols \cite{HybEntDist,LongDistNDampl} designed for establishing long-range links via optical and microwave mode squeezers, especially since such squeezers are still in their early stage of development \cite{MwOptEnt}. Furthermore, such CV entanglement schemes could find numerous applications in the areas of quantum cryptography \cite{QuanKeyDistNat,QuantCryptoTwoOrthStates,EXpQuantCrypto} and quantum illumination \cite{Quanillum}.   

\section*{ACKNOWLEDGMENTS}
We are grateful to Martin Sandberg for technical assistance, Nick Bronn, and Vincent Arena for printed circuit board design, Sarunya Bangsaruntip for device fabrication, and Luke Govia for helpful discussions.  

B.A. conceived and performed the experiments, developed the theory, and analyzed the data, W.S. developed the data-taking software, O.J. and J. R. made significant contributions to the experimental setup, J. O. supervised the project. B.A. wrote the paper with input from the other authors.

The authors declare no competing interests.
 
\appendix
 
\section{Nondegenerate two-mode squeezer}
	
	To obtain the output fields of the JM, we apply the unitary two-mode squeeze operator  $S=\exp{\left(re^{i\varphi_p}a^{\dagger}_{\rm{in}}b^{\dagger}_{\rm{in}}-re^{-i\varphi_p}a_{\rm{in}}b_{\rm{in}} \right) }$ to the input fields $a_{\rm{in}}$ and $b^{\dagger}_{\rm{in}}$ \cite{GenEntMwRadoverTL,CVreview}, 
	
	\begin{align}
		\begin{array}
			[c]{cc}%
			a_{\rm{out}}=S^{\dagger}a_{\rm{in}}S=\cosh\left(r \right) a_{\rm{in}}+e^{i\varphi_p}\sinh\left(r \right)b^{\dagger}_{\rm{in}}  ,\\
			b^{\dagger}_{\rm{out}}=S^{\dagger}b^{\dagger}_{\rm{in}}S=\cosh\left( r\right) b^{\dagger}_{\rm{in}}+e^{-i\varphi_p}\sinh\left( r\right) a_{\rm{in}},
		\end{array}
		\label{a_b_out_squeezer}
	\end{align}
	
	\noindent where $\varphi_p$ is the phase of the pump and the squeezing parameter $r\geq0$ relates to the device power gain by $G_J=\cosh^{2}(r)$.
	
	In particular, when acting on the vacuum state at the input, the two-mode squeeze operator turns it into a two-mode squeezed vacuum state at the output $\left|  \rm{TMS}\right\rangle=S\left| 0\right\rangle_{a}\left| 0\right\rangle_{b}=\cosh^{-1}\left( r \right)\sum\limits_{n}\left(\tanh \left( r\right)  \right)^n\left| n\right\rangle_{a}\left| n\right\rangle_{b}$, which represents a superposition of twin Fock states at the frequencies of modes a and b propagating on different physical channels. 
	
	Writing the field quadratures of the bosonic modes $a$ and $b$ in terms of the annihilation and creation operators, we have 
	
	\begin{align}
		I_{a}=\Re\left( a\right) =\dfrac{a+a^{\dagger}}{2}, 
		\quad Q_{a} =\Im\left( a\right)=\dfrac{a-a^{\dagger}}{2i},  
		\nonumber \\ 
		I_{b}=\Re\left( b\right) =\dfrac{b+b^{\dagger}}{2}, 
		\quad Q_{b} =\Im\left( b\right)=\dfrac{b-b^{\dagger}}{2i}, 	
		\label{IQxp_ab}
	\end{align}
	
	\noindent which are also equivalent to the dimensionless position and momentum coordinates of the harmonic oscillators associated with modes a and b or analogous to the same conjugate coordinates characterizing the motion of a pair of EPR particles 1 and 2
	
	\begin{align}
		x_{1}=\Re\left( a\right) =\dfrac{a+a^{\dagger}}{2}, 
		\quad p_{1} =\Im\left( a\right)=\dfrac{a-a^{\dagger}}{2i},
		\nonumber \\ \quad x_{2}=\Re\left( b\right) =\dfrac{b+b^{\dagger}}{2}, 
		\quad p_{2} =\Im\left( b\right)=\dfrac{b-b^{\dagger}}{2i},   \label{IQxp_12}
	\end{align}
	
	\noindent whose commutation relations for the input fields satisfy $[x_j,p_j]=i/2$, where $j=1,2$. It follows that  $\left\langle \left( \Delta x_j \right) ^2\right\rangle=\left\langle \left( \Delta p_j \right) ^2\right\rangle=1/4$ for input coherent states including vacuum and satisfy the Heisenberg limit $\left\langle \left( \Delta x_j \right) ^2\right\rangle \left\langle \left( \Delta p_j \right) ^2\right\rangle=1/16$, where  $\left\langle \left( \Delta x_j \right) ^2\right\rangle\equiv\left\langle x^{2}_j\right\rangle-\left\langle x_j\right\rangle^{2}$. In the expressions above, $\Re\left( .\right)$ and $\Im\left( .\right)$ refer to the real and imaginary parts.  
	
	While considering each output field separately, we obtain an amplified thermal state with variance $\left\langle \left( \Delta a_{\rm{out}}\right) ^{2}\right\rangle =\left\langle \left( \Delta x_{\rm{out,1}}\right) ^{2} \right\rangle+\left\langle \left( \Delta p_{\rm{out,1}}\right) ^{2}\right\rangle=\cosh\left( 2r\right)/2 $. We uncover, however, nonlocal two-mode squeezing effects when evaluating the combinations of the output fields given by
	
	\begin{equation}
		a_{\rm{out}}\pm e^{i\varphi_p}b^{\dagger}_{\rm{out}}=e^{\pm r}\left(a_{\rm{in}} \pm e^{i\varphi_p} b^{\dagger}_{\rm{in}} \right). \label{nonlocalsqueezing} 
	\end{equation}
	
	In particular for $\varphi_p=0$, the relations given by Eq.\,(\ref{nonlocalsqueezing}) exhibit cross correlations between $x_{\rm{out,1}}$ and $x_{\rm{out,2}}$, and $p_{\rm{out,1}}$ and $p_{\rm{out,2}}$ 
	
	\begin{align}
		\begin{array}
			[c]{cccc}%
			x_{\rm{out,1}}+x_{\rm{out,2}}=e^{r}\left( x_{\rm{in,1}}+x_{\rm{in,2}}\right)   ,\\
			x_{\rm{out,1}}-x_{\rm{out,2}}=e^{-r}\left( x_{\rm{in,1}}-x_{\rm{in,2}}\right)   ,\\
			p_{\rm{out,1}}+p_{\rm{out,2}}=e^{-r}\left( p_{\rm{in,1}}+p_{\rm{in,2}}\right)   ,\\
			p_{\rm{out,1}}-p_{\rm{out,2}}=e^{r}\left( p_{\rm{in,1}}-p_{\rm{in,2}}\right)   ,\\
		\end{array}
		\label{nonlocalsqueezing2}
	\end{align}
	
	\noindent which demonstrate squeezing in the amplitude of the relative position  $x_{\rm{out,1}}-x_{\rm{out,2}}$ and total momentum $p_{\rm{out,1}}+p_{\rm{out,2}}$ of the EPR pair and antisqueezing along the orthogonal combinations. Furthermore, if we calculate the variance for the relative position and total momentum $\left\langle \left(x_1-x_2\right) ^2\right\rangle=\left\langle \left(p_1+p_2\right) ^2\right\rangle =e^{-2r}/2$, we find that for any $r>0$, the noise variance is squeezed below the standard vacuum limit of $1/2$, while for $\left\langle \left(x_1+x_2\right) ^2\right\rangle=\left\langle \left(p_1-p_2\right) ^2\right\rangle =e^{2r}/2$ the noise variance is amplified above that limit. 
	
	In the phase-space spanned by the vector $\xi=\left(x_1,p_1,x_2,p_2 \right)$, the squeezing operation can be described by 
	
	\begin{align}
		S\left(r,\varphi_p \right)    =\left(
		\begin{array}
			[c]{cc}%
			\cosh\left( r\right)I_2  & \sinh\left( r\right)ZR\left(\varphi_p \right)  \\
			\sinh\left( r\right)ZR\left(\varphi_p \right) & \cosh\left( r\right)I_2 \\
		\end{array} 
		\right), \nonumber \\ \label{S_r_phi}
	\end{align}
	
	\noindent where 
	
	\begin{widetext}
		
		\begin{equation}
			I_{2}  =\left(
			\begin{array}
				[c]{cc}%
				1 & 0 \\
				0 & 1 
			\end{array}
			\right), 
			\quad Z  =\left(
			\begin{array}
				[c]{cc}%
				1 & 0 \\
				0 & -1 
			\end{array}
			\right), \quad \text{and} 
			\quad R\left(\varphi_p\right)   =\left(
			\begin{array}
				[c]{cc}%
				\cos\varphi_p & \sin\varphi_p \\
				-\sin\varphi_p & \cos\varphi_p 
			\end{array}
			\right) \cdot \label{IZRmat}
		\end{equation}
	\end{widetext}
	
	The resulting two-mode squeezed state which belongs to the class of Gaussian states can be expressed by the Wigner function, which takes the form of normalized Gaussian distributions \cite{CVreview}
	
	\begin{equation}
		W\left( \xi\right)=\dfrac{1}{4{\pi}^2\sqrt{\det{V}}}\exp{\left\lbrace-\dfrac{1}{2}\xi V^{-1} {\xi}^{\rm{{T}}} \right\rbrace }, \label{Wigner1}
	\end{equation}    
	
	\noindent which is fully characterized by its covariance matrix for the bipartite system denoted $V$, whose elements are given by 
	
	\begin{equation}
		V_{ij}=\dfrac{1}{2}\left\langle \xi_i \xi_j+ \xi_j \xi_i\right\rangle-\left\langle \xi_i\right\rangle \left\langle \xi_j\right\rangle, \label{Vij}
	\end{equation}
	
	\noindent where $\xi_{i}$, $\xi_{j}$ are components of $\xi$.  
	
	More explicitly, we can cast the two-mode squeezing covariance matrix $V$ in the form \cite{MwOptEnt}
	
	\begin{equation}
		V   =\left(
		\begin{array}
			[c]{cc}%
			V_{\rm{aa}} & V_{\rm{ab}} \\
			V^{\rm{T}}_{\rm{ab}} & V_{\rm{bb}} \\
		\end{array} 
		\right)=\left(
		\begin{array}
			[c]{cccc}%
			V_{\rm{11}} & 0 & \tilde{V}_{13} & \tilde{V}_{14}\\
			0 & V_{11} & \tilde{V}_{14} & -\tilde{V}_{13}\\
			\tilde{V}_{13} & \tilde{V}_{14} & V_{33} & 0\\
			\tilde{V}_{14} & -\tilde{V}_{13} & 0 & V_{33}\\
		\end{array}	\right) \label{V_TMS}
	\end{equation}
	
	\noindent where $V_{\rm{aa}}$, $V_{\rm{bb}}$, $V_{\rm{ab}}$ are real $2\times2$ matrices. The diagonal part of $V$, i.e., $V_{\rm{aa}}$, $V_{\rm{bb}}$ represent the variances of the quadratures of mode a and b (such as $x_{\rm{out,1}}$ and $p_{\rm{out,1}}$ for $V_{\rm{aa}}$), while the off-diagonal part, i.e., $V_{\rm{ab}}$ represents the covariance of the quadratures from both modes (such as $x_{\rm{out,1}}$ and $x_{\rm{out,2}}$). Since the nondegenerate two-mode squeezer is a phase preserving amplifier which equally amplifies both quadratures of the mode, we have $V_{11}=V_{22}$ and $V_{33}=V_{44}$ (in the ideal case $V_{11}=V_{22}=V_{33}=V_{44}$). Furthermore, due to the orthogonality of the $x$ and $p$ quadratures of the same mode, we have $V_{12}=V_{21}=V_{34}=V_{43}=0$. The off-diagonal part which embodies the two-mode squeezing can be expressed as
	
	\begin{equation}
		V_{\rm{ab}}=V_{13}\left[\cos\left(\varphi_p \right)Z+\sin\left( \varphi_p\right)X \right], \label{rawVab}
	\end{equation}     
	
	\noindent where $V_{13}\equiv\sqrt{\tilde{V}^2_{13}+\tilde{V}^2_{14}}$, $\tan\left(\varphi_p \right)\equiv\tilde{V}_{14}/\tilde{V}_{13}$, and
	
	\begin{equation}
		X  =\left(
		\begin{array}
			[c]{cc}%
			0 & 1 \\
			1 & 0 
		\end{array}
		\right) \cdot \label{Xmat}
	\end{equation}
	
	To further bring the covariance matrix into one of the standard forms and demonstrate the  inseparability of the two subsystems, we apply a unitary local operation $U_{\rm{loc}}=U_a \otimes U_b$, where $U_{a}$, $U_{b}$ are local operations on modes a and b, respectively. For our purpose, we choose $U_{\rm{loc}}$ that only rotates the b mode with phase $\varphi$ without altering mode a. This operation translates to a symplectic transformation $S\left(\varphi \right)=I_2\oplus R\left(\varphi \right) $, which acts on the off-diagonal part of the covariance matrix $V\left(\varphi \right)=S\left( \varphi\right) V S^{\rm{T}}\left( \varphi\right) $, yielding
	
	\begin{equation}
		V_{\rm{ab}}\left( \varphi \right) =V_{13}\left[\cos\left(\varphi_p-\varphi \right)Z+\sin\left( \varphi_p-\varphi\right)X \right]. \label{rotatedVab}
	\end{equation}  
	
	By choosing $\varphi_p-\varphi=m\pi$, where $m$ is an integer, the covariance matrix $V$ takes the standard form
	
	\begin{equation}
		V   =\left(
		\begin{array}
			[c]{cccc}%
			V_{\rm{11}} & 0 & V_{13} & 0\\
			0 & V_{11} & 0 & -V_{13}\\
			V_{13} & 0 & V_{33} & 0\\
			0 & -V_{13} & 0 & V_{33}\\
		\end{array}	\right). \label{V_TMS_standard}
	\end{equation}
	
	\subsection{Duan's criterion for entanglement}
	To demonstrate entanglement of modes a and b, we apply the sufficient condition presented in Refs. \cite{DuanSeparability,CVreview}, which states that for any separable system, there exists a lower bound for the total variance of a pair of EPR-type coordinates, such as $x_{-}=x_1-x_2$ and $p_{+}=p_1+p_2$. Thus, if for some local operation $U_{\rm{loc}}$ applied to the two subsystems, the inequality for the total variance given by 
	
	\begin{equation}
		\left\langle \left( \Delta x_{-} \right) ^2\right\rangle+\left\langle \left( \Delta p_{+} \right) ^2\right\rangle\geq1 \label{Duan1} 
	\end{equation}
	
	\noindent is violated, then the two subsystems are inseparable and therefore entangled. 
	
	Using the same $U_{\rm{loc}}$ introduced above, which only rotates mode b by $\varphi$ while leaving mode a unchanged, we get $x_{-}\left(\varphi \right) =\left( x_1-x_2\cos\varphi-p_2\sin\varphi \right)$ and $p_{+}\left(\varphi \right) \equiv\left( p_1-x_2\sin\varphi+p_2\cos\varphi\right)$ whose total variance can be written as
	
	\begin{align}
		\Delta_{\rm{EPR}}&=\left\langle \left( \Delta x_{-}(\varphi) \right) ^2\right\rangle+\left\langle \left( \Delta p_{+}(\varphi) \right) ^2\right\rangle \nonumber \\
		&=2\left(V_{11}+V_{33}-2V_{13}\cos\left( \varphi_p-\varphi\right)  \right),  \label{Duan2} 
	\end{align}
	
	\noindent where $\left\langle \left( \Delta x_{-}(\varphi) \right) ^2\right\rangle=\left\langle \left( \Delta p_{+}(\varphi) \right) ^2\right\rangle$.
	
	Note that $\Delta_{\rm{EPR}}$ is minimized for $\varphi_{-}=\varphi_p$ 
	
	\begin{equation}
		\Delta^{-}_{\rm{EPR}}=2\left(V_{11}+V_{33}-2V_{13} \right),  \label{EPRmin} 
	\end{equation}
	
	\noindent and corresponds to two-mode squeezing of modes a and b.
	
	Whereas, for $\varphi_{+}=\varphi-\pi$, $\Delta_{\rm{EPR}}$ is maximized
	
	\begin{equation}
		\Delta^{+}_{\rm{EPR}}=2\left(V_{11}+V_{33}+2V_{13} \right),  \label{EPRmax} 
	\end{equation}
	
	\noindent and corresponds to the case of anti-squeezing.
	
	\subsection{Simon criterion for entanglement}
	Another useful criterion for verifying the entanglement of two-mode Gaussian states is Simon's continuous-variable version of the Perez-Horodecki partial-transpose criterion \cite{PeresHorodeckiSeparability}, which for the standard form of the covariance matrix presented in Eq.\,(\ref{V_TMS_standard}) reads \cite{CVreview}
	
	\begin{equation}
		16\left(V_{11}V_{33}-V^2_{13}\right)^2\geq \left(V_{11}+V_{33}\right)+2V^2_{13}-\dfrac{1}{16}.  \label{Simon1} 
	\end{equation}
	
	\begin{figure*}
		[tb]
		\begin{center}
			\includegraphics[
			width=2\columnwidth 
			]%
			{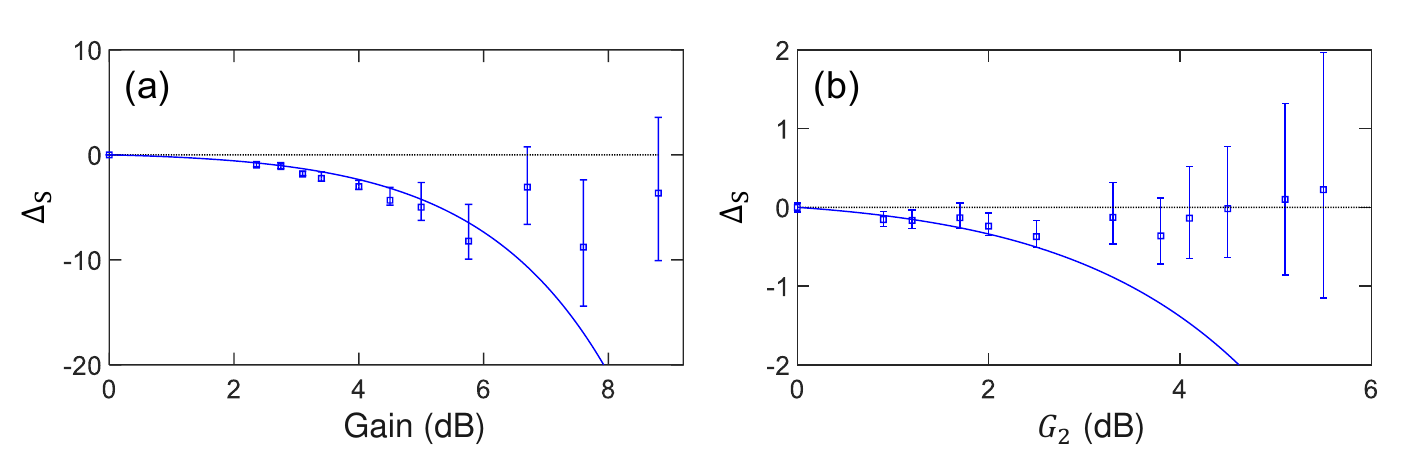}
			\caption{Simon's criterion for entanglement. (a) Violation of Simon's criterion (Eq.\,(\ref{Simon2})) for the TMSQ device presented in Fig.\,\ref{TwoModeSq}, plotted as a function of the device gain. The solid blue curve represents a theoretical calculation of $\Delta_{\rm{S}}$ which uses the entries of the covariance matrix in the presence of asymmetric loss (Eq.\,(\ref{V_TMS_standard_JM_asyloss})). (b) same as (a) but applies to the entanglement swapping experiment presented in Fig.\,\ref{EntSwapQuanVac}, plotted as a function of gain on Entangler 2 (while Entangler 1 gain is fixed at $G_1=1.4$ dB). Similarly, the solid blue curve represents a theoretical calculation of $\Delta_{\rm{S}}$, but uses the entries of the covariance matrix derived for the entanglement swapping case in the presence of loss (see Eqs.\,(\ref{mean_x_A_sq2})-(\ref{mean_pApB_sq2})).  
			}
			\label{SimonCr}
		\end{center}
	\end{figure*}
	
	Satisfying this inequality is a necessary and sufficient condition for the separability of any bipartite Gaussian states. Conversely, the violation of the inequality is a necessary and sufficient condition for inseparability and therefore entanglement of modes a and b.
	
	By defining a new variable
	
	\begin{align}
		\Delta_{\rm{S}}\equiv16\left(V_{11}V_{33}-V^2_{13}\right)^2 -\left(V_{11}+V_{33}\right)-2V^2_{13}+1/16, \nonumber \\  \label{DeltaS} 
	\end{align}  
	
	\noindent we can express the inequality \ref{Simon1} in a compact form 
	
	\begin{equation}
		\Delta_{\rm{S}}\geq 0.  \label{Simon2} 
	\end{equation}    
	
	In Fig.\,\ref{SimonCr} we exhibit two examples in which the above inequality is violated, thus indicating entanglement of the bipartite states. The examples presented in panels (a) and (b) belong to the TMSQ device and entanglement-swapping experiment whose two-mode squeezing and entanglement results are shown in Fig.\,\ref{TwoModeSq} and Fig.\,\ref{EntSwapQuanVac}, respectively. In panel (a) the parameter $\Delta_{\rm{S}}$ is plotted as a function of the JM gain, whereas in panel (b) it is plotted versus the gain of Entangler 2 while the gain on Entangler 1 is fixed at $1.4$ dB. The solid blue curves represent theoretical calculations of $\Delta_{\rm{S}}$ using the entries of the covariance matrix given by Eq.\,(\ref{V_TMS_standard_JM_asyloss}) for panel (a) and the entanglement-swapping covariance matrix given by Eqs.\,(\ref{mean_x_A_sq2})-(\ref{mean_pApB_sq2}) for panel (b). Note that in both cases Simon's criterion for entanglement is well aligned with the other entanglement measures applied in the respective cases, in particular the logarithmic negativity and EOF. 
	
	\subsection{Entanglement measures}
	
	\subsubsection{Logarithmic negativity}
	
	One of the important measures for quantifying entanglement of a bipartite state is the logarithmic negativity, which is directly related to the smallest symplectic eigenvalue of the partially transposed covariance matrix given by \cite{QuantumNode,ObsTwoModSqTWPA} 
	
	\begin{equation}
		\nu_{-}=\sqrt{\dfrac{\Delta\left(V \right)-\sqrt{\left( \Delta V\right)^2-4\det\left( V\right)  } }{2}},  \label{numinus} 
	\end{equation}
	
	\noindent where 
	
	\begin{equation}
		\Delta\left(V \right)=\det\left(V_{\rm{aa}} \right)+\det\left(V_{\rm{bb}}\right)-2\det\left( V_{\rm{ab}}  \right).  \label{DeltaV} 
	\end{equation}
	
	Using $\nu_{-}$, the logarithmic negativity for Gaussian states can be evaluated by
	
	\begin{equation}
		E_{N}=\max\left[  0,-\log_{2}\left( 4\nu_{-}\right) \right] . \label{Neglog1} 
	\end{equation}
	
	The two modes are thus entangled when $E_N>0$, which is satisfied when $4\nu_{-}<1$. The positivity of $E_N$ serves not only as a necessary and sufficient condition for entanglement, but also as an upper bound on the number of equivalent entangled bits (i.e., Bell pairs) that can be extracted from the entangled quantum state using distillation procedures.
	
	Note that in the symmetric case where $V_{11}=V_{33}$ the symplectic eigenvalue of the covariance matrix of Eq.\,(\ref{V_TMS_standard}), simplifies to $\nu_{-}=V_{11}-V_{13}$.
	
	\subsubsection{Entanglement of formation}
	
	Another important measure for characterizing entanglement is the entanglement of formation (EOF) \cite{GenEntMwRadoverTL,EF,EFArbtwomodeGauss} which quantifies the minimum amount of quantum bits (i.e., Bell pairs) required to create the quantum state as an ensemble of pure states. For a pure state, EOF reduces to the entanglement entropy which has an analytical solution for symmetric two-mode Gaussian states with a covariance matrix in the standard form of Eq.\,(\ref{V_TMS_standard}) and $V_{11}=V_{33}$. In this case, the EOF reads
	
	\begin{equation}
		E_{F}=\max\left[  0, h \left( \tilde{\nu}_{-} \right) \right], \label{EOF1} 
	\end{equation}   
	
	\noindent where $h \left( \tilde{\nu}_{-} \right)$ is defined as
	
	\begin{align}
		h\left(\tilde{\nu}_{-} \right)\equiv &\dfrac{\left(1+\tilde{\nu}_{-} \right)^2 }{4\tilde{\nu}_{-}}\log_{2}\left( \dfrac{\left(1+\tilde{\nu}_{-} \right)^2 }{4\tilde{\nu}_{-}}\right) \nonumber \\  &-\dfrac{\left(1-\tilde{\nu}_{-} \right)^2 }{4\tilde{\nu}_{-}}\log_{2}\left( \dfrac{\left(1-\tilde{\nu}_{-} \right)^2 }{4\tilde{\nu}_{-}}\right), \label{hx} 
	\end{align}   
	
	\noindent and $\tilde{\nu}_{-}=4\nu_{-}=4\left( V_{11}-V_{13}\right)$.
	
	\subsubsection{Purity of the entangled state}
	
	The purity of two-mode Gaussian state with covariance matrix $V$ reads \cite{LossAsymTWPA}
	
	\begin{equation}
		\mu=\dfrac{1}{16\sqrt{\det\left(V \right)}}, \label{purity} 
	\end{equation}
	
	\noindent which suggests that the purity condition for the Gaussian state ($\mu=1$) is met when $\det\left(V \right)=1/16^{2}$.  
	
	It is worth noting that unlike $E_N$ and $E_F$ which depend on both $\det \left( V\right) $ and $\Delta \left( V\right) $, which are invariant under the action of symplectic transformations, $\mu$ depends only on $\det \left( V\right)$. This implies that in general $E_N$ and $E_F$ provide a better characterization of the information content in the quantum state than $\mu$.
	
	\subsection{Loss modeling}
	
	In the lossless case, the covariance matrix of the nondegenerate Josephson mixer reads \cite{GenEntMwRadoverTL}
	
	\begin{widetext}
		
		\begin{equation}
			V   =S\left(r,0\right)V_{\rm{in}}S^{\rm{T}}\left(r,0\right)=\dfrac{1}{4} \left(
			\begin{array}
				[c]{cccc}%
				\cosh\left(2r \right)  & 0 & \sinh\left(2r \right) & 0\\
				0 & \cosh\left(2r \right) & 0 & -\sinh\left(2r \right)\\
				\sinh\left(2r \right) & 0 & \cosh\left(2r \right) & 0\\
				0 & -\sinh\left(2r \right) & 0 & \cosh\left(2r \right) \\
			\end{array}	\right), \label{V_TMS_standard_JM}
		\end{equation}
		
	\end{widetext}
	
	\noindent where $V_{\rm{in}}=I_4/4$ is the covariance matrix of the input vacuum noise for modes a and b.
	
	To quantify the effect of losses which the squeezed fields of modes a and b experience after leaving the device, we model the losses as beam-splitters which couple uncorrelated cold thermal noise to each mode, such that
	
	\begin{align}
		\begin{array}
			[c]{cc}%
			a_{\rm{out'}}=\sqrt{\bar{\alpha}}a_{\rm{out}}+\sqrt{\alpha}a_{\rm{th}}  ,\\
			b_{\rm{out'}}=\sqrt{\bar{\beta}}b_{\rm{out}}+\sqrt{\beta}b_{\rm{th}},
		\end{array}
		\label{a_b_out_BS}
	\end{align}
	
	\noindent where $\alpha$, $\beta$ are the effective lumped-element power losses in the path of modes a and b which satisfy the standard beam-splitter conditions $\bar{\alpha}=1-\alpha$, $\bar{\beta}=1-\beta$, and $a_{\rm{th}}$, $b_{\rm{th}}$ are the bosonic modes of the thermal baths at frequencies $f_a$ and $f_b$, respectively.   
	
	Thus, the covariance matrix in the presence of asymmetric loss becomes \cite{ObsTwoModSqTWPA}
	
	\begin{widetext}
		
		\begin{equation}
			V'   =\dfrac{1}{4} \left(
			\begin{array}
				[c]{cccc}%
				\bar{\alpha}\cosh\left(2r \right)+\alpha  & 0 & \sqrt{\bar{\alpha}\bar{\beta}}\sinh\left(2r \right) & 0\\
				0 & \bar{\alpha}\cosh\left(2r \right)+\alpha & 0 & -\sqrt{\bar{\alpha}\bar{\beta}}\sinh\left(2r \right)\\
				\sqrt{\bar{\alpha}\bar{\beta}}\sinh\left(2r \right) & 0 & \bar{\beta}\cosh\left(2r \right)+\beta & 0\\
				0 & -\sqrt{\bar{\alpha}\bar{\beta}}\sinh\left(2r \right) & 0 & \bar{\beta}\cosh\left(2r \right)+\beta \\
			\end{array}	\right). \label{V_TMS_standard_JM_asyloss}
		\end{equation}
		
	\end{widetext}
	
	While in the symmetric loss case (i.e., $\alpha=\beta$), Eq.\,(\ref{V_TMS_standard_JM_asyloss}) reads
	
	\begin{widetext}
		
		\begin{align}
			V'   =\dfrac{1-\alpha}{4} \left(
			\begin{array}
				[c]{cccc}%
				\cosh\left(2r \right)+\dfrac{\alpha}{1-\alpha}  & 0 & \sinh\left(2r \right) & 0\\
				0 & \cosh\left(2r \right)+\dfrac{\alpha}{1-\alpha} & 0 & -\sinh\left(2r \right)\\
				\sinh\left(2r \right) & 0 & \cosh\left(2r \right)+\dfrac{\alpha}{1-\alpha} & 0\\
				0 & -\sinh\left(2r \right) & 0 & \cosh\left(2r \right)+\dfrac{\alpha}{1-\alpha} \\
			\end{array}	\right), \label{V_TMS_standard_JM_symloss}
		\end{align}
		
	\end{widetext}

	\noindent which verifies entanglement since $4\nu_{-}=\left( 1-\alpha\right)e^{-2r}+\alpha<1$ for any gain $r>0$ and limited loss $\alpha<1$.
	
	\subsubsection{The effect of loss on the entanglement measures}
	
	In the presence of losses in modes a and b, the variance $\left\langle \left( \Delta x_{-} \right) ^2\right\rangle$ of the joint quadrature $x_{-}=x_1-x_2$ at its minimum value (i.e., squeezing) can be derived from Eq.\,(\ref{EPRmin}) by replacing the terms $V_{11}, V_{33}$, and $V_{13}$ with the corresponding matrix elements of $V'$ given by Eq.\,(\ref{V_TMS_standard_JM_asyloss}) and using the relation $\left\langle \left( \Delta x_{-} \right) ^2\right\rangle=\Delta^{-}_{\rm{EPR}}/2$  
	
	\begin{widetext}
		
		\begin{equation}
			\left\langle \left( \Delta x_{-} \right) ^2\right\rangle=\dfrac{1}{2}\left[\left(1-\dfrac{\bar{\alpha}+\bar{\beta}}{2} \right)+\dfrac{e^{-2r}}{4}\left(\sqrt{\bar{\alpha}}+\sqrt{\bar{\beta}} \right)^2  +\dfrac{e^{2r}}{4}\left(\sqrt{\bar{\alpha}}-\sqrt{\bar{\beta}} \right)^2   \right], \label{asy_loss_Xm}
		\end{equation}
		
	\end{widetext}
	
	\noindent which is equivalent to the expression derived in Ref. \cite{LossAsymTWPA}.
	
	Similarly, we can obtain the variance $\left\langle \left( \Delta x_{+} \right) ^2\right\rangle$ of the orthogonal joint quadrature $x_{+}=x_1+x_2$ at its maximum value (i.e., antisqueezing) by substituting the corresponding matrix elements of $V'$ in Eq.\,(\ref{EPRmax}) and using the relation $\left\langle \left( \Delta x_{+} \right) ^2\right\rangle=\Delta^{+}_{\rm{EPR}}/2$    
	
	\begin{widetext}
		
		\begin{equation}
			\left\langle \left( \Delta x_{+} \right) ^2\right\rangle=\dfrac{1}{2}\left[\left(1-\dfrac{\bar{\alpha}+\bar{\beta}}{2} \right)+\dfrac{e^{2r}}{4}\left(\sqrt{\bar{\alpha}}+\sqrt{\bar{\beta}} \right)^2+\dfrac{e^{-2r}}{4}\left(\sqrt{\bar{\alpha}}-\sqrt{\bar{\beta}} \right)^2   \right]. \label{asy_loss_Xp}
		\end{equation}
		
	\end{widetext}
	
	Likewise, to calculate $E_N$ and $E_F$ in the presence of asymmetric losses in modes a and b, we evaluate Eq.\,(\ref{Neglog1}) and Eq.\,(\ref{EOF1}) while using the covariance matrix $V'$ (i.e., Eq.\,(\ref{V_TMS_standard_JM_asyloss})) to evaluate the symplectic eigenvalue $\nu_{-}$ of Eq.\,(\ref{numinus}). We also use $V'$ in Eq.\,(\ref{purity}) to evaluate the purity of the two-mode Gaussian state.    
	
	In the case of asymmetric losses and in the limit of large $r$ the logarithmic negativity approaches \cite{LossAsymTWPA}
	
	\begin{align}
		E_N\approx-\log_{2}\left[e^{-2r}+\epsilon\left(1-e^{-2r}\right)+\tanh\left( r\right)\epsilon^2\delta^2  \right], \nonumber \\ \label{EN_asym_loss2}
	\end{align}
	
	\noindent and the purity takes the approximate analytical form \cite{LossAsymTWPA}
	
	\begin{align}
		\mu \approx \dfrac{1}{1+2\left(1-\epsilon \right) \epsilon\left[\cosh\left( 2r\right)-1 \right]}-\left[\dfrac{\epsilon\delta}{2\left(1-\epsilon \right) \epsilon} \right] ^2 e^{-2r}, \nonumber \\ \label{mu_asym_loss2}
	\end{align}
	
	\noindent where $\epsilon=\left(\alpha+\beta \right)/2$ represents the average loss, while $\delta=\left(\alpha-\beta \right)/\left(\alpha+\beta\right)$ corresponds to the relative asymmetry in the loss. 
	
	In the symmetric loss case (i.e., $\alpha=\beta$), Eq.\,(\ref{asy_loss_Xm}) and Eq.\,(\ref{asy_loss_Xp}) reduce to
	
	\begin{equation}
		\left\langle \left( \Delta x_{-} \right) ^2\right\rangle=\dfrac{1}{2}\left[1-\bar{\alpha}+\bar{\alpha}e^{-2r} \right], \label{sym_loss_Xm}
	\end{equation}
	
	\noindent and 
	
	\begin{equation}
		\left\langle \left( \Delta x_{+} \right) ^2\right\rangle=\dfrac{1}{2}\left[1-\bar{\alpha}+\bar{\alpha}e^{2r} \right]. \label{sym_loss_Xp}
	\end{equation}
	
	Furthermore, the logarithmic negativity and state purity simplify to 
	
	\begin{equation}
		E_N=-\log_{2}\left[e^{-2r}+\alpha\left(1-e^{-2r}\right) \right], \label{EN_sym_loss}
	\end{equation}
	
	\noindent and 
	
	\begin{equation}
		\mu=\dfrac{1}{1+2\bar{\alpha}\alpha\left[\cosh\left( 2r\right)-1 \right]}. \label{mu_sym_loss}
	\end{equation}
	
	\subsection{Reconstruction of the covariance matrix}
	
	Since the measured histograms of the output fields contain thermal noise added by the output chain, we can reconstruct the two-mode squeezing covariance matrix at the output of the device using the relation \cite{QuantumNode,ObsTwoModSqTWPA} 
	
	\begin{equation}
		V=V^{\rm{on}}-V^{\rm{off}}+V_0, \label{TMS_V}
	\end{equation}
	
	\noindent where $V_0=I_4/4$ corresponds to the covariance matrix for the input vacuum noise, and $V^{\rm{on}}$, $V^{\rm{off}}$ correspond to the covariance matrices referenced back to the output of the two-mode squeezer, measured when the pump is on and off, respectively, 
	
	\begin{align}
		\begin{array}
			[c]{cc}%
			V^{\rm{on}}=V+V^{\rm{th}}  ,\\
			V^{\rm{off}}=V_0+V^{\rm{th}},
		\end{array}
		\label{V_on_off}
	\end{align}
	
	\noindent where $V^{\rm{th}}$ represents the diagonal covariance matrix of the uncorrelated thermal noise added by the output chain
	
	\begin{widetext}
		
		\begin{equation}
			V^{\rm{th}}   =\dfrac{1}{4} \left(
			\begin{array}
				[c]{cccc}%
				\left(1+2n^{\rm{th,sys}}_a\right)   & 0 & 0 & 0\\
				0 & \left(1+2n^{\rm{th,sys}}_a\right) & 0 & 0\\
				0 & 0 & \left(1+2n^{\rm{th,sys}}_b\right) & 0\\
				0 & 0 & 0 & \left(1+2n^{\rm{th,sys}}_b\right) \\
			\end{array}	\right), \label{V_thermal}
		\end{equation}
		
	\end{widetext}
	
	\noindent and $n^{\rm{th,sys}}_a=\left\langle h^{\dagger}_ah_a\right\rangle $, $n^{\rm{th,sys}}_b=\left\langle h^{\dagger}_bh_b\right\rangle $.
	
	\subsection{Which-path information eraser}
	
	The entanglement between the output fields of the nondegenerate two-mode squeezer originates from its action as a which-path information erasor for the input fields impinging on its ports which are spectrally and physically distinct. To demonstrate this fact we start with the input and output scattering relations of the JM which read \cite{QuantBAScience}
	
	\begin{align}
		\begin{array}
			[c]{cc}%
			a_{\rm{out}}=\sqrt{G_J}a_{\rm{in}}+\sqrt{G_J-1}b^{\dagger}_{\rm{in}}  ,\\
			b^{\dagger}_{\rm{out}}=\sqrt{G_J}b^{\dagger}_{\rm{in}}+\sqrt{G_J-1}a_{\rm{in}},
		\end{array}
		\label{a_b_out_eraser}
	\end{align}
	
	\noindent where we assume, without loss of generality, that $\varphi_p=0$. Using Eq.\,(\ref{a_b_out_eraser}) and Eq.\,(\ref{IQxp_ab}), the output \textit{I} and \textit{Q} quadratures of mode a when referred back to the input read
	
	\begin{align}
		\begin{array}
			[c]{cc}%
			\tilde{I}_{\rm{out,a}}=I_{\rm{out,a}}/\sqrt{G_J}=I_{\rm{in,a}}+\left( \sqrt{G_J-1}/\sqrt{G_J}\right) I_{\rm{in,b}}  ,\\
			\tilde{Q}_{\rm{out,a}}=Q_{\rm{out,a}}/\sqrt{G_J}=Q_{\rm{in,a}}-\left( \sqrt{G_J-1}/\sqrt{G_J}\right) Q_{\rm{in,b}},
		\end{array}
		\nonumber \\ \label{I_Q_out_eraser}
	\end{align}
	
	\noindent which in the limit of large gain $G_J\gg1$ reduces to
	
	\begin{align}
		\begin{array}
			[c]{cc}%
			\tilde{I}_{\rm{out,a}}=I_{\rm{in,a}}+I_{\rm{in,b}}  ,\\
			\tilde{Q}_{\rm{out,a}}=Q_{\rm{in,a}}-Q_{\rm{in,b}}.
		\end{array}
		\label{I_Q_out_eraser1}
	\end{align}
	
	Equation\,(\ref{I_Q_out_eraser1}) shows that in the high-gain limit, the quadratures of the output field, which constitute the measured physical observables of the field, contain symmetric combinations of the quadratures of its input fields. Hereby, measurement of the output field at either port does not convey information about the origin of the input signals. Also, note that $\tilde{I}_{\rm{out,a}}$ and $\tilde{Q}_{\rm{out,a}}$ commute, i.e., $\left[\tilde{I}_{\rm{out,a}},\tilde{Q}_{\rm{out,a}} \right] =0$. This means they can be measured simultaneously with arbitrary precision. A similar relation can be derived for the output quadratures $I$ and $Q$ of mode b.
	
	\section{Quantum teleporation of coherent states with nondegenerate two-mode squeezers}
	
	The calculations carried out in this section and the next show how the scheme of Fig.\,\ref{TeleportationScheme} performs a quantum teleportation of an unknown coherent state from Alice to Bob. 
	
	We start with the entangled fields at the output of the Entangler given by  
	
	\begin{align}
		\begin{array}
			[c]{cc}%
			a_{\rm{out},\textit{E}}=\cosh\left( r_{E}\right) a_{\rm{in},\textit{E}}+e^{i\varphi_E}\sinh\left( r_{E}\right) b^{\dagger}_{\rm{in},\textit{E}}  ,\\
			b^{\dagger}_{\rm{out},\textit{E}}=\cosh\left( r_{E}\right) b^{\dagger}_{\rm{in},\textit{E}}+e^{-i\varphi_E}\sinh\left( r_{E}\right)a_{\rm{in},\textit{E}}.
		\end{array}
		\label{a_b_out_E}
	\end{align}
	
	The Entangler field that reaches Alice reads
	
	\begin{equation}
		a_{\rm{in},\textit{A}}=\sqrt{\bar{\alpha}}a_{\rm{out},\textit{E}}+\sqrt{\alpha}a_{\rm{th},\textit{E}}  ,\label{EfieldatAlice}\\
	\end{equation}
	
	\noindent where we use  beam-splitter interactions with thermal baths to model the effect of losses experienced by modes a and b in their paths shown in Fig.\,\ref{TeleportationScheme}. Namely, we define $\bar{\beta}\equiv1-\beta$, $\bar{\beta}_c\equiv1-\beta_c$, $\bar{\alpha}\equiv1-\alpha$, $\bar{\beta}_f\equiv1-\beta_f$, and use $a_{\rm{th}}, b_{\rm{th}}$ to represent the bosonic modes of the thermal baths introduced by the loss in the paths of modes a and b.   
	
	Let $b^{\dagger}_{\rm{in},A}$ be the unknown input quantum state which Alice wants to teleport to Bob. The output of the effective projective joint measurement that Alice performs on the unknown input state and the Entangler field (sent to her by the EPR source) reads
	
	\begin{align}
		\begin{array}
			[c]{cc}%
			a_{\rm{out},\textit{A}}=\cosh\left( r_{A}\right) a_{\rm{in},\textit{A}}+e^{i\varphi_A}\sinh\left( r_{A}\right) b^{\dagger}_{\rm{in},\textit{A}}  ,\\
			b^{\dagger}_{\rm{out},\textit{A}}=\cosh\left( r_{A}\right) b^{\dagger}_{\rm{in},\textit{A}}+e^{-i\varphi_A}\sinh\left( r_{A}\right)a_{\rm{in},\textit{A}}.
		\end{array}
		\label{a_b_out_A}
	\end{align}
	
	Substituting Eq.\,(\ref{EfieldatAlice}) in Eqs.\,(\ref{a_b_out_A}) while using Eqs.\,( \ref{a_b_out_E}) yields
	
	\begin{align}
		b^{\dagger}_{\rm{out},\textit{A}} = &\cosh\left( r_{A}\right) b^{\dagger}_{\rm{in},\textit{A}} \nonumber \\ &+e^{-i\varphi_A}\sqrt{\bar{\alpha}}\sinh\left( r_{A}\right)\cosh\left( r_E\right)  a_{\rm{in},\textit{E}} \nonumber \\
		&+e^{i\varphi_{EA}}\sqrt{\bar{\alpha}}\sinh\left( r_A\right)\sinh\left( r_E\right)b^{\dagger}_{\rm{in},\textit{E}} \nonumber \\
		&+e^{-i\varphi_{A}}\sqrt{\alpha}\sinh\left( r_A\right)  a_{\rm{th},\textit{E}},  \label{b_out_sub}
	\end{align}
	
	\noindent and
	
	\begin{align}
		a_{\rm{out},\textit{A}} = & \sqrt{\bar{\alpha}}\cosh\left( r_{A}\right)\cosh\left(  r_E\right)  a_{\rm{in},\textit{E}} \nonumber \\
		&+e^{i\varphi_E}\sqrt{\bar{\alpha}}\cosh\left( r_{A}\right)\sinh\left( r_E\right)  b^{\dagger}_{\rm{in},\textit{E}}
		\nonumber \\
		&+e^{i\varphi_{A}}\sinh\left(r_A\right)b^{\dagger}_{\rm{in},\textit{A}}+\sqrt{\alpha}\cosh\left( r_A\right) a_{\rm{th},\textit{E}}, \nonumber \\ \label{a_out_sub}
	\end{align}
	
	\noindent where $\varphi_{EA}\equiv\varphi_{E}-\varphi_{A}$. 
	
	The field at Bob's output, which represents the teleported state, is given by
	
	\begin{align}
		b^{\dagger}_{B} = & \sqrt{\beta_c}\left[ \sqrt{\bar{\beta}_f}b^{\dagger}_{\rm{out},\textit{A}}+\sqrt{\beta_f}b^{\dagger}_{\rm{th},\textit{A}}\right] \nonumber \\
		&+\sqrt{\bar{\beta_c}}\left[ \sqrt{\bar{\beta}}b^{\dagger}_{\rm{out},\textit{E}}+\sqrt{\beta}b^{\dagger}_{\rm{th},\textit{E}}\right]  .\label{bOutatBob}
	\end{align}
	
	The first term in the above expression represents the projective measurement result, which Alice feedforwards to Bob, while the second represents the field sent to Bob by the Entangler (i.e., the EPR source).
	
	Substituting Eqs.\,(\ref{a_b_out_E}) and (\ref{b_out_sub}) in Eq.\,(\ref{bOutatBob}) gives
	
	\begin{widetext}
		\begin{align}
			b^{\dagger}_{B}
			= & \sqrt{\beta_c\bar{\beta}_f}\cosh\left(r_A \right)b^{\dagger}_{\rm{in},A} 
			+e^{-i\varphi_E}\left[\sqrt{\beta_c\bar{\beta}_f\bar{\alpha}}e^{i\varphi_{EA}}\sinh\left( r_A\right)\cosh\left(r_E \right)+\sqrt{\bar{\beta}\bar{\beta}_c}\sinh\left( r_E\right)    \right]a_{\rm{in},\textit{E}} \nonumber \\
			& +\left[ \sqrt{\beta_c\bar{\beta}_f\bar{\alpha}}e^{i\varphi_{EA}}\sinh\left(r_A \right)\sinh\left(r_E \right)+\sqrt{\bar{\beta}\bar{\beta}_c}\cosh\left( r_E\right)    \right]b^{\dagger}_{\rm{in},\textit{E}} 
			+\sqrt{\beta_c\bar{\beta}_f\alpha}e^{-i\varphi_A}\sinh\left(r_A \right)a_{\rm{th},\textit{E}}+\sqrt{\beta_c\beta_f}b^{\dagger}_{\rm{th},\textit{A}}+\sqrt{\beta\bar{\beta_c}}b^{\dagger}_{\rm{th},\textit{E}}.  \nonumber \\ \label{bOutatBobsub}
		\end{align}
	\end{widetext}
	
	Under the condition of unity transmission in the feedforward channel $\sqrt{\beta_c\bar{\beta}_f}\cosh\left(r_A \right)=1$, large gain at Alice $\cosh\left( r_A\right) \gg1$ (mimicking a projective measurement) and small coupling at Bob $\beta_c\ll1$, we obtain  
	
	\begin{align}
		b^{\dagger}_{B}
		= & b^{\dagger}_{\rm{in},\textit{A}} \nonumber\\
		&+e^{-i\varphi_E}\left[\sqrt{\bar{\alpha}}e^{i\varphi_{EA}}\cosh\left( r_E\right)+\sqrt{\bar{\beta}}\sinh\left(r_E \right)   \right]a_{\rm{in},\textit{E}} \nonumber\\
		&+\left[\sqrt{\bar{\alpha}}e^{i\varphi_{EA}}\sinh\left( r_E\right) +\sqrt{\bar{\beta}}\cosh\left( r_E\right)  \right]b^{\dagger}_{\rm{in},\textit{E}} \nonumber\\
		&+\sqrt{\alpha}e^{-i\varphi_A}a_{\rm{th},\textit{E}}+\sqrt{\beta}b^{\dagger}_{\rm{th},\textit{E}}    ,\label{bOutatBob2}
	\end{align}
	
	\noindent where the approximations $\sqrt{\beta_c\bar{\beta}_f}\sinh\left(r_A \right)\cong1$ and $\bar{\beta}_c\backsimeq1$ have been used.  
	
	In the limit of vanishing insertion losses $\bar{\alpha},\bar{\beta}\longrightarrow1$ ($\alpha, \beta\longrightarrow0$), Eq.\,(\ref{bOutatBob2}) simplifies to
	
	\begin{align}
		b^{\dagger}_{B}=&b^{\dagger}_{\rm{in},\textit{A}}+e^{-i\varphi_E}\left[e^{i\varphi_{EA}}\cosh\left( r_E\right)+\sinh\left(r_E \right)   \right]a_{\rm{in},\textit{E}} \nonumber \\
		& +\left[e^{i\varphi_{EA}}\sinh\left( r_E\right) +\cosh\left( r_E\right)  \right]b^{\dagger}_{\rm{in},\textit{E}}    .\label{bOutatBob3}
	\end{align}
	
	In the special case when the relative pump phase between the Entangler and Alice's amplifier is set to $\varphi_{EA}=0$ (or integer multiples of $2\pi$), we get the original input state $b^{\dagger}_{\rm{in},\textit{A}}$ superimposed with an amplified input vacuum noise of the Entangler that is enhanced by an amplitude gain factor of $e^{r_E}$ 
	
	\begin{equation}
		b^{\dagger}_{B}=b^{\dagger}_{\rm{in},\textit{A}}+e^{r_E}\left[e^{-i\varphi_E}a_{\rm{in},\textit{E}}+b^{\dagger}_{\rm{in},\textit{E}}\right]. \label{bOutatBobAmp}\\
	\end{equation}
	
	On the other hand, when the relative pump phase between the Entangler and Alice's amplifier is set to $\varphi_{EA}=\pi$ (or odd integer multiples of $\pi$), we get the original input state $b^{\dagger}_{\rm{in},\textit{A}}$ superimposed with a squeezed input vacuum noise of the Entangler that is diminished by the same amplitude gain factor $e^{r_E}$ 
	
	\begin{equation}
		b^{\dagger}_{B}=b^{\dagger}_{\rm{in},\textit{A}}+e^{-r_E}\left[e^{-i\varphi_E}a_{\rm{in},\textit{E}}+b^{\dagger}_{\rm{in},\textit{E}}\right]. \label{bOutatBobSq}\\
	\end{equation}
	
	In turn, in the limit of large two-mode squeezing of the Entangler $r_E\gg0$, we obtain at Bob's output a complete and faithful teleportation of the unknown coherent state input at Alice, that is   
	
	\begin{equation}
		b^{\dagger}_{B}=b^{\dagger}_{\rm{in},\textit{A}}. \label{bOutatBobComplete}\\
	\end{equation}
	
	In comparison, when classical teleportation is conducted in which no entanglement is shared between Alice and Bob ($r_E=0$), Eq.\,(\ref{bOutatBob3}) becomes
	
	\begin{equation}
		b^{\dagger}_{B}=b^{\dagger}_{\rm{in},\textit{A}}+e^{-i\varphi_E}a_{\rm{in},\textit{E}}+b^{\dagger}_{\rm{in},\textit{E}}. \label{bOutatBobClassical}\\
	\end{equation}
	
	From these results, we see that while in quantum teleportation, it is possible to fully teleport the unknown coherent quantum state from Alice to Bob (Eq.\,( \ref{bOutatBobComplete})), in the classical case even in the absence of losses, the input state is teleported with unavoidable excess noise corresponding to vacuum noise added by Alice's amplifier (measurement device) and vacuum noise added at Bob's location (which are respectively represented by the second and third terms in Eq.\,(\ref{bOutatBobClassical})). 
	
	\section{Noise power of the teleported coherent state}
	
	The noise power measured with a spectrum analyzer is proportional to the symmetrized variance of the field operator
	
	\begin{equation}
		\left( \Delta b^{\dagger}_{B}\right) ^2=\dfrac{\left\langle\left\lbrace b_{B},b^{\dagger}_{B}\right\rbrace  \right\rangle }{2}-\left|\left\langle b^{\dagger}_{B}\right\rangle  \right| ^2, \label{bBNoisepower}\\
	\end{equation}
	
	\noindent where $\left\lbrace .,.\right\rbrace $ denotes the anticommutator.
	
	Under the conditions of unity transmission in the feedforward classical channel, i.e., $\sqrt{\beta_c\bar{\beta}_f}\cosh\left(r_A \right)=1$, large gain approximation at Alice $\sqrt{\beta_c\bar{\beta}_f}\sinh\left(r_A \right)\cong1$, and small coupling at Bob, i.e., $\beta_c\ll1$, substituting Eq.\,(\ref{bOutatBob2}) in Eq.\,(\ref{bBNoisepower}) yields
	
	\begin{align}
		\left(\Delta b^{\dagger}_{B}\right) ^2
		=&\left( \Delta b^{\dagger}_{\rm{in},\textit{A}}\right) ^2 \nonumber\\
		&+\left|\sqrt{\bar{\alpha}}e^{i\varphi_{EA}}\cosh\left( r_E\right)+\sqrt{\bar{\beta}}\sinh\left(r_E \right)   \right|^2\left(\Delta a_{\rm{in},\textit{E}}\right)^2 \nonumber\\
		&+\left|\sqrt{\bar{\alpha}}e^{i\varphi_{EA}}\sinh\left( r_E\right) +\sqrt{\bar{\beta}}\cosh\left( r_E\right)  \right|^2\left(\Delta b^{\dagger}_{\rm{in},\textit{E}}\right)^2 \nonumber\\
		&+\alpha \left( \Delta a_{\rm{th},\textit{E}}\right)^2 +\beta \left( \Delta b^{\dagger}_{\rm{th},\textit{E}}\right)^2, \label{bOutatBobNoisePow1}
	\end{align}
	
	\noindent which expands into
	
	\begin{widetext}
		
		\begin{align}
			\left(\Delta b^{\dagger}_{B}\right) ^2&
			=\left( \Delta b^{\dagger}_{\rm{in},\textit{A}}\right) ^2+\left[\bar{\alpha}\cosh^2\left(r_E\right) +\bar{\beta}\sinh^2\left(r_E \right)   +\sqrt{\bar{\alpha}\bar{\beta}}\sinh\left(2r_E \right)\cos\left(\varphi_{EA}\right)  \right]\left(\Delta a_{\rm{in},\textit{E}}\right)^2 \nonumber \\
			&+\left[\bar{\alpha}\sinh^2\left(r_E\right) +\bar{\beta}\cosh^2\left(r_E \right)   +\sqrt{\bar{\alpha}\bar{\beta}}\sinh\left(2r_E \right)\cos\left(\varphi_{EA}\right)\right] \left(\Delta b^{\dagger}_{\rm{in},\textit{E}}\right)^2 \nonumber\\
			&+\alpha \left( \Delta a_{\rm{th},\textit{E}}\right)^2 +\beta \left( \Delta b^{\dagger}_{\rm{th},\textit{E}}\right)^2. \label{bOutatBobNoisePow2}
		\end{align}
	\end{widetext}  
	
	Assuming balanced losses in the path of the entangled fields, i.e., $\bar{\alpha}=\bar{\beta}$ we get
	
	\begin{widetext}  
		
		\begin{align}
			\left(\Delta b_{B}\right) ^2
			=\left( \Delta b^{\dagger}_{\rm{in},\textit{A}}\right) ^2 +\bar{\alpha}\left[\cosh\left(2r_E\right) +\sinh\left(2r_E \right)\cos\left(\varphi_{EA}\right)  \right]\left[\left(\Delta a_{\rm{in},\textit{E}}\right)^2+\left(\Delta b^{\dagger}_{\rm{in},\textit{E}}\right)^2 \right]+\alpha \left[\left( \Delta a_{\rm{th},\textit{E}}\right)^2+\left( \Delta b^{\dagger}_{\rm{th},\textit{E}}\right)^2\right]. \nonumber \\ \label{bOutatBobNoisePow3}
		\end{align}
		
	\end{widetext}

	\begin{figure*}
		[tb]
		\begin{center}
			\includegraphics[
			width=2\columnwidth 
			]%
			{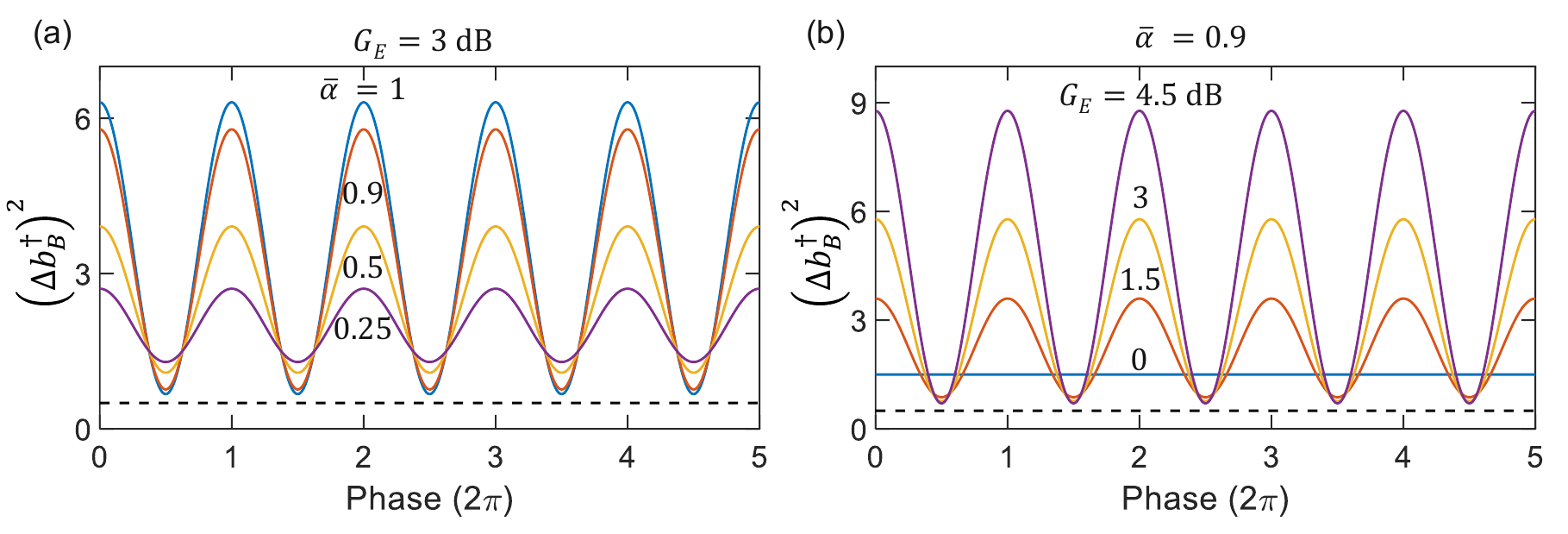}
			\caption{Calculated noise at Bob (in photons) as a function of the relative pump phase between the Entangler and Alice. The curves in (a) and (b) are calculated in response to a vacuum noise at Alice's input in the quantum teleportation setup (depicted in Fig.\,\ref{TeleportationScheme}). Quantum teleportation occurs at the minima points of the curves corresponding to odd multiples of $\pi$, where the teleported state approaches the vacuum state (i.e., $1/2$) indicated by the dashed black line. In both (a) and (b) the transmission loss is assumed to be symmetric, i.e., $\bar{\alpha}=\bar{\beta}$. While the curves in (a) are calculated for a constant $G_E=3$ dB with different transmission losses $\bar{\alpha}$, the curves in (b) are calculated for a constant $\bar{\alpha}=0.9$ with different Entangler gains $G_E$. The blue constant line in (b) corresponds to the classical teleportation case for which $G_E=0$ dB and equals $3/2$ as expected.   
			}
			\label{TelCVvacthy}
		\end{center}
	\end{figure*}
	
	When the relative phase between the pumps is set to $\varphi_{EA}=0$ (or integer multiples of $2\pi$), we get
	
	\begin{align}
		\left(\Delta b^{\dagger}_{B}\right) ^2
		= & \left( \Delta b^{\dagger}_{\rm{in},\textit{A}}\right) ^2+\bar{\alpha}e^{2r_E}\left[\left(\Delta a_{\rm{in},\textit{E}}\right)^2+\left(\Delta b^{\dagger}_{\rm{in},\textit{E}}\right)^2 \right] \nonumber \\
		&+\alpha \left[ \left( \Delta a_{\rm{th},\textit{E}}\right)^2+\left( \Delta b^{\dagger}_{\rm{th},\textit{E}}\right)^2\right]. \label{bOutatBobNoisePow4}
	\end{align}
	
	In this case the noise power of the teleported state $\left(\Delta b^{\dagger}_{B}\right) ^2$ exceeds that of the original input state $\left( \Delta b^{\dagger}_{\rm{in},\textit{A}}\right) ^2$ and the added noise by Alice and Bob is amplified by the factor $e^{2r_E}$.  
	
	On the other hand, when the relative phase between the pumps is set to $\varphi_{EA}=\pi$ (or odd integer multiples of $\pi$), we obtain
	
	\begin{align}
		\left(\Delta b^{\dagger}_{B}\right) ^2
		= & \left( \Delta b^{\dagger}_{\rm{in},\textit{A}}\right) ^2+\bar{\alpha}e^{-2r_E}\left[\left(\Delta a_{\rm{in},\textit{E}}\right)^2 +\left(\Delta b^{\dagger}_{\rm{in},\textit{E}}\right)^2 \right] \nonumber \\
		&+\alpha \left[\left( \Delta a_{\rm{th},\textit{E}}\right)^2+\left( \Delta b^{\dagger}_{\rm{th},\textit{E}}\right)^2\right],\label{bOutatBobNoisePow5}
	\end{align}
	
	\noindent which in this case exhibits suppression of the added noise by Alice and Bob by the same factor $e^{2r_E}$.
	
	If we further assume lack of losses in the paths of the entangled fields $\bar{\alpha}=1$ ($\alpha=0$), then in the limit of large squeezing by the Entangler $r_E\gg0$, the relation (\ref{bOutatBobNoisePow5}) reduces into
	
	\begin{equation}
		\left(\Delta b^{\dagger}_{B}\right) ^2
		=\left( \Delta b^{\dagger}_{\rm{in},\textit{A}}\right) ^2. \label{bOutatBobNoisePowQuant}
	\end{equation}
	
	This result demonstrates that in the quantum domain it is possible to teleport the coherent state faithfully from one place to another without adding any noise throughout the process. 
	
	In contrast, in classical teleportation, when Alice and Bob do not share any entangled fields from an EPR source (such as the Entangler), corresponding to the condition $r_E=0$, we obtain in the ideal lossless case
	
	\begin{equation}
		\left(\Delta b^{\dagger}_{B}\right) ^2
		=\left( \Delta b^{\dagger}_{\rm{in},\textit{A}}\right)^2+\left(\Delta a_{\rm{in},\textit{E}}\right)^2+\left(\Delta b^{\dagger}_{\rm{in},\textit{E}}\right)^2.\label{bOutatBobNoisePowClassical}
	\end{equation}
	
	Equation\,(\ref{bOutatBobNoisePowClassical}) shows that classical teleportation inevitably adds excess noise to the teleported state manifested by the second and third terms of the expression, which can be attributed to  vacuum noise added by Alice during measurement and Bob during construction of the state, respectively.  
	
	To illustrate the general behavior of the noise variance at Bob resulting from the teleportation setup (shown in Fig.\,\ref{TeleportationScheme}), we plot in Fig.\,\ref{TelCVvacthy} the variance $\left(\Delta b^{\dagger}_{B}\right)^2$ given by Eq.\,(\ref{bOutatBobNoisePow3}) as a function of the relative pump phase between the Entangler and Alice. The different curves depicted in panel (a) correspond to different power transmission parameters, i.e., $\bar{\alpha}$, while the Entangler gain is fixed $G_E=3$ dB, whereas in Fig.\,\ref{TelCVvacthy}(b) the curves correspond to different $G_E$, while $\bar{\alpha}=0.9$ is kept constant. The dashed black lines in both panels correspond to the vacuum variance equal to $1/2$. The horizontal blue line in Fig.\,\ref{TelCVvacthy}(b) corresponds to the classical teleportation scenario in which no entanglement is shared, i.e., $G_E=0$ dB, which yields $3/2$ in agreement with the result of Eq.\,(\ref{bOutatBobNoisePowClassical}). As seen in Fig.\,\ref{TelCVvacthy}, the noise variance is periodic with the pump phase. It is maximized (minimized) at even (odd) number multiples of $\pi$, where teleportation with maximum fidelity occurs at the minima points. Also, as expected, the achievable minimum of the teleported noise decrease as the loss in the setup is decreased (Fig.\,\ref{TelCVvacthy}(a)) or as the Entangler gain is increased (Fig.\,\ref{TelCVvacthy}(b)).
	
	\section{Entanglement swapping with nondegenerate two-mode squeezers}

	The calculation carried out in this section demonstrates how the entanglement swapping scheme of Fig.\,\ref{EntSwapQuanVac}(a) performs its operation. 
	
	We start with the entangled fields at the output of Entangler 1 given by  
	
	\begin{align}
		\begin{array}
			[c]{cc}%
			a_{\rm{out},1}=\cosh\left( r_1\right) a_{\rm{in},1}+e^{i\varphi_1}\sinh\left( r_1\right) b^{\dagger}_{\rm{in},1}  ,\\
			b^{\dagger}_{\rm{out},1}=\cosh\left( r_1\right) b^{\dagger}_{\rm{in},1}+e^{-i\varphi_1}\sinh\left( r_1\right)a_{\rm{in},1}.
		\end{array}
		\label{a_b_out_E1}
	\end{align}
	
	Likewise, the entangled fields at the output of Entangler 2 are given by
	
	\begin{align}
		\begin{array}
			[c]{cc}%
			a_{\rm{out},2}=\cosh\left( r_2\right) a_{\rm{in},2}+e^{i\varphi_2}\sinh\left( r_2\right) b^{\dagger}_{\rm{in},2}  ,\\
			b^{\dagger}_{\rm{out},2}=\cosh\left( r_2\right) b^{\dagger}_{\rm{in},2}+e^{-i\varphi_2}\sinh\left( r_2\right)a_{\rm{in},2}.
		\end{array}
		\label{a_b_out_E2}
	\end{align}
	
	The input fields at Claire's amplifier can be written as
	
	\begin{align}
		\begin{array}
			[c]{cc}%
			a_{\rm{in},3}=\sqrt{\bar{\alpha}_2}a_{\rm{out},2}+ \sqrt{\alpha_2}a_{\rm{th},2}  ,\\
			b^{\dagger}_{\rm{in},3}=\sqrt{\bar{\beta}_1} b^{\dagger}_{\rm{out},1}+\sqrt{\beta_1}b^{\dagger}_{\rm{th},1},
		\end{array}
		\label{a_b_in_Claire}
	\end{align}
	
	\noindent which in turn yield the following entangled fields at the output 
	
	\begin{align}
		\begin{array}
			[c]{cc}%
			a_{\rm{out},3}=\cosh\left( r_3\right) a_{\rm{in},3}+e^{i\varphi_3}\sinh\left( r_3\right) b^{\dagger}_{\rm{in},3}  ,\\
			b^{\dagger}_{\rm{out},3}=\cosh\left( r_3\right) b^{\dagger}_{\rm{in},3}+e^{-i\varphi_3}\sinh\left( r_3\right)a_{\rm{in},3}.
		\end{array}
		\label{a_b_out_Claire}
	\end{align}
	
	Substituting Eqs.\,(\ref{a_b_out_E1}), (\ref{a_b_out_E2}), (\ref{a_b_in_Claire}) in Eq.\,(\ref{a_b_out_Claire}) gives
	
	\begin{widetext}
		\begin{align}
			a_{\rm{out},3}
			=&\sqrt{\bar{\alpha}_2}\cosh\left(r_2 \right)\cosh\left(r_3 \right)a_{\rm{in},2}+\sqrt{\bar{\alpha}_2}e^{i\varphi_2}\sinh\left(r_2 \right)\cosh\left(r_3 \right)b^{\dagger}_{\rm{in},2}+\sqrt{\bar{\beta}_1}e^{i\varphi_3}\cosh\left( r_1\right)\sinh\left(r_3 \right)b^{\dagger}_{\rm{in},1} \nonumber \\
			&+\sqrt{\bar{\beta}_1}e^{i\varphi_{31}}\sinh\left(r_1 \right)\sinh\left(r_3 \right)a_{\rm{in},1}+\sqrt{\alpha_2}\cosh\left(r_3 \right)a_{\rm{th},2}+\sqrt{\beta_1}e^{i\varphi_3}\sinh\left(r_3 \right)b^{\dagger}_{\rm{th},1}.             \label{aOut3_1}
		\end{align}
	\end{widetext}
	
	The output field at Alice is given by
	
	\begin{align}
		a_{A}=& \sqrt{\beta_c}\left[ \sqrt{\bar{\alpha}_f}a_{\rm{out},3}+\sqrt{\alpha_f}a_{\rm{th},3}\right] \nonumber \\
		&+ \sqrt{\bar{\beta}_c}\left[ \sqrt{\bar{\alpha}_1}a_{\rm{out},1}+\sqrt{\alpha_1}a_{\rm{th},1}\right], \label{a_A1}
	\end{align}
	
	\noindent while the output field at Bob reads
	
	\begin{equation}
		b^{\dagger}_{B}=\sqrt{\bar{\beta}_2}b^{\dagger}_{\rm{out},2}+\sqrt{\beta_2}b^{\dagger}_{\rm{th},2}. \label{b_dagger_B1}
	\end{equation}
	
	Substituting Eq.\,(\ref{a_b_out_E1}) and Eq.\,(\ref{aOut3_1}) in Eq.\,(\ref{a_A1}) gives
	
	\begin{widetext}
		
		\begin{align}
			a_{A}
			=& \sqrt{\beta_c \bar{\alpha}_f \bar{\alpha}_2}\cosh\left(r_2 \right)\cosh\left( r_3\right)a_{\rm{in},2}+\sqrt{\beta_c \bar{\alpha}_f \bar{\alpha}_2}e^{i\varphi_2}\sinh\left( r_2\right)\cosh\left( r_3\right)b^{\dagger}_{\rm{in},2} \nonumber \\
			& +e^{i\varphi_1}\left[\sqrt{\beta_c \bar{\alpha}_f \bar{\beta}_1}e^{i\varphi_{31}}\cosh\left(r_1 \right)\sinh\left( r_3 \right)+\sqrt{\bar{\beta}_c \bar{\alpha}_1}\sinh\left( r_1\right)  \right]b^{\dagger}_{\rm{in},1} \nonumber \\
			& +\left[\sqrt{\beta_c \bar{\alpha}_f \bar{\beta}_1}e^{i\varphi_{31}}\sinh\left(r_1 \right)\sinh\left( r_3 \right)+\sqrt{\bar{\beta}_c \bar{\alpha}_1}\cosh\left( r_1\right)  \right]a_{\rm{in},1}  \nonumber \\
			& +\sqrt{\beta_c \bar{\alpha}_f \alpha_2}\cosh\left(r_3 \right)a_{\rm{th},2}+\sqrt{\beta_c \bar{\alpha}_f \beta_1}e^{i\varphi_3}\sinh\left( r_3 \right)b^{\dagger}_{\rm{th},1}+\sqrt{\beta_c \alpha_f}a_{\rm{th},3}+\sqrt{\bar{\beta}_c \alpha_1}a_{\rm{th},1}.               \label{a_A2}
		\end{align}
		
	\end{widetext}
	
	Requiring unity transmission through the feedforward classical channel, given by $\sqrt{\beta_c \bar{\alpha}_f}\cosh\left(r_3 \right)=1 $ and large gain at Claire given by $\sqrt{\beta_c \bar{\alpha}_f}\sinh\left(r_3 \right)\cong1 $, yields
	
	\begin{widetext}
		\begin{align}
			a_{A}
			= & \sqrt{\bar{\alpha}_2}\cosh\left(r_2 \right)a_{\rm{in},2}+\sqrt{\bar{\alpha}_2}e^{i\varphi_2}\sinh\left( r_2\right)b^{\dagger}_{\rm{in},2}  +e^{i\varphi_1}\left[\sqrt{ \bar{\beta}_1}e^{i\varphi_{31}}\cosh\left(r_1 \right)+\sqrt{\bar{\beta}_c \bar{\alpha}_1}\sinh\left( r_1\right)  \right]b^{\dagger}_{\rm{in},1} \nonumber \\
			& +\left[\sqrt{ \bar{\beta}_1}e^{i\varphi_{31}}\sinh\left(r_1 \right)+\sqrt{\bar{\beta}_c \bar{\alpha}_1}\cosh\left( r_1\right)  \right]a_{\rm{in},1}  
			+\sqrt{ \alpha_2}a_{\rm{th},2}+\sqrt{ \beta_1}e^{i\varphi_3}b^{\dagger}_{\rm{th},1}+\sqrt{\beta_c \alpha_f}a_{\rm{th},3}+\sqrt{\bar{\beta}_c \alpha_1}a_{\rm{th},1}.               \label{a_A3}
		\end{align}
	\end{widetext}
	
	In the absence of loss in the path between Entanglers 1 and 2 and Claire, i.e., $\bar{\alpha}_{1,2}=1$, $\bar{\beta}_{1,2}=1$ ($\alpha_{1,2}=0$, $\beta_{1,2}=0$), and small coupling at Alice $\beta_c\backsimeq0$ ($\bar{\beta}_c\backsimeq1$), Eq.\,(\ref{a_A3}) reduces to
	
	\begin{equation}
		a_{A}=a_{\rm{out},2}+\left[a_{\rm{out},1}+e^{i\varphi_3}b^{\dagger}_{\rm{out},1} \right],                \label{a_A4}
	\end{equation}
	
	\noindent which can also be cast in the form  
	
	\begin{align}
		a_{A}=&a_{\rm{out},2}+\left[\cosh\left(r_1 \right) +e^{i\varphi_{31}}\sinh\left( r_1\right) \right] a_{\rm{in},1} \nonumber \\
		&+\left[e^{i\varphi_{31}}\cosh\left(r_1 \right) +\sinh\left( r_1\right) \right] e^{i\varphi_1}b^{\dagger}_{\rm{in},1}.                \label{a_A5}
	\end{align}
	
	Setting $\varphi_{31}$, the relative phase of the pumps between Claire and Entangler 1, to $0$ (or integer multiples of $2\pi$), gives
	
	\begin{equation}
		a_{A}=a_{\rm{out},2}+e^{r_1}\left[a_{\rm{in},1} +e^{i\varphi_{1}}b^{\dagger}_{\rm{in},1} \right], \label{a_Aamp}
	\end{equation}
	
	\noindent whereas setting $\varphi_{31}$ to $\pi$ (or odd integer multiples of $\pi$), yields
	
	\begin{equation}
		a_{A}=a_{\rm{out},2}+e^{-r_1}\left[a_{\rm{in},1} -e^{i\varphi_{1}}b^{\dagger}_{\rm{in},1} \right]. \label{a_ASQ}
	\end{equation}
	
	In the limit of large gain on Entangler 1, corresponding to $r_1\gg0$, Eq.\,(\ref{a_ASQ}) becomes
	
	\begin{equation}
		a_{A}=a_{\rm{out},2}. \label{a_ASQlarge}
	\end{equation}
	
	This means that the output field of mode a of Entangler 2 is fully teleported to Alice. Moreover, since the output field at Bob in the lossless case is equal to $b^{\dagger}_{\rm{out},2}$ (see Eq.\,(\ref{b_dagger_B1})), the output fields of Alice and Bob become entangled. Thus, even though the fields of Entangler 1 ($a_{\rm{out},1}$) and 2 ($b^{\dagger}_{\rm{out},2}$) never interact directly, they become entangled through the joint measurement of $b^{\dagger}_{\rm{out},1}$ and $a_{\rm{out},2}$ performed by Claire and the classical signal she sends to Alice containing the measurement outcome. Therefore, this scheme effectively passes the entanglement generated at Claire between the received fields on to their distant non-interacting counterparts.
	
	\subsection{Covariance matrix of entanglement swapping} 
	
	Here we derive the covariance matrix that characterizes the correlations between the output fields at Alice and Bob upon applying the entanglement swapping operation depicted in Fig.\,\ref{EntSwapQuanVac}(a). 
	
	To calculate the covariance matrix elements given by Eq.\,(\ref{Vij}), we express the generalized position $x_{A}$, $x_{B}$ and momentum $p_{A}$, $p_{B}$ coordinates associated with the output fields at Alice and Bob as 
	
	\begin{align}
		x_{A}=\dfrac{a_{A}+a^{\dagger}_{A}}{2}, 
		\quad p_{A} =\dfrac{a_{A}-a^{\dagger}_{A}}{2i}, \nonumber \\ x_{B}=\dfrac{b_{B}+b^{\dagger}_{B}}{2}, 
		\quad p_{B} =\dfrac{b_{B}-b^{\dagger}_{B}}{2i},   \label{x_p_A_B}
	\end{align}
	
	\noindent where $a_{A}$ is given by Eq.\,(\ref{a_A3}) (assuming a large gain is applied at Claire and unity transmission in the classical channel) and $b^{\dagger}_{B}$ by Eq.\,(\ref{b_dagger_B1}). 
	
	Further assuming for simplicity that the phases of the pumps feeding the two entanglers and Claire satisfy $\varphi_1=0$, $\varphi_2=0$, and $\varphi_3=\pi$ and that the noise entering the two entanglers is uncorrelated and has zero mean value, we get  
	
	\begin{widetext}
		\begin{align}
			V_{11}=&\left\langle x^2_A\right\rangle \nonumber
			\\
			=&\bar{\alpha}_2\cosh^2\left( r_2\right)\left\langle x^2_{\rm{in,a2}}\right\rangle+\bar{\alpha}_2\sinh^2\left( r_2\right)\left\langle x^2_{\rm{in,b2}}\right\rangle + \left[\sqrt{\bar{\beta}_c\bar{\alpha}_1}\sinh\left( r_1\right)-\sqrt{\bar{\beta}_1}\cosh\left( r_1\right)   \right]^2 \left\langle x^2_{\rm{in,b1}}\right\rangle \nonumber\\
			&+\left[\sqrt{\bar{\beta}_c\bar{\alpha}_1}\cosh\left( r_1\right)-\sqrt{\bar{\beta}_1}\sinh\left( r_1\right)   \right]^2 \left\langle x^2_{\rm{in,a1}}\right\rangle +\alpha_2\left\langle x^2_{\rm{th,a2}}\right\rangle +\beta_1\left\langle x^2_{\rm{th,b1}}\right\rangle +\beta_c\alpha_f\left\langle x^2_{\rm{th,a3}}\right\rangle+\bar{\beta}_c\alpha_1\left\langle x^2_{\rm{th,a1}}\right\rangle,   \label{mean_x_A_sq}
			\\
			V_{33}=&\left\langle x^2_B\right\rangle \nonumber
			\\
			=&\bar{\beta}_2\cosh^2\left( r_2\right)\left\langle x^2_{\rm{in,b2}}\right\rangle+\bar{\beta}_2\sinh^2\left( r_2\right)\left\langle x^2_{\rm{in,a2}}\right\rangle+\beta_2\left\langle x^2_{\rm{th,b2}}\right\rangle,   \label{mean_x_B_sq}
			\\
			V_{13}=&\left\langle x_Ax_B\right\rangle \nonumber
			\\
			=&\sqrt{\bar{\beta}_2\bar{\alpha}_2}\cosh\left( r_2\right)\sinh\left( r_2\right)\left\langle x^2_{\rm{in,b2}}\right\rangle +\sqrt{\bar{\beta}_2\bar{\alpha}_2}\cosh\left( r_2\right)\sinh\left( r_2\right)\left\langle x^2_{\rm{in,a2}}\right\rangle,   \label{mean_xAxB_sq}
			\\
			V_{24}=&\left\langle p_Ap_B\right\rangle \nonumber
			\\
			=&-\sqrt{\bar{\beta}_2\bar{\alpha}_2}\cosh\left( r_2\right)\sinh\left( r_2\right)\left\langle p^2_{\rm{in,b2}}\right\rangle -\sqrt{\bar{\beta}_2\bar{\alpha}_2}\cosh\left( r_2\right)\sinh\left( r_2\right)\left\langle p^2_{\rm{in,a2}}\right\rangle,   \label{mean_pApB_sq}
		\end{align}
	\end{widetext}
	
	\noindent where the subscripts of the noise variances, e.g., $\left\langle x^2_{\rm{in,b2}}\right\rangle$, indicate the origin of the noise, i.e. input (in), thermal bath (th), its mode frequency a or b, and whether it traces back to fields from Entangler 1, 2 or Claire 3. Similarly, it is straightforward to show that $V_{22}=\left\langle p^2_A\right\rangle=V_{11}$, $V_{44}=\left\langle p^2_B\right\rangle=V_{33}$, $V_{31}=\left\langle x_Bx_A\right\rangle=V_{13}$, $V_{42}=\left\langle p_Bp_A\right\rangle=V_{24}$ and that $V_{12}=V_{21}=V_{23}=V_{32}=V_{14}=V_{41}=V_{34}=V_{43}=0$. 
	
	Considering vacuum noise at the input of Entanglers 1, 2 and the same added by the beamsplitters representing the losses between the stages, we obtain  
	
	\begin{align}
		V_{11}=&\dfrac{1}{4}\bar{\alpha}_2\cosh\left( 2r_2\right)+\dfrac{1}{4}\left[\bar{\beta}_1+\bar{\beta}_c\bar{\alpha}_1 \right]\cosh\left(2r_1 \right) \nonumber \\
		&-\dfrac{1}{2}\sqrt{\bar{\beta}_c\bar{\alpha}_1\bar{\beta}_1} \sinh\left(2r_1 \right) \nonumber \\
		&+\dfrac{1}{4}\left[\alpha_2+\beta_1+\beta_c\alpha_f+\bar{\beta}_c\alpha_1 \right],   \label{mean_x_A_sq2}
		\\
		V_{33}=&\dfrac{1}{4}\bar{\beta}_2\cosh\left( 2r_2\right)+\dfrac{1}{4}{\beta}_2,   \label{mean_x_B_sq2}
		\\
		V_{13}=&\dfrac{1}{4}\sqrt{\bar{\beta}_2\bar{\alpha}_2}\sinh\left( 2r_2\right),   \label{mean_xAxB_sq2}
		\\
		V_{24}=&-\dfrac{1}{4}\sqrt{\bar{\beta}_2\bar{\alpha}_2}\sinh\left( 2r_2\right). \label{mean_pApB_sq2}
	\end{align}
	
	In the absence of loss, i.e., $\beta_{1,2}=0,\alpha_{1,2,f}=0$, and having a small coupling $\beta_c\rightarrow0$, Eqs.\,(\ref{mean_x_A_sq2})-(\ref{mean_pApB_sq2}) become
	
	\begin{align}
		V_{11}&=\dfrac{1}{4}\cosh\left( 2r_2\right)+\dfrac{e^{-2r_1}}{2},   \label{mean_x_A_sq3}
		\\
		V_{33}&=\dfrac{1}{4}\cosh\left( 2r_2\right),   \label{mean_x_B_sq3}
		\\
		V_{13}&=\dfrac{1}{4}\sinh\left( 2r_2\right),   \label{mean_xAxB_sq3}
		\\
		V_{24}&=-\dfrac{1}{4}\sinh\left( 2r_2\right). \label{mean_pApB_sq3}
	\end{align}
	
	Equations\,(\ref{mean_x_A_sq3})-(\ref{mean_pApB_sq3}) show that: (1) in the limit of infinite gain on Entangler 1, i.e., $r_1\rightarrow\infty$, the covariance matrix associated with the entanglement swapping becomes indistinguishable from the covariance matrix of an ideal two-mode squeezer (see Eq.\,(\ref{V_TMS_standard_JM})) with a squeezing parameter $r_2$ of Entangler 2, (2) when Entangler 1 is idle, i.e. $r_1=0$, corresponding to classical teleportation of mode a of Entangler 2 to Alice, both quadratures $x_A$ and $p_A$ acquire excess noise of half a photon as seen in Eq.\,(\ref{mean_x_A_sq3}), representing $V_{11}$ and $V_{22}$, which is due to both Claire and Alice adding half a photon of noise to the processed signal as required by classical teleportation, (3) for finite two-mode squeezing on Entangler 1, such as small $r_1$, entanglement swapping scheme exhibits certain degradation in performance compared to an ideal two-mode squeezer even in the absence of losses as implied by the presence of the second term in Eq.\,(\ref{mean_x_A_sq3}).   
	
	\begin{figure*}
		[tbh]
		\begin{center}
			\includegraphics[
			width=2\columnwidth 
			]%
			{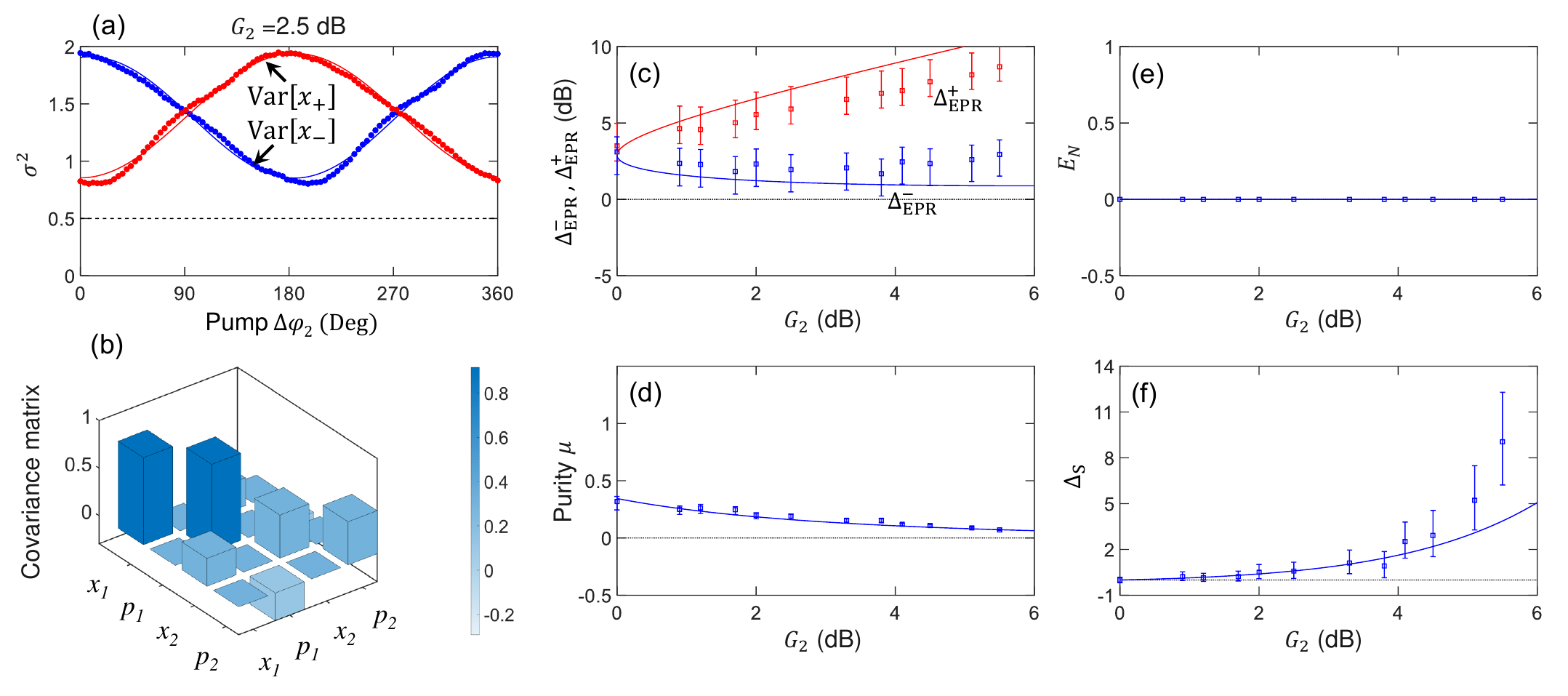}
			\caption{Entanglement swapping measurement with Entangler 1 off. (a)-(f) show the lack entanglement and squeezing when only one Entangler in the entanglement swapping scheme is on ($G_2>0$ dB), while the other is off ($G_1=0$ dB). (a) The variance of EPR-like coordinates $x_{-}$ (blue filled circles) and $x_{+}$ (red filled circles) are plotted as a function of the pump phase of Entangler 2 relative to Claire's. The relative pump phase axis is shifted to yield a minimum of $\rm{Var}[x_{-}]$ at $180$ degrees. Solid blue and red lines represent sinusoidal fits.  Dashed black line represents the variance of vacuum noise. Since the minimum of $\rm{Var}[x_{-}]$ is above the vacuum noise, there is no two-mode squeezing between Alice and Bob. (b) Bar graph representing the covariance matrix elements of the modes a and b measured at Alice and Bob for the minimum $\rm{Var}[x_{-}]$ working point in (a). (c)-(f) exhibit various measured figures of merit for characterizing the remote entanglement (or lack of in this case) between Alice and Bob as a function of Entangler 2 gain ($G_2$). (c) Duan's criterion for entanglement. (d)-(f) display the state purity $\mu$, logarithmic negativity $E_N$, and Simon's criterion for entanglement. The solid curves in (c)-(f) are calculated fits based on an entanglement swapping scheme model which incorporates intermediate losses. (c), (e), (f) show the absence of any two-mode entanglement between Alice and Bob when there is no shared entanglement between Alice and Claire.  		
			}
			\label{EntSwapClassicalVac}
		\end{center}
	\end{figure*}
	
	To characterize the resulting entanglement between the output fields in the entanglement swapping scheme, we apply the same measures introduced earlier for two-mode squeezers. Equipped with Eqs.\,(\ref{mean_x_A_sq2})-(\ref{mean_pApB_sq2}), we can evaluate Simon's criterion, logarithmic negativity, entanglement of formation, and state purity using Eqs.\,(\ref{Simon1}), (\ref{Neglog1}), (\ref{EOF1}), and (\ref{purity}), respectively.   
	
	Revisiting Duan's criterion given by Eq.\,(\ref{Duan1}), we recalculate here $\Delta^{-}_{\rm{EPR}}$ and $\Delta^{+}_{\rm{EPR}}$ given by Eq.\,(\ref{EPRmin}) and Eq.\,(\ref{EPRmax}), which yield for the case of lossless entanglement swapping the simple result
	
	\begin{align}
		\Delta^{-}_{\rm{EPR}}&=e^{-2r_2}+e^{-2r_1},   \label{EPRmEntSwap}
		\\
		\Delta^{+}_{\rm{EPR}}&=e^{2r_2}+e^{-2r_1}.   \label{EPRpEntSwap}
	\end{align}   
	
	What is notable about Eq.\,(\ref{EPRmEntSwap}) is that in the case of classical teleportation, i.e., $r_1=0$, we get $\Delta^{-}_{\rm{EPR}}>1$ regardless of $r_2$. Also, for any positive $r_1$, there exists a threshold value for $r_2$ above which $\Delta^{-}_{\rm{EPR}}<1$.  
	
	In Fig.\,\ref{EntSwapClassicalVac} we show another crucial test of the entanglement swapping setup shown in Fig.\,\ref{EntSwapQuanVac}(a). Specifically, we show that two-mode squeezing and entanglement are extinguished when one of the Entanglers in the setup is turned off. In this measurement, we turn off Entangler 1, whose role is to enable the quantum teleportation of one-half of the entangled state produced by Entangler 2. Thus, when it is turned off, only classical teleportation is allowed to take place. Like Fig.\,\ref{EntSwapQuanVac}(f) we show in Fig.\,\ref{EntSwapClassicalVac}(a) the dependence of $\rm{Var}[x_{-}]$ and $\rm{Var}[x_{+}]$ on the relative pump phase $\Delta \varphi_2$ for $G_2=2.5$ dB. But in contrast to Fig.\,\ref{EntSwapQuanVac}(f), the result here does not exhibit two-mode squeezing since both variances do not drop below the vacuum state noise variance indicated by the dashed black line. The solid curves in the plot represent sinusoidal fits. Similarly, in reference to Fig.\,\ref{EntSwapQuanVac}(g), we show in Fig.\,\ref{EntSwapClassicalVac}(b) a bar graph representation of the covariance matrix elements of the modes at Alice and Bob, obtained for the same gain as Fig.\,\ref{EntSwapClassicalVac}(a) and relative pump phase that minimizes $\rm{Var}[x_{-}]$. In Figs.\,\ref{EntSwapClassicalVac}(c)-(f), we plot various measures of two-mode entanglement as a function of $G_2$, i.e., Duan's criterion of entanglement (Fig.\,\ref{EntSwapClassicalVac}(c)), $\mu$ (Fig.\,\ref{EntSwapClassicalVac}(d)), $E_N$ (Fig.\,\ref{EntSwapClassicalVac}(e)), and Simon's criterion of entanglement (Fig.\,\ref{EntSwapClassicalVac}(f)). The lack of generated two-mode entanglement is particularly obvious in this case since both Duan's and Simon's inequalities are not violated $\Delta^{-}_{\rm{EPR}}>0$ dB (Fig.\,\ref{EntSwapClassicalVac}(c)) and $\Delta_{\rm{S}}\geq0$ (Fig.\,\ref{EntSwapClassicalVac}(f)), and $E_N$ is zero (Fig.\,\ref{EntSwapClassicalVac}(e)). The solid curves in Figs.\,\ref{EntSwapClassicalVac}(c)-(f) represent theoretical calculations of the various measures based on the covariance matrix derived for the entanglement swapping case. 
	
	\section{Experimental setup}
	
		\begin{figure*}
		[tbh]
		\begin{center}
			\includegraphics[
			width=1.7\columnwidth 
			]%
			{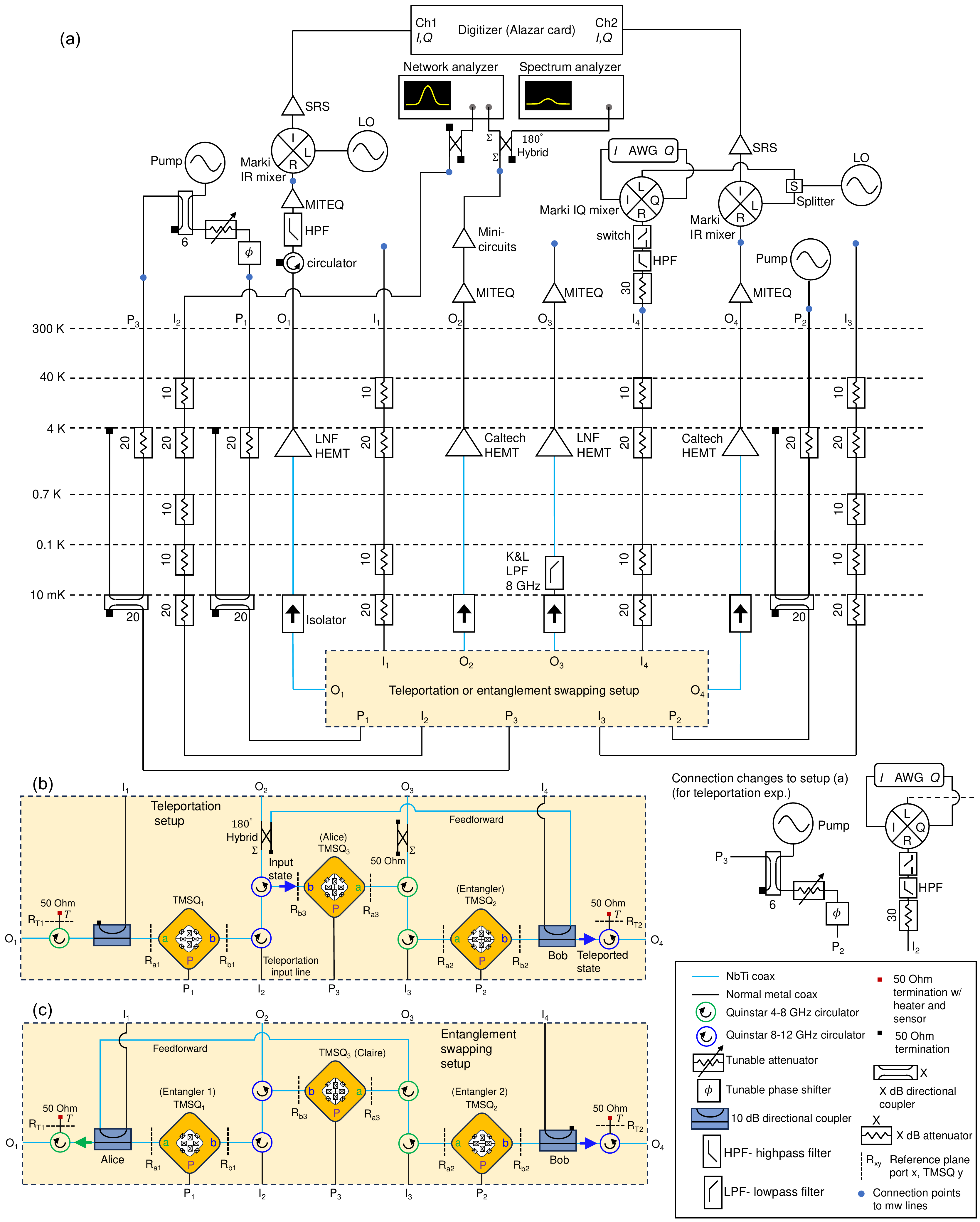}
			\caption{Teleportation and entanglement swapping setup. (a) The fridge and room temperature measurement setups used in the teleportation and entanglement swapping experiments. The bottom right side of panel (a) shows two changes to the room temperature setup relevant to the teleportation experiments. (b) Circuit components and configuration at base used in the teleportation experiments. (c) Circuit components and configuration at base used in the entanglement swapping experiments. 
			}
			\label{TeleportationAndEntSwapSetup}
		\end{center}
	\end{figure*}
	
	The data sets presented in this work which correspond to two-mode squeezing, teleportation of coherent states, and entanglement swapping were taken separately in the course of three cooldowns. Certain modifications were made to the setup between cooldowns. 
	
	A schematic diagram of the experimental setup used in the teleportation and entanglement swapping experiments is shown in Fig.\,\ref{TeleportationAndEntSwapSetup}. The experimental setup above the mixing chamber was largely unchanged between the three cooldowns, which is mostly shown in panel (a). The main changes were introduced in the components and connections below the mixing chamber, which are respectively depicted in panels (b) and (c) for the teleportation and entanglement swapping experiments.     
	
	The setup includes four input and output lines denoted I1, I2, I3, I4, and O1, O2, O3, O4, respectively. It also includes three pump lines denoted P1, P2, P3, which are used for driving the three JMs. Both input and output lines incorporate standard commercial microwave components that are commonly used in superconducting qubit experiments such as, cryogenic attenuators and low-noise HEMT amplifiers mounted inside the fridge, and IQ mixers and fast switches located outside the fridge. 
	
	The pump lines include wideband $20$ dB directional couplers at the base stage, which attenuate the pump tones that drive the JMs. The attenuation is realized by directing a large portion of the pump power towards the $4$K stage where it gets dissipated in a $50$ Ohm termination.   
	
	The blue dots on the input, output, and pump lines shown at the room-temperature stage, mark the locations at which the various lines get connected or disconnected to microwave generators and measuring devices, such as a vector network analyzer (VNA), spectrum analyzer (SA), and hetrodyne setup that is commonly used in dispersive readout of superconducting qubits. All microwave sources and measurement devices are phase locked to a $10$ MHz reference oscillator of a rubidium atomic clock.
	
	With the exception of the measurement results shown in Fig.\,\ref{TelQuantVac} and some in Fig.\,\ref{GsysNsysCalibGvsT}, which are taken with the SA, all other results are taken using the heterodyne setup, in which output noise or coherent states exiting the fridge are downconverted, sampled, and digitized using a dual-channel Alazar card.
	
	The black lines connecting the various components inside and outside the fridge correspond to normal-metal coax cables, whereas the cyan lines inside the fridge correspond to NbTi superconducting coax cables. The latter lines are primarily employed to minimize insertion losses between JMs and at sections of the output lines that precede the HEMTs which are known to degrade CV entanglements and the SNR of output lines.
	
	The three JMs used in this work are mounted on the same OFHC copper bracket at the bottom of the dilution fridge and reside deep inside a 4-inch diameter cryoperm magnetic shield can (not shown). While the commercial cryogenic circulators, which connect between the JMs or between the JMs and the input and output lines are mounted on a different bracket located outside the cryoperm magentic shield to minimize magnetic interference between the circulators and the JMs. 
	
	Likewise, to save space and decrease the footprint of the setup, we mount the directional couplers inside the magnetic shield can near the top and mount the $180^{\circ}$ hybrids on the same bracket as the circulators. Thus, it is worth emphasizing that the approximate cable lengths between the various components or parties in our experiments (indicated in Fig.\,\ref{TeleportationScheme} and Fig.\,\ref{EntSwapQuanVac}(a)) are largely set by the relative locations of these components rather than by  performance considerations. 
	
	Small superconducting magnetic coils (not shown) are attached to the OFHC copper packages of the JMs, which enable us to match the resonance frequencies of mode a and b of the JMs by flux tuning each JM individually. 
	
	The red $50$ Ohm terminations shown in the figure at the mixing chamber in the teleportation and entanglement swapping setups, correspond to cryogenic $50$ Ohm terminations that are thermalized with tight OFHC copper clamps that are attached to the base plate of the mixing chamber via thin narrow copper braids fixed by small washers and screws at both ends. On the surface of the clamp holding the termination, we mount with screws a heater in the form of a $500$ Ohm resistor with voltage and current leads soldered across it and a $\rm{RuOx_{2}}$ thermometer, which we thermally calibrated in a separate cooldown. 
	
	Using this configuration, we are able to vary in-situ the steady-state temperature of the $50$ Ohm terminations by applying a dc current through the heaters and record the resultant temperature using the thermal sensors. We connect these variable-temperature $50$ Ohm terminations to the third port of auxiliary circulators that we mount following the directional couplers at the output of port a of Entangler 1 and port b of Entangler 2. The $50$ Ohm terminations are connected to the circulators via a 4-inch coax cables. The length of the braid connecting the clamps to the base plate is set experimentally to be long enough such that it provides some thermal isolation so that the termination can be heated locally without considerably raising the mixing chamber plate temperature, but also short enough so that the termination can be cooled back down relatively quickly once we turn off the dc current flowing through the heater. 
	
	We use these variable-temperature $50$ Ohm terminations to vary the thermal noise at the bottom of the fridge, and measure the resultant noise power emitted from the output lines $O_1$ and $O_4$. The JMs in these measurement are turned off, thus acting as perfect mirrors, which reflect the generated thermal noise into the output chains. 
	
	Such noise measurements enable us to extract the total gain and effective added noise associated with the output chains and verify, after accounting for the intermediate losses between the circulators and the JM planes, that the extracted parameters are in agreement with the values obtained by varying the gain of the JMs without heating the terminations. 
	
	Finally, the setup used for taking the two-mode squeezing data is very similar to the teleportation setup, with four main differences, (1) the cables connecting the different components are normal-metal coaxes, (2) JM1 measured in the experiment is mounted where JM2 is shown in the schematic, (3) the hetrodyne setup is connected to the output lines $O_3$ and $O_4$, and (4) an additional isolator is incorporated in each output line.

	\begin{table*}[tbh]
		\centering
		\begin{tabular}{c c c c c  c  c  c  c  c } 
			\hline
			Qubit & $f_{a,\rm{max}}$ (GHz)  & $f_{b,\rm{max}}$ (GHz)  & $\gamma_{a,\rm{m}}/2\pi$ (MHz)  & $\gamma_{b,\rm{m}}/2\pi$ (MHz) & $\gamma_{a}/2\pi$ (MHz)  & $\gamma_{b}/2\pi$ (MHz)  \\ [0.5ex] 
			\hline
			\textbf{JM1} & 7.363 & 9.968 & 103 & 80 & 103 & 78  \\ 
			\textbf{JM2} & 7.364 & 9.967 & 107 & 79 & 97 & 90  \\
			\textbf{JM3} & 7.36 & 9.964 & 103 & 85 & 78 & 84   \\
			\hline 
		\end{tabular}
		\caption{Bandwidths and maximum resonance frequencies of the JMs. $f_{a,\rm{max}}$ and $f_{b,\rm{max}}$ are the measured maximum resonance frequencies of modes a and b of the JM. $\gamma_{a,\rm{m}}/2\pi$ and $\gamma_{b,\rm{m}}/2\pi$ are the linear bandwidths of modes a and b of the JM measured at their maximum frequencies, while $\gamma_{a}/2\pi$ and $\gamma_{b}/2\pi$ are the bandwidths measured at the teleportation and entanglement swapping working points in the main text.}
		\label{JMmacroParams}
	\end{table*}
	
	\section{TMSQ parameters} 
	
	The three JMs employed in this work, which we denote as JM1, JM2 and JM3, are nominally identical. They share the same design and fabrication process. In Table\,\ref{JMmacroParams}, we list the maximum resonance frequencies for resonator a and b, i.e., $f_{a,\rm{max}}$ and $f_{b,\rm{max}}$, measured for the three JMs as well as their linear bandwidths, i.e., $\gamma_{a,\rm{m}}/2\pi$ and $\gamma_{b,\rm{m}}/2\pi$, measured at the maximum resonance frequencies and the bandwidths at the working points employed in the main text, i.e., $\gamma_{a}/2\pi$ and $\gamma_{b}/2\pi$, which correspond to the resonance frequencies $f_a=7.23$ GHz and $f_b=9.707$ GHz. Furthermore, for documentation purposes, we specify in Table\,\ref{JMvsExp} the role played by each JM in the different experimental setups reported in the main text. 
	
	The dynamical bandwidth of the JM at power gain $G$, which we denote as $B=\gamma/2\pi$ is given by \cite{JPCreview}
	
	\begin{equation}
		\gamma=\dfrac{\gamma_{0}}{\sqrt{G}},   \label{DynB}
	\end{equation}
	
	\noindent where $\gamma_{0}=2\gamma_{a}\gamma_{b}/\left( \gamma_{a}+\gamma_{b}\right) $. Also, as mentioned in the main text the JMs are flux tunable. Thus, by applying external flux to the JRM loop, the resonance frequencies of mode a and b can be tuned over about $500$ MHz.
	 	
	\begin{table*}[tbh]
		\centering
		\begin{tabular}{c c c c c  c  c  c  c  c } 
			\hline
			JM & TMSQ Exp. & Teleportation Exp. & Entanglement swapping Exp. \\ [0.5ex] 
			\hline
			\textbf{JM1} & TMSQ (Fig.\,\ref{TwoModeSq}) & - & Entangler 1 (Fig.\,\ref{EntSwapQuanVac})   \\ 
			\textbf{JM2} & - & Entangler (Fig.\,\ref{TeleportationScheme}) & Entangler 2 (Fig.\,\ref{EntSwapQuanVac})  \\
			\textbf{JM3} & - & Alice (Fig.\,\ref{TeleportationScheme}) & Claire (Fig.\,\ref{EntSwapQuanVac})  \\
			\hline 
		\end{tabular}
		\caption{The roles played by the three JMs in the experiments of the main text.}
		\label{JMvsExp}
	\end{table*} 
	
	\section{Calibration of the gain and noise of the output chain}
	
	The thermal photon population corresponding to angular frequency $\omega$ and temperature $T$ at the bottom of the output line is given by the Bose-Einstein statistics given by $1/2\coth\left(\hbar\omega/2k_BT \right)$. The measured noise power of the output chain at room temperature is given by $P_{N,k}=\left( \sigma^{2}_{I,k}+\sigma^{2}_{Q,k}\right) /R$ \cite{ObsTwoModSqTWPA,BroadbandCVKerrJMetamaterial}, where the index $k=a,b$ corresponds to mode a and b of the JM, $\sigma^{2}_{I,k}$ and $\sigma^{2}_{Q,k}$ are the voltage noise variance of the $I$ and $Q$ quadratures of mode $k$, and $R=50$ Ohm is the input resistance of the measurement device. 
	
	When the JM is off and the temperature $T$ on the cryogenic termination is varied, we get 
	
	\begin{align}
		P_{N,k}\left(T, G_J=1 \right)&=  P_{0,k} \times N_{ph,k}\left(T\right)   \nonumber
		\\ 
		&=P_{0,k} \times \left(\dfrac{1}{2}\coth\left(\dfrac{\hbar\omega_k}{2k_BT}\right) +N_{sys,k}\right),  \nonumber \\ \label{G_Nsys0}
	\end{align}
	
	\noindent where $P_{0,k}=G_{sys,k} \times BW  \times \hbar \omega_{k}$, $G_{sys,k}$ is the total gain/loss of the output chain for mode $k$, $BW$ is the measurement bandwidth, $\omega_k$ is the angular frequency of mode $k$, $T$ is the temperature of the variable-temperature $50$ Ohm termination at the input of the output chain, $N_{sys,k}$ is the total added noise photons by the output chain for mode $k$, and $N_{ph,k}\left(T\right)$ is the total output noise equivalent photon number for mode $k$ as a function of $T$. 
	
	\begin{figure*}
		[tb]
		\begin{center}
			\includegraphics[
			width=2\columnwidth 
			]%
			{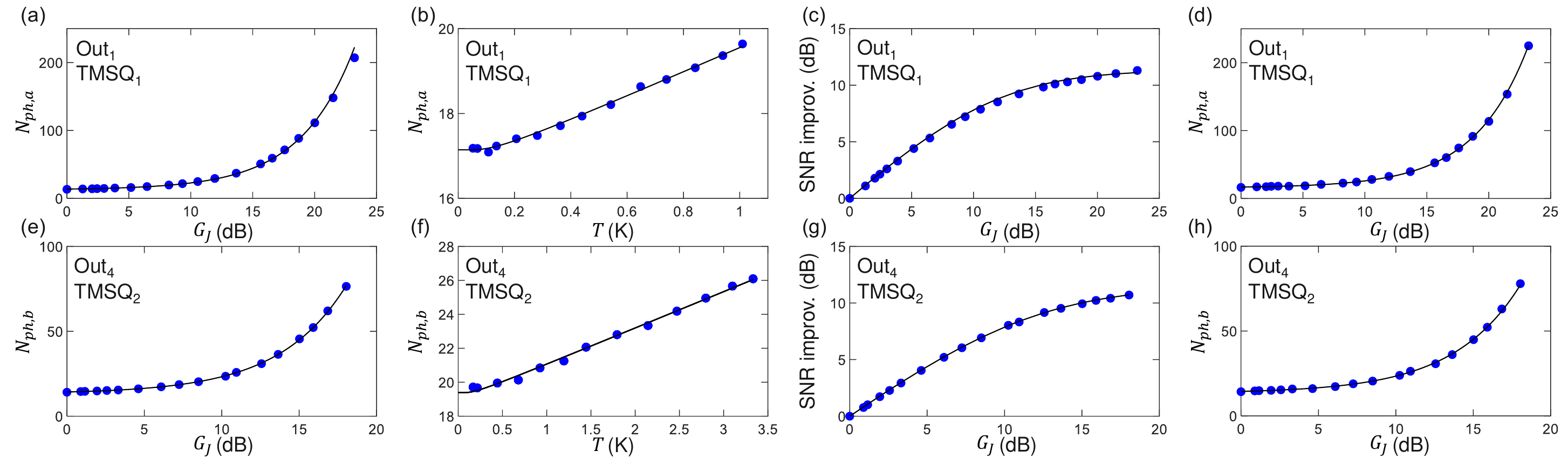}
			\caption{Noise power measurements for extracting the total gain and added noise of output lines 1 and 4. The blue filled circles in (a) and (b)  represent the noise equivalent photon number $N_{ph}$ of mode a measured using the digitizer card setup as a function of $\rm{TMSQ_1}$ gain ($G_J$) and the local temperature of the termination ($T$), respectively. The solid black curves are fits to the data which employ Eqs.\,(\ref{G_Nsys1}) and (\ref{G_Nsys0}), respectively. (c) exhibits SNR improvement data for mode a of $\rm{TMSQ_1}$ versus $G_J$. The fitting curve employs Eq.\,(\ref{SNR_improv}) and uses the same $N_{sys}$ parameter extracted in (a). (d) $N_{ph}$ of mode a measured at the output of line 1 versus $G_J$ taken using a spectrum analyzer. (e)-(h) same as (a)-(d), taken for output 4 and mode b of $\rm{TMSQ_2}$. The extracted parameters from the fits, presented in Table\,\ref{GsysNphGT}, are employed in the calibration of the entanglement swapping experiment.    
			}
			\label{GsysNsysCalibGvsT}
		\end{center}
	\end{figure*}
	
	In the high-frequency/cold-temperature limit $\hbar \omega \gg K_BT$, we obtain $1/2\coth\left(\hbar\omega/2k_BT \right)\cong1/2$,  which is the default case when the termination is not heated, since the temperatures of the base stage and termination are $10$ mK and $52$ mK, respectively, and the mode frequencies are $\omega_a/2\pi=7.23$ GHz, $\omega_b/2\pi=9.707$ GHz.
	
	Similarly, if the termination is kept cold while the gain on the JM is varied, we can write the measured noise power of the output chain at room temperature as
	
	\begin{align}
		P_{N,k}\left(T\cong0, G_J \right)&=P_{0,k}\times N_{ph,k}\left(G_J\right) \nonumber
		\\
		& =P_{0,k} \times \left(G_J\dfrac{1}{2}+\left(G_J-1\right)\dfrac{1}{2} +N_{sys,k}\right), \label{G_Nsys1}
	\end{align}
	
	\noindent where $G_J$ is the power gain of the JM and $N_{ph,k}\left(G_J\right)$ is the total output noise equivalent photon number for mode $k$ as a function of $G_J$.  
	
	\begin{table*}[tbh]
		\centering
		\begin{tabular}{|c|cccc|cccc|cccc|} 
			
			\hline
			Output,mode & Fig. & $G_{sys}$ & $N_{sys}$ & $T_{sys}$ (K) & Fig. & $G_{sys}$ & $N_{sys}$ & $T_{sys}$ (K) & Fig. &$G_{sys}$ & $N_{sys}$ & $T_{sys}$ (K)\\ 
			\hline
			\textbf{O1,a} &  \ref{GsysNsysCalibGvsT}(a) & $3.5\cdot10^6$ & 13.2 & 4.6 & \ref{GsysNsysCalibGvsT}(b) & $2.7\cdot10^6$ & 16.6 & 5.8 & \ref{GsysNsysCalibGvsT}(d) & $4.4\cdot10^5$ & 16.4 & 5.7 \\ 
			\textbf{O4,b} &  \ref{GsysNsysCalibGvsT}(e) & $1.8\cdot10^7$ & 13.8 & 6.4 & \ref{GsysNsysCalibGvsT}(f) & $1.3\cdot10^7$ & 18.9 & 8.8 & \ref{GsysNsysCalibGvsT}(h) & $2.3\cdot10^6$ & 13.9 & 6.5 \\ 
			\hline 
		\end{tabular}
		\caption{Estimated total gain and added noise for output lines 1 and 4. $G_{sys}$ and $N_{sys}$ are extracted from fits to the data shown in Fig.\,\ref{GsysNsysCalibGvsT}. $T_{sys}$ is calculated using $T_{sys,k}=N_{sys,k}\hbar \omega_{k}/k_B$. The extracted parameters are employed in the calibration of the entanglement swapping experiment. The uncertainty in evaluating $G_{sys}$ and $N_{sys}$ is less than $\pm10\%$.} 
		\label{GsysNphGT}
	\end{table*} 
	
	To find $G_{sys,k}$ and $N_{sys,k}$ for each output chain, we measure the output noise power for varying $T$ or $G_J$ and extract their values by fitting the data using Eq.\,(\ref{G_Nsys0}) or Eq.\,(\ref{G_Nsys1}). While varying $T$ of a variable-temperature termination is one of the widely used and trusted methods for extracting $G_{sys,k}$ and $N_{sys,k}$ of output chains, it has several drawbacks: (1) it requires adding auxiliary components that do not play an active role in the experiment, such as cold terminations, dedicated heaters, thermal sensors, extra dc wiring for applying heat and reading out the temperature, and circulators to direct the generated thermal noise. In experiments that employ multiple output lines, applying such overhead to all output lines could be challenging. (2) It introduces some undesired loss in the signal path due to the need to couple the variable-temperature termination to the output chain, for example via a circulator. And most important, (3) it adds uncertainties to the extracted values of $G_{sys,k}$ and $N_{sys,k}$ due to the unavoidable separation and added intermediate loss between the variable-temperature termination and the reference plane of the device under test. For these reasons, we demonstrate in Fig.\,\ref{GsysNsysCalibGvsT} that the two methods, i.e., varying $T$ or $G_J$, can be used to extract $G_{sys,k}$ and $N_{sys,k}$ for two output lines, i.e., O1 and O4 (shown in Fig.\,\ref{TeleportationAndEntSwapSetup}). Furthermore, we show below that the extracted values obtained using the two methods are in agreement after accounting for the intermediate losses between the reference planes of the variable-temperature terminations and the JMs.
	
	\begin{figure*}
		[tb]
		\begin{center}
			\includegraphics[
			width=2\columnwidth 
			]%
			{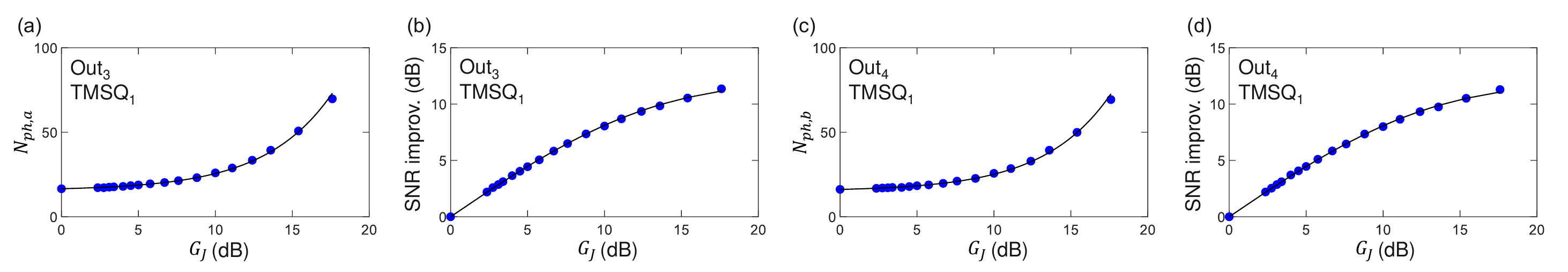}
			\caption{Noise power measurements for calibrating the output lines in the two-mode squeezing experiment. (a) ((c)) exhibits the noise equivalent photon number $N_{ph}$ of mode a (b) measured using the digitizer card setup as a function of $\rm{TMSQ_1}$ gain $G_J$. The solid black curves are fits to the data which employ Eq.\,(\ref{G_Nsys1}). (b) ((d)) plots SNR improvement data for mode a (b) of $\rm{TMSQ_1}$ versus $G_J$. The fitting curve employs Eq.\,(\ref{SNR_improv}) and uses the same $N_{sys}$ parameter extracted in (a) ((c)). The extracted parameters from the fits, presented in Table \ref{GsysNphTMSQ}, are employed in the calibration of the two-mode squeezing experiment.    
			}
			\label{GsysNsysCalibGTMSQ}
		\end{center}
	\end{figure*}

	\begin{table*}[tbh]
		\centering
		\begin{tabular}{|c|cccc|} 
			
			\hline
			Output,mode & Fig. & $G_{sys}$ & $N_{sys}$ & $T_{sys}$ (K) \\ 
			\hline
			\textbf{O3,a} &  \ref{GsysNsysCalibGTMSQ}(a) & $6.8\cdot10^6$ & 16.1 & 5.6 \\ 
			\textbf{O4,b} &  \ref{GsysNsysCalibGTMSQ}(c) & $1.3\cdot10^7$ & 15.7 & 7.3  \\ 
			\hline 
		\end{tabular}
		\caption{Estimated total gain and added noise for output lines used in the two-mode squeezing experiment. $G_{sys}$ and $N_{sys}$ are extracted from fits to the data shown in Fig.\,\ref{GsysNsysCalibGTMSQ}. The uncertainty in evaluating $G_{sys}$ and $N_{sys}$ is less than $\pm10\%$.} 
		\label{GsysNphTMSQ}
	\end{table*} 
	
	In the experiment, we measure the noise power using two measuring devices: an analog-to-digital converter (ADC) digitizer card and a spectrum analyzer. Despite having certain differences in their amplification and mixing components at room temperature, both measuring devices yield similar results for $N_{sys,k}$ of the output chains. It is worth noting that when measuring the output noise power using a spectrum analyzer the limiting bandwidth is set by the resolution bandwidth, whereas when measuring using the digitizer card, it corresponds to the inverse of the integration time $1/T_{\rm{int}}$.  
	
	Another useful measure that we consider is the improvement in the signal-to-noise ratio of the output chain due to the JM given by $G_J/G_{N,k}$, where $G_{N,k}$ is the output noise ratio (or noise rise) corresponding to the JM being on versus off
	
	\begin{align}
		G_{N,k}&=\dfrac{P_{N,k}\left(G_J \right) }{P_{N,k}\left(G_J=1 \right) } \nonumber \\
		&=\dfrac{T_{sys,k}+G_JT_{Q,k}+\left( G_J-1\right)T_{Q,k} }{T_{sys,k}+T_{Q,k}}, \label{G_N}
	\end{align}
	
	\noindent where $T_{Q,k}=\hbar \omega_k/2k_B$ is the equivalent temperature of the zero-point fluctuations at mode $k$, and $T_{sys,k}=N_{sys,k}\hbar \omega_{k}/k_B$ is the effective noise temperature of the output chain at mode $k$ in the absence of the JM. Note that the figure $G_{N,k}$, unlike $P_{N,k}$, is independent of the measurement bandwidth and system gain. It mainly depends on the total noise added by the system. 
	
	Using Eq.\,(\ref{G_N}), the signal-to-noise ratio improvement due to the JM reads
	
	\begin{equation}
		\dfrac{G_J}{G_{N,k}}=\dfrac{T_{sys,k}+T_{Q,k}}{T_{sys,k}/G_J+T_{Q,k}\left[1+\left(G_J-1\right)/G_J\right] }. \label{SNR_improv}
	\end{equation}
	
	While the relations \ref{G_Nsys1} and \ref{SNR_improv} are not independent, we, nevertheless, apply the extracted system noise from the fit of Eq.\,(\ref{G_Nsys1}) into Eq.\,(\ref{SNR_improv}) and confirm that the resulting curve fits the signal-to-noise ratio improvement data as well.
	
	In Fig.\,\ref{GsysNsysCalibGvsT}, we exhibit calibration measurements of output lines O1 and O4 used in the entanglement swapping experiment. Figures\,\ref{GsysNsysCalibGvsT}(a)-(d) exhibit measurement results for mode a taken for O1. The filled blue circles represent measured data, while the solid black curves are theory fits. Figures\,\ref{GsysNsysCalibGvsT}(a)-(c) show noise power measurements taken using the digitizer setup. Figure\,\ref{GsysNsysCalibGvsT}(a) and Fig.\,\ref{GsysNsysCalibGvsT}(b) display $N_{ph,a}$ as a function of $G_J$ and $T$, respectively. Figure\,\ref{GsysNsysCalibGvsT}(c) plots the SNR improvement versus $G_J$. Figure\,\ref{GsysNsysCalibGvsT}(d) exhibits $N_{ph,a}$ measured as a function of $G_J$ using a spectrum analyzer. Figures\,\ref{GsysNsysCalibGvsT}(e)-(h) are same as Figs.\,\ref{GsysNsysCalibGvsT}(a)-(d), measured instead for O4 and mode b. The extracted parameters from the fits, i.e., $G_{sys}$ and $N_{sys}$, are listed in Table\,\ref{GsysNphGT}. By calculating the total gain ratio for mode a (b) between $G_{sys}$ extracted using the variable-temperature and the variable-gain methods, we obtain an estimate for the insertion loss between the reference planes $\rm{R_{T1}}$ and $\rm{R_{a1}}$ ($\rm{R_{T2}}$ and $\rm{R_{b2}}$), outlined in Fig.\,\ref{TeleportationAndEntSwapSetup}, of about $-1.1$ dB ($-1.4$ dB), which corresponds to a gain ratio of $0.77$ ($0.72$). Such estimates for the insertion losses between the variable-temperature terminations and JMs match the expected total loss of the intermediate microwave components, which include a short normal-metal cable, circulator, directional coupler, and a few normal-metal connectors. 
	
	Likewise, Fig.\,\ref{GsysNsysCalibGTMSQ} depicts the calibration data and fits for the output lines used in the two-mode squeezing experiment. The extracted parameters from the fits, i.e., $G_{sys}$ and $N_{sys}$, are listed in Table\,\ref{GsysNphTMSQ}.
	
	It is worth noting that similar calibration procedures and measurements are carried out for the teleportation experiment as well (data not shown).
	
	\subsection{Estimation of intermediate losses in the setup}
	
	To get an estimate for the intermediate losses between JM stages connected in-series and attached to the same output line, we use an in-situ calibration method in which we measure the noise power of the output chain as a function of the gain on each JM separately. Using these two measurements, we extract two different $G_{sys,k}$ parameters for the two JMs whose ratio gives an estimate for the amount of loss present between their respective output reference planes denoted as $R_{\rm{x,y}}$ in Fig.\,\ref{TeleportationAndEntSwapSetup}, where index $\rm{x}$ refers to port a or b and $\rm{y}$ identifies the JM.  
	
	To better see this relation, we rewrite Eq.\,(\ref{G_Nsys1}) for two JMs connected in series with an intermediate loss characterized by a power transmission amplitude $\bar{\alpha}$ between their output reference planes. The output noise power measured for the top JM with subscript $p$ (closer to the output line) is given by  
	
	\begin{widetext}
		
		\begin{equation}
			P_{N,k,p}\left(G_{J,p} \right)=G_{sys,k,p} \times BW  \times \hbar \omega_{k} \times \left(G_{J,p}\dfrac{1}{2}+\left(G_{J,p}-1\right)\dfrac{1}{2} +N_{sys,k,p}\right), \label{G_Nsystop}
		\end{equation}
		
	\end{widetext}

	\noindent whereas, the output noise power measured for the bottom JM with subscript $m$ (farther away from the output line) is given by
	
	\begin{widetext}
		
		\begin{equation}
			P_{N,k,m}\left(G_{J,m} \right)=G_{sys,k,m} \times BW  \times \hbar \omega_{k} \times \left(G_{J,m}\dfrac{1}{2}+\left(G_{J,m}-1\right)\dfrac{1}{2} +N_{sys,k,m}\right), \label{G_Nsysbottom}
		\end{equation}
		
	\end{widetext}
	
	\noindent where $G_{sys,k,m}=\bar{\alpha}G_{sys,k,p}$, and $N_{sys,k,m}=N_{sys,k,p}/\bar{\alpha}+\left( 1-\bar{\alpha}\right) /2\bar{\alpha}$. The second term in $N_{sys,k,m}$ assumes vacuum noise is introduced by the cold dissipation. 
	
	To demonstrate how we apply this calibration method to get estimates for intermediate losses, we give here two examples related to the teleportation experiment whose base-temperature stage configuration is shown in Fig.\,\ref{TeleportationAndEntSwapSetup}(b). In the first example, we estimate the insertion loss between the reference planes $\rm{R_{a2}}$ and $\rm{R_{a3}}$ denoted $\bar{\alpha}$, by measuring the noise power emitted from output line 3 twice. The first versus the gain on $\rm{TMSQ_3}$ while $\rm{TMSQ_2}$ is off and the second versus the gain on $\rm{TMSQ_2}$ while $\rm{TMSQ_3}$ is off. By fitting the data of the two measurements, we extract $G_{sys,a,p}=7.3\cdot10^6$ and $G_{sys,a,m}=4.5\cdot10^6$ obtained using $\rm{TMSQ_3}$ and $\rm{TMSQ_2}$, respectively. Calculating their ratio yields the estimate  $\bar{\alpha}=G_{sys,a,m}/G_{sys,a,p}=0.61$ (corresponding to $-2.1$ dB). In the second example, we use the same procedure to estimate the insertion loss of the feedforward channel in the teleportation experiment, denoted as $\bar{\beta}_f$. This time, however, we measure the noise power for mode b emitted from output line 4 as a function of the gain on $\rm{TMSQ_2}$ while $\rm{TMSQ_3}$ is off and as a function of the gain on $\rm{TMSQ_3}$ while $\rm{TMSQ_2}$ is off. After extracting the corresponding parameters $G_{sys,b,p}=2.1\cdot10^7$ and $G_{sys,b,m}=8.2\cdot10^5$, we get  $G_{sys,b,m}/G_{sys,b,p}=0.04$, which corresponds to $14$ dB of loss. After subtracting $10$ dB of coupling loss introduced by the directional coupler, we obtain an estimate of $\bar{\beta}_f=0.4$ (corresponding to $-4$ dB).   
	
	\subsection{Calibration of the measured quadratures}
	
	As explained in Appendix A.5, to reconstruct the two-mode squeezing covariance matrix and calculate the various entanglement measures, we must refer the output field quadratures measured using the room-temperature electronics back to the output of the device. To do so, we employ the total gain and noise of the output chain measured for the different modes and lines (detailed in the previous subsections).
	
	The experimentally measured four dimensionless quadratures of the output fields corresponding to mode a and b at room temperature $X_a, P_a, X_b, P_b$ can be written as \cite{ObsTwoModSqTWPA}    
	
	\begin{equation}
		X_{k}=\dfrac{A_{k}+A^{\dagger}_{k}}{2}, \quad P_{k}=\dfrac{A_{k}-A^{\dagger}_{k}}{2i}. \label{RT_XP_quad}
	\end{equation}
	
	\noindent which are obtained from the measured raw quadratures in \textit{Volts}, $X^{\rm{raw}}_{a}, P^{\rm{raw}}_{a}, X^{\rm{raw}}_{b}, P^{\rm{raw}}_{b}$ using the conversion relations
	
	\begin{equation}
		X_{k}=\sqrt{\gamma_{k}}X^{\rm{raw}}_{k}, \quad P_{k}=\sqrt{\gamma_{k}}P^{\rm{raw}}_{k}. \label{RT_XPraw_quad}
	\end{equation}
	
	\noindent with the conversion factor $\gamma_{k}$ given by 
	
	\begin{equation}
		\gamma_{k}=\dfrac{T_{\rm{int}}}{RE^{\rm{ph}}_{k}}, \label{convFactor}
	\end{equation}
	
	\noindent where $R=50$ Ohm, $T_{\rm{int}}$ is the integration time of the measurement, and $E^{\rm{ph}}_{k}=\hbar\omega_{k}$ is the photon energy of mode $k$.
	
	Finally, we refer the measured quadratures $X_a, P_a, X_b, P_b$ back to the output of the device using the amplitude scaling 
	
	\begin{equation}
		x_{k}=X_{k}/\sqrt{G_{sys,k}}, \quad p_{k}=P_{k}/\sqrt{G_{sys,k}}. \label{xp_quad}
	\end{equation}
	
\section{Measurement parameters}
	
	In all 2D histogram data shown in the paper, we measure $N=10^5$ data points. In the digitizer setup, we apply $1$ $\mu$s measurement pulses and integration times $T_{\rm{int}}$ of the same duration. We also use a $10$ MHz frequency offset in the upconversion and downconversion process. 
	
	For the teleportation measurement of Fig.\,\ref{TelQuantVac}, which we measure using a spectrum analyzer, we measure the output at $9.707$ GHz with $1001$ data points, a sweep time of $0.1$ s, resolution bandwidth and video bandwidth of $470$ kHz, and $300$ averages. We vary the pump phase by offsetting the frequency of the pump on one of the JMs by $50$ Hz, while triggering the spectrum analyzer with $10$ Hz, $50\%$ duty cycle square pulses originating from an arbitrary wave generator.  
	
	\section{Analysis of measurement errors}
	
	Errors in the measured entanglement and squeezing figures, namely $\Delta^{\pm}_{\rm{EPR}}$, $\mu$, $E_N$, $E_F$, $\Delta_S$, stem from statistical and systematic errors associated with the measurement and reconstruction of the covariance matrix of the TMSQ device and the entanglement swapping apparatus. 
	
	The statistical errors, which are dominant in our experiment, originate from uncertainties in the calculation of variances and covariance of a bivariate variable, which arise from having a finite number of observations of a random sample.  
	
	Thus, for a general bivariate covariance matrix of the form \cite{MwOptEnt}
	
	\begin{equation}
		\Sigma  =\left(
		\begin{array}
			[c]{cc}%
			\sigma^2_{11} & \rho\sigma_{11}\sigma_{22} \\
			\rho\sigma_{11}\sigma_{22} & \sigma^2_{22} 
		\end{array}
		\right),
	\end{equation}
	
	\noindent the variance of the covariance matrix reads
	
	\begin{align}
		\rm{Var}\left( \Sigma\right)   =\frac{1}{N-1}\left(
		\begin{array}
			[c]{cc}%
			2\sigma^4_{11} & \left( 1+\rho^2\right) \sigma^2_{11}\sigma^2_{22} \\
			\left( 1+\rho^2\right) \sigma^2_{11}\sigma^2_{22} & 2\sigma^4_{22} 
		\end{array}
		\right), \nonumber \\
	\end{align}
	
	\noindent where $\sigma^2_{ii}$ is the variance of sample distribution sampled from a Gaussian variable and $N$ is the number of samples. 
	
	When compared to the covariance matrices of the TMSQ device and the entanglement swapping apparatus presented in Appendix A and D, we have $V_{11}=\sigma^2_{11}$, $V_{33}=\sigma^2_{22}$, and $\rho=V_{13}/\sqrt{V_{11}V_{33}}$.
	
	The systematic errors on the other hand, which play a lesser role in our system, originate from calibration errors of the vacuum noise levels, or more specifically from estimation errors of the total gains and added noise of the output chains. 
	
	However, since the effect of systematic errors on the calculation of the covariance matrices is nontrivial, we employed a worst-case scenario approach for calculating the total errors, i.e., systematic and statistical, which is similar to the one employed in Ref. \cite{MwOptEnt}. In this approach, we repeat the full calculation of the covariance matrix for the upper and lower bounds of the vacuum noise levels, while accounting for the various statistical errors ($\pm1$ standard deviation) in each run. We then calculate the various two-mode squeezing and entanglement measures using the different covariance matrices obtained in the previous step and set the total error bars of each quantity based on the maximum achievable range.
	
	\section{Teleportation fidelity}
	
\begin{figure*}
		[tb]
		\begin{center}
			\includegraphics[
			width=1.8\columnwidth 
			]%
			{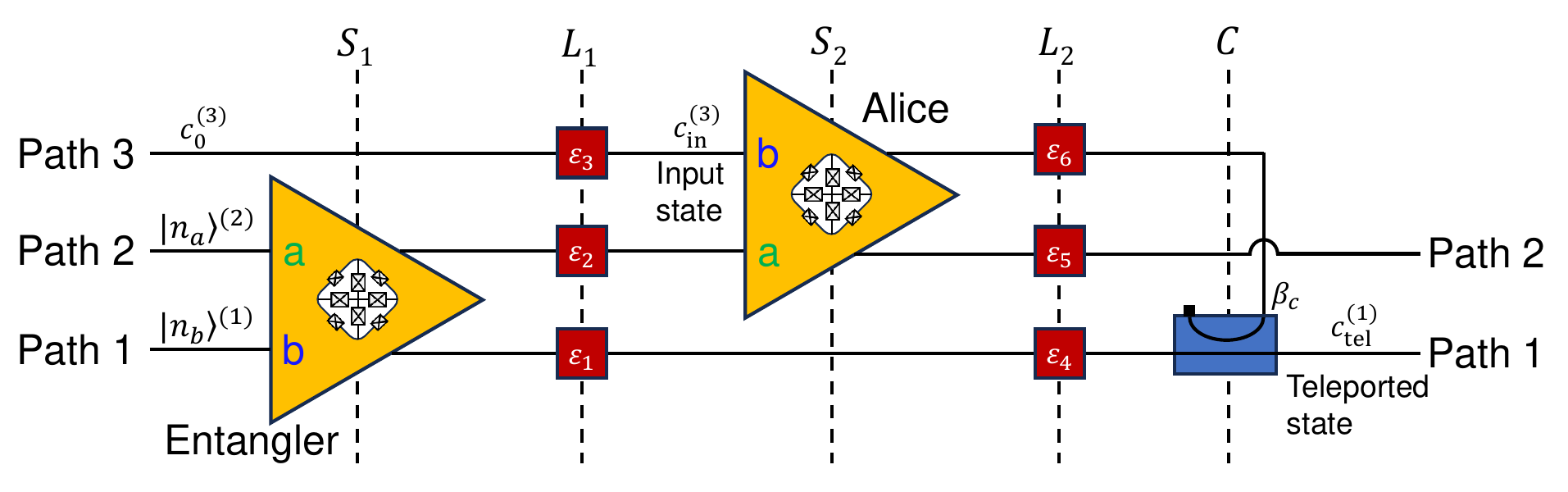}
			\caption{Schematic diagram for calculating the teleported state and the teleportation fidelity of coherent states. The teleportation process is modeled as a sequence of operations, i.e., two-mode squeezing $S$, loss $L$, and coupling $C$, which act on signals propagating along paths, 1, 2, and 3. The inputs to the Entangler are typically vacuum states. At the end of the sequence, coherent states input on path 3 within the bandwidth of mode b of the Entangler and Alice, get teleported to the output of path 1. 
			}
			\label{TelOpSeq}
		\end{center}
	\end{figure*}

	The teleportation operation presented in Fig.\,\ref{TeleportationScheme} can be broken down into a sequence of quantum operators acting on propagating signals along three paths as shown in Fig.\,\ref{TelOpSeq}. In this representation, (1) the input and output fields are expressed as column vectors with 3 pairs of rows (6 in total), where the first, second and third pairs correspond to the $I$ and $Q$ quadratures of the microwave signals in paths 1, 2, and 3, respectively, and (2) the quantum operators acting on the $I$, $Q$ quadratures of these signals, indicated by the dashed vertical lines, are expressed as matrices of size $6\times 6$. The main three operators playing a role in this scheme are the two-mode squeeze ($S$), insertion loss ($L$) represented by beam-splitter interaction, and coupling ($C$). Using this representation, adapted from Ref. \cite{ExpQuantTelMw}, allows us to calculate the output fields and added noise through multiplication of vectors and matrices, and in turn to calculate the teleportation fidelity by evaluating the overlap between the input and output states.    
	
	In Fig.\,\ref{TelOpSeq}, $c^{(3)}_{0}$ represents a coherent state propagating along path 3, while the input states $\left| n_{b}\right\rangle ^{(1)}$, $\left| n_{a}\right\rangle ^{(2)}$ entering the Entangler correspond to the thermal noise in modes b and a that are present in paths 1 and 2, respectively, with average noise photon numbers given by $n_{b,a}=1/2\coth\left(\hbar\omega_{b,a}/2k_BT \right)$. Note that in the quantum limit applicable here, i.e., $\hbar\omega_i/2k_BT\gg1$, these input states correspond to the vacuum. Thus, the general covariance matrix for the input thermal noise of the system can be cast as
	
	\begin{align}
		V_0  =\frac{1}{4}\left(
		\begin{array}
			[c]{ccc}%
			\left(1+2n^{\rm{th}}_b\right)I_2 & 0_2 & 0_2 \\
			0_2 & \left(1+2n^{\rm{th}}_a\right)I_2 & 0_2 \\
			0_2 & 0_2 & \left( 1+2n^{\rm{th}}_b\right)I_2  
		\end{array}  
		\right), \nonumber \\ \label{V0}
	\end{align}
	
	\noindent where $0_{2}$ is a zero matrix of dimension 2x2, and $n^{\rm{th}}_{a,b}=1/\left(e^{\left( \hbar\omega_{a,b}/2k_BT\right) }-1\right)$.  
	
	Using Eq.\,(\ref{S_r_phi}), the two-mode squeezing operations at the Entangler and Alice can be, respectively, written as   
	
	\begin{align}
		S_1   =\left(
		\begin{array}
			[c]{ccc}%
			\cosh\left(r_E\right)I_2 & \sinh\left(r_E\right)ZR\left(\varphi_E\right) & 0_2 \\
			\sinh\left(r_E\right)ZR\left(\varphi_E\right) & \cosh\left(r_E\right)I_2 & 0_2 \\
			0_2 & 0_2 & I_2  
		\end{array} 
		\right), \nonumber \\ \label{S1}
	\end{align}
	
	\noindent and
	
	\begin{align}
		S_2   =\left(
		\begin{array}
			[c]{ccc}%
			I_2 & 0_2 & 0_2 \\
			0_2 & \cosh\left(r_A\right)I_2 & \sinh\left(r_A\right)ZR\left(\varphi_A\right) \\
			0_2 & \sinh\left(r_A\right)ZR\left(\varphi_A\right) & \cosh\left(r_A\right)I_2
		\end{array} 
		\right). \nonumber \\ \label{S2}
	\end{align}
	
	As to the loss operation, we can express the diagonal matrix representing the transmission amplitudes in the signal path as   
	
	\begin{equation}
		L_{1,2}   =\left(
		\begin{array}
			[c]{ccc}%
			\sqrt{\bar{\varepsilon}_{1,4}}I_2 & 0_2 & 0_2 \\
			0_2 & \sqrt{\bar{\varepsilon}_{2,5}}I_2 & 0_2 \\
			0_2 & 0_2 & \sqrt{\bar{\varepsilon}_{3,6}}I_2
		\end{array} \label{L12_mat}
		\right), 
	\end{equation}
	
	\noindent where $\bar{\varepsilon_i}=1-\varepsilon_i$, and $\varepsilon_i$ corresponds to the power loss in element $i$ of the circuit. From this result, it follows that the added noise due to losses in the signal paths (modeled as finite couplings to thermal baths) can be expressed as  
	
	\begin{widetext}
		
		\begin{equation}
			A_{1,2}   =\frac{1}{4}\left(
			\begin{array}
				[c]{ccc}%
				\left(1+2n^{\rm{th}}_b\right){\varepsilon}_{1,4}I_2 & 0_2 & 0_2 \\
				0_2 & \left(1+2n^{\rm{th}}_a\right){\varepsilon}_{2,5}I_2 & O_2 \\
				0_2 & 0_2 & \left(1+2n^{\rm{th}}_b\right){\varepsilon}_{3,6}I_2
			\end{array} \label{A12_mat}
			\right). 
		\end{equation}
		
	\end{widetext}

	Note that some of the losses shown in the sequence of Fig.\,\ref{TelOpSeq} are redundant, but they are nevertheless included to express the loss matrix in a standard generic form that applies to the different paths and loss stages. In particular, the losses $\varepsilon_3$, $\varepsilon_4$, $\varepsilon_5$ can be zero without loss of generality. We also note that using the notations of Fig.\,\ref{TeleportationScheme}, we have $\varepsilon_1=\beta$, $\varepsilon_2=\alpha$, and $\varepsilon_6=\beta_f$.     
	
	Lastly, the directional coupler operation can be expressed as
	
	\begin{equation}
		C   =\left(
		\begin{array}
			[c]{ccc}%
			\sqrt{\bar{\beta}_c}I_2 & 0_2 & \sqrt{\beta_c}I_2 \\
			0_2 & I_2 & 0_2 \\
			-\sqrt{\beta_c}I_2 & 0_2 & \sqrt{\bar{\beta}_c}I_2
		\end{array} \label{C_mat}
		\right). 
	\end{equation}
	
	\begin{figure*}
		[tb]
		\begin{center}
			\includegraphics[
			width=2\columnwidth 
			]%
			{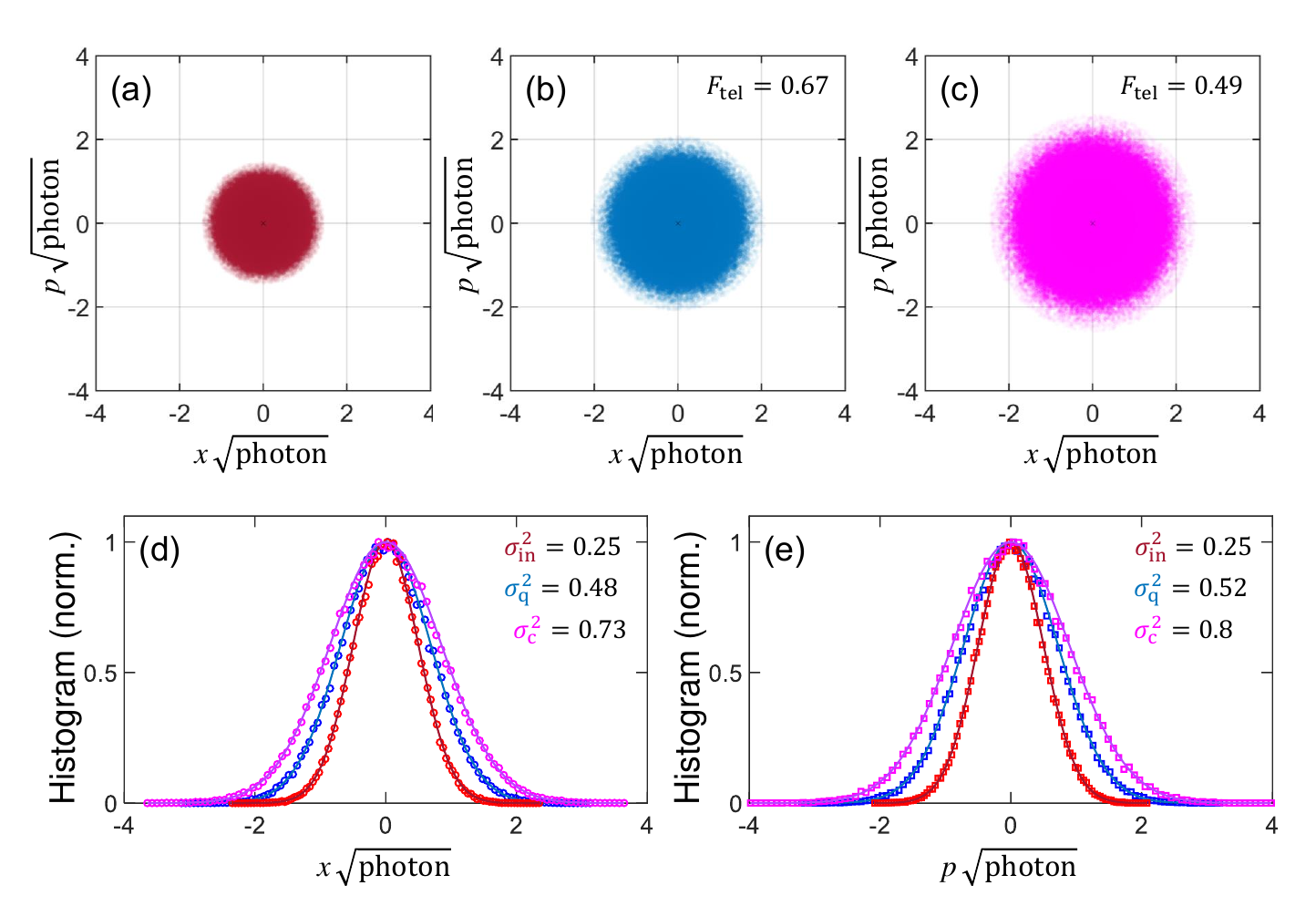}
			\caption{Quantum and classical teleportation of vacuum state. (a) \textit{I-Q} (\textit{x-p}) quadrature representation of the vacuum state at the input of port b of Alice to be teleported to Bob via the teleportation scheme depicted in Fig.\,\ref{TeleportationScheme}. (b) and (c) \textit{I-Q} (\textit{x-p}) quadrature representation of the quantum-mechanically ($G_E=2.1$ dB) and classically ($G_E=0$ dB) teleported vacuum states at Bob, respectively. (d) and (e) Normalized histograms of the scatter data in (a) (red), (b) (blue), (c) (magenta) taken along the position and momentum axis, respectively. The solid curves in (d) and (e) are Gaussian fits.     
			}
			\label{TelQuantVacADC}
		\end{center}
	\end{figure*}
	
	\begin{figure}
		[tb]
		\begin{center}
			\includegraphics[
			width=\columnwidth 
			]%
			{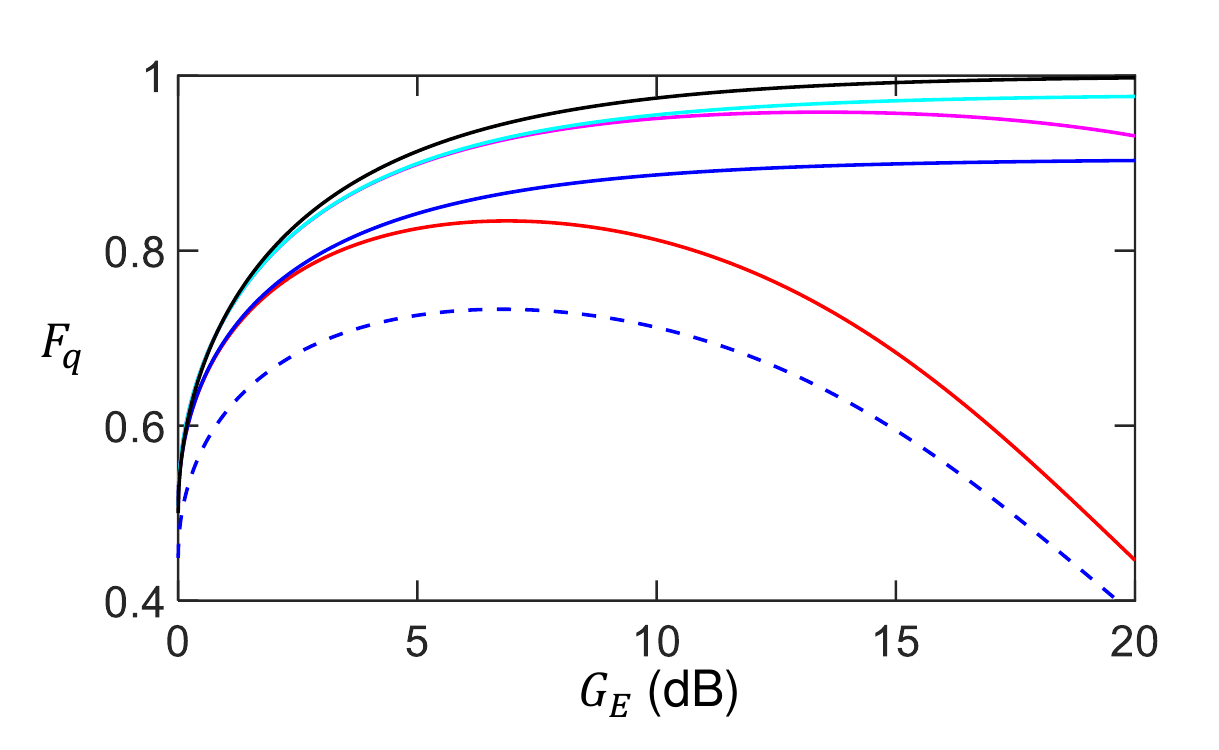}
			\caption{Calculated fidelity of quantum teleportation of vacuum states versus Entangler gain. The solid curves represent cases, where $k=1$ (i.e., unity-transmission classical channel, in particular $G_A=15$ dB and $\beta_c=-15$ dB). The black solid curve corresponds to the lossless case. The solid cyan and blue curves are calculated for the cases of symmetric loss of $0.1$ dB and $0.5$ dB, respectively, where the loss is between the Entangler and Alice and the Entangler and Bob. The solid magenta and red curves are calculated for the cases of asymmetric loss of $0.2$ dB and $1$ dB, respectively, where the loss is only between the Entangler and Bob. The dashed blue curve represents an amplification case, where $k>1$ (i.e., $G_A=16$ dB and $\beta_c=-15$ dB), and has a symmetric loss of $0.5$ dB between the Entangler and Alice and the Entangler and Bob.
			}
			\label{TelVackeq1andbigger}
		\end{center}
	\end{figure}
	
	Using these matrices, we can calculate the teleported state and teleportation fidelity of a coherent state propagating along path 3. 
	
	Starting with the following column vector representing an input coherent state with an average photon number $n_s$ and phase $\theta_s$
	
	\begin{equation}
		c_0  =\left(
		\begin{array}
			[c]{cccccc}%
			0, & 0, & 0, & 0, & \sqrt{n_s}\cos\left(\theta_s \right), & \sqrt{n_s}\sin\left(\theta_s \right)  	
		\end{array}
		\right)^{\rm{T}},  \label{c_0_vec}
	\end{equation}
	
	\noindent we can express the input coherent state at Alice as
	
	\begin{equation}
		c_{\rm{in}}=L_1S_1c_0, \label{c_vec}
	\end{equation}
	
	\noindent and the corresponding covariance matrix of the input state as
	
	\begin{equation}
		V_{\rm{in}}=L_1S_1V_0S^{\dagger}_1L^{\dagger}_1. \label{V_c}
	\end{equation}
	
	Similarly, to get the output vector, denoted $c_{\rm{tel}}$ (which contains the teleported state), and the covariance matrix of the output, denoted $V_{\rm{tel}}$, we calculate 
	
	\begin{equation}
		c_{\rm{tel}}=Tc_0, \label{c_tel_vec}
	\end{equation}
	
	\noindent where $T$ is the teleportation matrix given by the product
	
	\begin{equation}
		T=CL_2S_2L_1S_1, \label{T_mat}
	\end{equation}
	
	\noindent and 
	
	\begin{equation}
		V_{\rm{tel}}=TV_0T^{\dagger}+A, \label{V_tel}
	\end{equation}
	
	\noindent where $A$ represents the added noise of the whole sequence, which reads
	
	\begin{equation}
		A=CL_2S_2A_1S^{\dagger}_2L^{\dagger}_2C^{\dagger}+CA_2C^{\dagger}. \label{A}
	\end{equation}
	
	Focusing on the displacement coordinates of the input and teleported coherent states denoted $c^{(3)}_{\rm{in}}$ and $c^{(1)}_{\rm{tel}}$ in paths 3 and 1, respectively, and on the bottom-right $2\times2$ block of $V_{\rm{in}}$, which we denote $V^{(3)}_{\rm{in}}$ and the top-left $2\times2$ block of $V_{\rm{tel}}$ which we denote $V^{(1)}_{\rm{tel}}$, the teleportation fidelity between the input and teleported Gaussian states can be written as 
	
	\begin{align}
		F_q\left(c^{(3)}_{\rm{in}}, V^{(3)}_{\rm{in}}, c^{(1)}_{\rm{tel}}, V^{(1)}_{\rm{tel}}\right)=\dfrac{1}{2}\dfrac{\exp{\left[-\beta^{\rm{T}}_{\rm{tel}}\left(V^{(3)}_{\rm{in}}+V^{(1)}_{\rm{tel}} \right) ^{-1}\beta_{\rm{tel}} \right] }}{\sqrt{\Lambda+\Delta}-\sqrt{\Delta}}, \nonumber \\ \label{F_Uhlmann}
	\end{align}
	
	\noindent where $\beta_{\rm{tel}}=c^{(3)}_{\rm{in}}-c^{(1)}_{\rm{tel}}$, $\Lambda=\det{\left(V^{(3)}_{\rm{in}}+ V^{(1)}_{\rm{tel}}\right) }$, and $\Delta=16\left(\det{V^{(3)}_{\rm{in}}} -1/16\right)\left( \det{V^{(1)}_{\rm{tel}}}-1/16\right)$. 
	
	In the special case of vacuum-state teleportation, where $\beta_{\rm{tel}}=0$, $\Delta=0$, Eq.\,(\ref{F_Uhlmann}) reduces to 
	
	\begin{equation}
		F=\frac{2}{\sqrt{\left(1+4\sigma^2_x \right) \left(1+4\sigma^2_p\right) }}, \label{Quant_F_vac_reduced}
	\end{equation}
	
	\noindent where $\sigma^2_x=\left\langle \left(\Delta x \right) ^2 \right\rangle $, $\sigma^2_p=\left\langle \left(\Delta p \right) ^2 \right\rangle$ are the variances of the $I$ and $Q$ quadratures of the teleported field. 
	
	Furthermore, in the common case of unsqueezed vacuum, where $\sigma^2_x=\sigma^2_p=\sigma^2$, we obtain
	
	\begin{equation}
		F=\frac{2}{\left(1+4\sigma^2 \right)}. \label{Quant_F_vac_reduced2}
	\end{equation}

	In Fig.\,\ref{TelQuantVacADC} we exhibit the results of quantum and classical teleportation of vacuum states obtained for $G_E=2.1$ dB, which are quantitatively similar to the results of Fig.\,\ref{TelQuantCoh} obtained for coherent states with nonzero mean. In Fig.\,\ref{TelQuantVacADC}(a), we plot the $x-p$ quadrature representation of the the vacuum field at port b of Alice that is teleported to Bob via the teleportation setup shown in Fig.\,\ref{TeleportationScheme}. Likewise, in Fig.\,\ref{TelQuantVacADC}(b) and Fig.\,\ref{TelQuantVacADC}(c) we plot the teleported fields measured at Bob, which correspond to quantum ($G_E=2.1$ dB) and classical ($G_E=0$ dB) teleportation operations, respectively. While the teleportation fidelity achieved by the latter is $0.49$, a much higher fidelity $0.67$ is achieved  using quantum teleportation, which notably beats the classical bound of $0.5$ for the teleportation of coherent states. Furthermore, in Fig.\,\ref{TelQuantVacADC}(d) and Fig.\,\ref{TelQuantVacADC}(e), we plot normalized histograms of the scatter data shown in Fig.\,\ref{TelQuantVacADC}(a),(b),(c) plotted using red, blue, and magneta markers taken along the position $x$ and momentum $p$ quadratures, respectively. We also list on the sides of Fig.\,\ref{TelQuantVacADC}(d) and Fig.\,\ref{TelQuantVacADC}(e) the histogram variances extracted from the corresponding Guassian fits (solid curves). 
	
	\section{Upper bound on teleportation fidelity}
	
	Here we derive analytical expressions for the teleportation fidelity in the ideal lossless case. 
	
	Without loss of generality we set $\varphi_E=0$ for the Entangler and $\varphi_A=\pi$ for Alice's amplifier. For these phases, the rotation matrices for the Entangler and Alice become $R\left( \varphi_E\right) =I_2$ and $R\left( \varphi_A\right) =-I_2$. Thus, the two-mode squeezing matrices for the Entangler ($S_1$) and Alice ($S_2$) become 
	
	\begin{equation}
		S_1   =\left(
		\begin{array}
			[c]{ccc}%
			\cosh\left(r_E\right)I_2 & \sinh\left(r_E\right)ZI_2 & 0_2 \\
			\sinh\left(r_E\right)ZI_2 & \cosh\left(r_E\right)I_2 & 0_2 \\
			0_2 & 0_2 & I_2  
		\end{array} \label{S1_ideal}
		\right), 
	\end{equation}   
	
	\noindent and
	
	\begin{equation}
		S_2   =\left(
		\begin{array}
			[c]{ccc}%
			I_2 & 0_2 & 0_2 \\
			0_2 & \cosh\left(r_A\right)I_2 & -\sinh\left(r_A\right)ZI_2 \\
			0_2 & -\sinh\left(r_A\right)ZI_2 & \cosh\left(r_A\right)I_2
		\end{array} \label{S2_ideal}
		\right). 
	\end{equation} 
	
	Using Eqs.\,(\ref{S1_ideal}), (\ref{S2_ideal}) and (\ref{C_mat}), we calculate the teleportation matrix $T=CS_2S_1$ whose block elements read
	
	\begin{align}
		\begin{array}
			[c]{c}%
			T_{11}= \sqrt{\bar{\beta_c}}\cosh\left( r_E\right)I_2-\sqrt{\beta_c}\sinh\left(r_A\right) \sinh\left( r_E\right)I_2   ,\\
			T_{12}= \sqrt{\bar{\beta_c}}\sinh\left( r_E\right)Z-\sqrt{\beta_c}\sinh\left(r_A\right) \cosh\left( r_E\right)Z,\\
			T_{13}=\sqrt{\beta_c}\cosh\left(r_A\right)I_2,\\
			T_{21}=\cosh\left(r_A\right)\sinh\left(r_E\right)Z,\\
			T_{22}=\cosh\left(r_A\right)\cosh\left(r_E\right)I_2,\\
			T_{23}=-\sinh\left(r_A\right)Z,\\
			T_{31}=-\sqrt{\beta_c}\cosh\left( r_E\right)I_2-\sqrt{\bar{\beta_c}}\sinh\left(r_A\right) \sinh\left( r_E\right)I_2   ,\\
			T_{32}=-\sqrt{\beta_c}\sinh\left( r_E\right)Z-\sqrt{\bar{\beta_c}}\sinh\left(r_A\right) \cosh\left( r_E\right)Z   ,\\
			T_{33}=\sqrt{\bar{\beta}_c}\cosh\left(r_A\right)I_2.\\
		\end{array}
		\label{T_ideal}
	\end{align}
	
	Using the relation Eq.\,(\ref{c_tel_vec}) to calculate the teleported state vector, we get for the input coherent state $c_0$ defined in Eq.\,(\ref{c_0_vec}),

	\begin{widetext}
		
		\begin{equation}
			c_{\rm{tel}}  =\left(
			\begin{array}
				[c]{cccccc}%
				\left[  \sqrt{\beta_c}\cosh\left( r_A\right) \right]  \sqrt{n_s}\cos\left(\theta_s \right), & \left[  \sqrt{\beta_c}\cosh\left( r_A\right) \right] \sqrt{n_s}\sin\left(\theta_s \right), & * , & * , & * , & *   	
			\end{array}
			\right)^{\rm{T}},  \label{c_tel_vec_ideal}
		\end{equation}
		
	\end{widetext}
	
	\noindent where we only listed the two elements associated with the teleported state in path 1.  
	
	The covariance matrix for the input coherent state is $V_0=I_6/4$, where $I_6$ is the unity matrix of size $6\times6$. Hence, the covariance matrix of the teleported state is given by $V_{\rm{tel}}=TV_0T^{\dagger}=TT^{\dagger}/4$.
	
	The covariance elements associated with the teleported state in path 1 correspond to the top-left $2\times2$ block of $V_{\rm{tel}}$ which read
	
	\begin{widetext}
		\begin{equation}
			V^{(1)}_{\rm{tel}}=\frac{1}{4}\left[\bar{\beta}_c\cosh\left( 2r_E\right)+\beta_c\sinh^{2}\left( r_A\right) \cosh\left( 2r_E\right)-2\sqrt{\bar{\beta}_c\beta_c}\sinh\left( r_A\right)\sinh\left( 2r_E\right)  +\beta_c\cosh^{2}\left( r_A\right) \right]I_2. \label{V_tel_expanded}
		\end{equation} 
		
	\end{widetext}

	In the limit of unity gain for the classical feedforward signal, i.e., $\beta_c\cosh^{2}\left( r_A\right)=1$, large gain in Alice's which-path information eraser, i.e., $G_A=\cosh^{2}\left(r_A \right)\gg1$, small coupling in the directional coupler, i.e., $\beta_c\approx0$, and the approximation $\beta_c\sinh^{2}\left( r_A\right)\cong1$, we obtain for the top-left $2\times2$ block of the teleported covariance matrix representing path $1$
	
	\begin{equation}
		V^{(1)}_{\rm{tel}}=\frac{1}{4}\left(2e^{-2r_E}+1\right)I_2. \label{V_tel_q_reduced}
	\end{equation}
	
	We also get $\beta_{\rm{tel}}=0$ since $c^{(3)}_0=c^{(1)}_{\rm{tel}}$, and $\Delta=0$ because $\det{V^{(3)}_0}=1/16$ for the input coherent state. Substituting these parameters and $\Lambda=\det{\left(V^{(3)}_{\rm{in}}+ V^{(1)}_{\rm{tel}}\right) }=\left(e^{-2r_E}+1 \right)^2/4$ in Eq.\,(\ref{F_Uhlmann}), reduces the quantum teleportation fidelity into 
	
	\begin{equation}
		F_q=\frac{1}{e^{-2r_E}+1}. \label{Quant_F_reduced}
	\end{equation}
	
	This simple result demonstrates that the quantum teleportation fidelity in the lossless case, (1) depends only on the two-mode squeezing parameter $r_E$, (2) $F_q$ can reach unity in the limit of large squeezing $r_E\gg0$, and more importantly, (3) $F_q$ exceeds $1/2$ for any nonzero two-mode squeezing $r_E>0$, thus beating the upper bound on the achievable fidelity in the classical case, i.e., $F_c=1/2$, attained when $r_E=0$.
	
	\subsection{Teleportation fidelity with nonunity-gain classical channel}
	
	\begin{figure}
		[tb]
		\begin{center}
			\includegraphics[
			width=1\columnwidth 
			]%
			{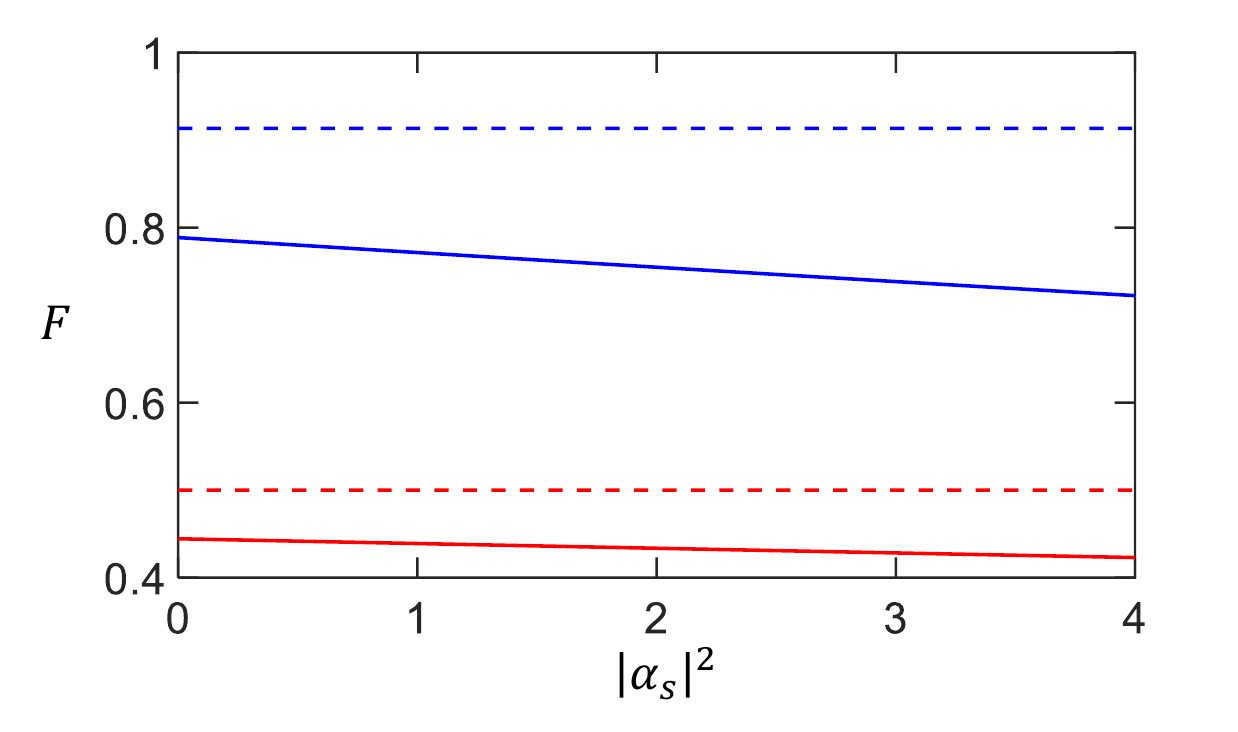}
			\caption{Calculated teleportation fidelity of coherent states versus their average photon number in the absence of loss. The dashed blue and red line represent the quantum and classical teleportation fidelity in the case of unity-transmission classical channel, i.e., $k=1$. Whereas, the solid blue and red line represent the quantum and classical teleportation fidelity in the case of nonunity-transmission classical channel, i.e., $k>1$, where in this numerical example $G_A=10$ and $\beta_c=1/8$ .In the quantum teleportation scenarios (blue lines) the fidelity is calculated for $G_E=5$ dB. Unlike the $k>1$ cases, the fidelities for $k=1$ are independent of the average photon number of the input coherent state. 
			}
			\label{TelCohvsnsklarger1thy}
		\end{center}
	\end{figure}
	
	In this subsection, we calculate the teleportation fidelity for the lossless case when the unity-gain requirement for the classical channel is not met, i.e., $\beta_c\cosh^{2}\left( r_A\right)\neq1$. 
	
	Starting with the classical teleportation scenario for which no entanglement is shared between Alice and Bob. Substituting $r_E=0$ in Eq.\,(\ref{V_tel_expanded}) gives
	
	\begin{equation}
		V^{(1)}_{\rm{tel}}=\frac{1}{4}\left[1+2\beta_c\left( G_A-1\right)  \right] I_2, \label{V_tel_c_reduced}
	\end{equation}
	
	\noindent where $G_A\equiv\cosh^{2}\left( r_A\right)$. Since  $\beta_{\rm{tel}}=c^{(3)}_0-c^{(1)}_{\rm{tel}}=\left(1-\sqrt{\beta_c}\cosh\left( r_A\right) ,1-\sqrt{\beta_c}\cosh\left( r_A\right) \right) ^{\rm{T}}$, $\Delta=0$, and $\left(V^{(3)}_{\rm{in}}+ V^{(1)}_{\rm{tel}}\right)^{-1}=I_2/\sqrt{\Lambda}$, where $\Lambda=\left[1+\beta_c\left(G_A-1 \right)  \right] ^2/4$, the classical teleportation fidelity can be written as 
	
	\begin{equation}
		F_c=\dfrac{\exp{\left[-\left|\alpha_s \right|^2 \dfrac{B\left(\beta_c,G_A \right)}{A\left(\beta_c,G_A \right)}\right]}}{2A\left(\beta_c,G_A \right) }, \label{F_c_nonunity}
	\end{equation}
	
	\noindent where $\left|\alpha_s \right|^2=n_s$, $A\left(\beta_c,G_A \right)=\left[1+\beta_c\left( G_A-1\right)  \right] /2$ and $B\left(\beta_c,G_A \right)=\beta_cG_A-2\sqrt{\beta_cG_A}+1$. 
	
	Note that in the unity-gain limit $\beta_cG_A=1$ and low coupling $\beta_c\cong0$ we recover the expected upper bound on classical teleportation fidelity for coherent states $F_c=1/2$. Also, for vacuum noise without displacement $\left|\alpha_s \right|^2=0$ and Alice's amplifier off ($G_A=1$), i.e., not performing any measurement, we get $F_c=1$ as expected. Lastly, in the high-amplification limit of the classical channel $\beta_cG_A\gg1$, the teleportation fidelity diminishes $F_c\longrightarrow0$ due to increase in the added noise originating from the amplification process.  
	
	Turning now to the case of quantum teleportation, we rewrite Eq.\,(\ref{V_tel_expanded}) as 
	
	\begin{widetext}
		
		\begin{equation}
			V^{(1)}_{\rm{tel}}=\frac{1}{4}\left[\left( 1-2\beta_c+k\right)\cosh\left( 2r_E\right)-2\sqrt{k-k\beta_c-\beta_c+\beta_c^2} \sinh\left( 2r_E\right)+k\right]I_2, \label{V_tel_expanded2}
		\end{equation}
		
	\end{widetext}
	
	\noindent and $\beta_{\rm{tel}}=c^{(3)}_0-c^{(1)}_{\rm{tel}}=\left(1-\sqrt{k} ,1-\sqrt{k} \right) ^{\rm{T}}$, where $k\equiv\beta_cG_A$. Using these relations in combination with $\left(V^{(3)}_{\rm{in}}+ V^{(1)}_{\rm{tel}}\right)^{-1}=I_2/\sqrt{\Lambda}$ and $\Delta=0$, we arrive at the following quantum teleportation fidelity 
	
	\begin{equation}
		F_q=\dfrac{\exp{\left[-\left|\alpha_s \right|^2 \dfrac{B\left(k \right) }{C\left(r_E,k \right)}\right]}}{2C\left(r_E,k \right) }, \label{F_q_nonunity}
	\end{equation}
	
	\noindent where $B\left(k \right) =\left(\sqrt{k}-1 \right)^2$ and $C\left(r_E,k \right)\equiv\sqrt{\Lambda}$ given by
	
	\begin{widetext}
		
		\begin{equation}
			C\left(r_E,k \right)=\frac{1}{4}\left[1+k+\left( 1-2\beta_c+k\right)\cosh\left( 2r_E\right)-2\sqrt{k-k\beta_c-\beta_c+\beta_c^2} \sinh\left( 2r_E\right)\right]. \label{C_r_k_expanded}
		\end{equation}
		
	\end{widetext}

	In the limit of very small coupling $\beta_c\longrightarrow0$ and large gain at Alice $G_A\gg1$, Eq.\,(\ref{C_r_k_expanded}) simplifies to 
	
	\begin{align}
		C\left(r_E,k \right)=\frac{1}{2}\left[\left( 1+k\right)\cosh^{2}\left( r_E\right)-\sqrt{k}\sinh\left( 2r_E\right)\right]. \nonumber \\ \label{C_r_k_expanded2}
	\end{align}
	
	Substituting $r_E=0$ in Eq.\,(\ref{F_q_nonunity}) and using Eq.\,(\ref{C_r_k_expanded2}) allows us to rewrite the classical teleportation fidelity in the form
	
	\begin{equation}
		F_c=\dfrac{\exp{\left[-\left|\alpha_s \right|^2 \dfrac{2\left(\sqrt{k}-1 \right)^2}{\left(1+k \right)}\right]}}{\left(1+k \right) }. \label{F_c_nonunity2}
	\end{equation}
	
	Note that for $k=1$, i.e., corresponding to unity gain in the classical channel, Eq.\,(\ref{F_c_nonunity2}) yields the expected bound on classical teleportation, i.e., $F_c=1/2$.
	
	In Fig.\,\ref{TelVackeq1andbigger}, we plot the fidelity of quantum teleportation of vacuum states as a function of $G_E$, calculated using Eq.\,(\ref{F_Uhlmann}) for multiple cases of interest. The solid curves correspond to cases where the classical channel has unity transmission $k=1$. The solid black curve corresponds to the lossless case, where the fidelity simplifies to Eq.\,(\ref{Quant_F_reduced}). The solid cyan and blue curves correspond to cases of symmetric loss of $0.1$ dB and $0.5$ dB, respectively, present between the Entangler and Alice and the Entangler and Bob. The solid magenta and red curves correspond to cases of asymmetric loss of $0.2$ dB and $1$ dB, respectively, where the loss is only present between the Entangler and Bob. Finally, the dashed blue curve corresponds to the case in which the classical channel amplifies $k>1$ (i.e., $G_A=16$ dB and $\beta_c=-15$ dB) and the two paths from the Entangler to Alice and Bob have equal losses of $0.5$ dB.
	
	By inspecting these test cases, we observe that: (1) $F_q$ is monotonic and increases with $G_E$ for unity-transmission classical channels and symmetric losses, (2) $F_q$ peaks for finite $G_E$ in the cases of asymmetric losses combined with unity-transmission classical channels or symmetric losses combined with amplifying classical channels, and (3) the maximum achievable fidelities decrease with increased losses.    
	
	Finally, in Fig.\,\ref{TelCohvsnsklarger1thy}, we illustrate the effect of the transmission magnitude of the classical channel, i.e., $k$, on the quantum and classical teleportation fidelity of coherent states as a function of their average photon number $\left|\alpha_s \right|^2$. In particular, we consider the lossless case which we have analytical expressions for. The dashed blue and red lines represent the calculated quantum and classical teleportation fidelities, corresponding to $G_E=5$ dB and $G_E=0$ dB, respectively, for the case of unity-transmission classical channel, i.e., $k=1$. As expected from Eq.\,(\ref{F_q_nonunity}) and Eq.\,(\ref{F_c_nonunity2}), the fidelities are constant and independent of $\left|\alpha_s \right|^2$. Whereas, in the amplifying case in which $k>1$, where we assume in this example $G_A=10$ and $\beta_c=1/8$, the fidelities decrease with $\left|\alpha_s \right|^2$ and their maximum values are much lower than in the $k=1$ case.

\end{document}